\documentclass[12pt]{article}
\usepackage{amssymb,amsmath}
\usepackage{epsfig}
\usepackage{epstopdf}
\usepackage{latexsym}
\usepackage{graphicx}
\usepackage{subfigure}
\usepackage{booktabs}
\usepackage{bbm}
\usepackage{enumitem}
\usepackage[numbers,compress]{natbib}

\usepackage[margin=20pt,small]{caption}

\usepackage[toc]{appendix}

\usepackage{color}
\usepackage{datetime}
\usepackage[
      colorlinks=false,
      linkcolor=darkblue,  
      urlcolor=blue,    
      filecolor=blue,     
      citecolor=red,
linktocpage=true,
      pdfstartview=FitV,
      bookmarksopen=true    
      ]{hyperref}

\ifpdf
\fi

\newcommand*{\boxedcolor}{red}
\makeatletter
\renewcommand{\boxed}[1]{\textcolor{\boxedcolor}{%
  \fbox{\normalcolor\m@th$\displaystyle#1$}}}
\makeatother

\definecolor{cardinal}{rgb}{0.6,0,0}
\definecolor{darkgreen}{rgb}{0,0.5,0}
\definecolor{golden}{rgb}{0.92, 0.7, 0}
\definecolor{midnight}{rgb}{0, 0, 0.5}
\definecolor{darkblue}{rgb}{0.2, 0, 0.8}

 \def\Im{{\rm Im}}

\def\coeff#1#2{\relax{\textstyle {#1 \over #2}}\displaystyle}
\def\ds{\displaystyle}

\def\IR{\mathbb{R}}
\def\ZZ{\mathbb{Z}}

\def\cA{{\cal A}}
\def\cB{{\cal B}}
\def\cD{{\cal D}}

\def\cK{{\cal K}}
\def\cM{{\cal M}}
\def\cN{{\cal N}}
\def\cO{{\cal O}}
\def\cP{{\cal P}}

\def\cR{{\cal R}}

\def\cV{{\cal V}}
\def\Neql#1{{\cal N}\!=\!{#1}}

\def\eql{=}

\def\eql{=}

\def\cW{{\cal W}}
\def\cL{{\cal L}}

\def\bfs#1{{\boldsymbol #1}}

\def\cals#1{\mathcal{#1}}

\def\cA{{\cals A}}

\def\oPi{\overset{_{\circ}}{\Pi}{}}

\begin{document}  

\begin{titlepage}
 \begin{flushright}
IPhT-T13/259
 \end{flushright}
 
\medskip
\begin{center} 
{\Large \bf  Supersymmetric Janus Solutions in Four Dimensions}

\bigskip
\bigskip

{\bf Nikolay Bobev,${}^{(1)}$  Krzysztof Pilch,${}^{(2)}$  and Nicholas P. Warner${}^{(2,3,4)}$ \\ }
\bigskip
${}^{(1)}$
Perimeter Institute for Theoretical Physics \\
31 Caroline Street North, ON N2L 2Y5, Canada
\vskip 5mm
${}^{(2)}$ Department of Physics and Astronomy \\
University of Southern California \\
Los Angeles, CA 90089, USA  \\
\bigskip
$^{(3)}$ Institut de Physique Th\'eorique, CEA Saclay \\
CNRS-URA 2306, 91191 Gif sur Yvette, France\\
\bigskip
$^{(4)}$ Institut des Hautes Etudes Scientifiques \\
Le Bois-Marie, 35 route de Chartres \\
Bures-sur-Yvette, 91440, France \\
\bigskip
nbobev@perimeterinstitute.ca,~pilch@usc.edu,~warner@usc.edu  \\
\end{center}

\begin{abstract}

\noindent  
\end{abstract}

\noindent We use maximal 
gauged supergravity in four dimensions to construct the gravity dual of a class of supersymmetric conformal interfaces in the theory on the world-volume of multiple M2-branes. We study three classes of examples in which the $(1+1)$-dimensional defects preserve  $(4,4)$, $(0,2)$ or $(0,1)$ supersymmetry. Many of the solutions have the maximally supersymmetric $AdS_4$ vacuum dual to the $\Neql8$ ABJM theory on both sides of the interface. We also find new special classes of solutions including one that interpolates between the maximally supersymmetric vacuum and a conformal fixed point with $\Neql1$ supersymmetry and $G_2$ global symmetry. We find another solution that interpolates between two distinct conformal fixed points with $\Neql1$ supersymmetry and $G_2$ global symmetry. In eleven dimensions,  this $G_2$ to $G_2$ solution corresponds  to a domain wall across which a magnetic flux reverses orientation. 


\end{titlepage}


\setcounter{tocdepth}{2}
\tableofcontents

\section{Introduction}

Defects in conformal field theories have long been recognized as useful probes of interesting physics and have been broadly used in all areas where CFTs are ubiquitous, ranging from condensed matter and statistical physics to particle theory. In this paper we will study, holographically,  superconformal interface defects in the maximally supersymmetric theory on the world-volume of multiple M2-branes. This theory was constructed in \cite{Aharony:2008ug} and is a Chern-Simons matter theory with two $SU(N)$ gauge groups of equal and opposite Chern-Simons levels $k$  with $N$ being the number of M2-branes.\footnote{Since we are interested in a limit where the number of M2-branes is large we will use the ABJM theory. See \cite{Bagger:2007jr,Gustavsson:2007vu} for earlier work on the problem of finding the world volume theory of multiple M2-branes.} For $k=1,2$ the theory preserves $\cN=8$ superconformal symmetry and, at large $N$, has a holographic description in terms of eleven-dimensional supergravity on the background $AdS_4\times S^7$. For $k>2$, the supersymmetry is broken to $\cN=6$ and the gravity dual background is $AdS_4\times S^7/\mathbb{Z}_k$, where the $\mathbb{Z}_k$ acts on the Hopf fiber of $S^7$ written as a $U(1)$ bundle over $\mathbb{CP}^3$. 

There are two types of codimension-one defects in conformal field theory: the ones that only support degrees of freedom present in the bulk and ones that support new degrees of freedom confined to the defect.  Here we study  the first kind of defects and refer to them as interfaces or Janus configurations.  Such Janus configurations have been constructed before for $\cN=4$ SYM theory in four dimensions. Indeed, the holographic description of Janus configurations was initiated in \cite{Bak:2003jk} where a non-supersymmetric Janus solution was constructed directly in IIB supergravity. The field theory interpretation of this interface was clarified in detail in \cite{Clark:2004sb}, and it was shown in \cite{Papadimitriou:2004rz} how to calculate correlation functions in the presence of this interface holographically. This construction was later generalized and a number of supersymmetric and superconformal interfaces in $\cN=4$ SYM were found in field theory \cite{Clark:2004sb,D'Hoker:2006uv,Gaiotto:2008sd}. 

The supergravity duals of some of these defects were constructed in \cite{Clark:2005te,D'Hoker:2006uu,D'Hoker:2007xy,Suh:2011xc}. There also had been  a number of studies on codimension-one defects in $\cN=4$ SYM which support extra degrees of freedom, see, for example, \cite{DeWolfe:2001pq} for a holographic approach to such defects and \cite{Gaiotto:2008sa,Gaiotto:2008ak} for a detailed field theory analysis. It will be very interesting to study such generalizations of our Janus configurations both from the point of view of the field theory and in supergravity. In particular the low-energy theory for the well-known M2-M5 intersection will be described by such an interface with $(4,4)$ supersymmetry.\footnote{See the recent work \cite{Berman:2009kj,Berman:2009xd,Fujimori:2010ec,Faizal:2011cd,Okazaki:2013kaa} for a discussion on supersymmetric boundary conditions with various amounts of supersymmetry in supersymmetric Chern-Simons theories coupled to matter.}  However we will not study these types of defects in the current work.

The dual gravitational description of the lowest dimension operators in the spectrum of the $\Neql8$ ABJM theory is given by $\Neql8$, $SO(8)$ gauged supergravity in four dimensions \cite{de Wit:1982ig}. Since we are interested in describing Janus configurations that support only such low-dimension ABJM operators (or degrees of freedom), this supergravity theory will be our main tool for constructing the gravity dual solutions to superconformal interfaces. We employ the usual Janus Ansatz of \cite{Bak:2003jk} with its domain-wall metric having an $AdS_3$ slicing.\footnote{See \cite{Karch:2000gx} for early work on holography for asymptotically $AdS_{D+1}$ solutions with $AdS_{D}$ slicing.} The metric of the Janus solutions is asymptotically $AdS_4$ and the only other non-trivial fields of the $\Neql8$ supergravity theory will be the scalars that vary as a function of the domain-wall radial variable. 

Using this Ansatz and solving the BPS equations of the $\Neql8$ supergravity theory we find Janus solutions  that preserve conformal invariance in $(1+1)$ dimensions and  $1/2$, $1/8$ or $1/16$ of the maximal $(8,8)$ supersymmetry. In particular we  find a Janus configuration with  $(4,4)$ supersymmetry and an $SO(4)\times SO(4)$ $\cR$-symmetry, a $(0,2)$ defect with $SU(3)\times U(1)$ global symmetry with $U(1)$ $\cR$-symmetry as well as a $(0,1)$ defect with $G_2$ global symmetry.  

Our $1/2$-BPS $SO(4)\times SO(4)$ Janus solutions can be uplifted, using existing technology, to a solution of eleven-dimensional supergravity and they represent a one-parameter generalization of the Janus solution found in \cite{D'Hoker:2009gg}.    It is also interesting to note that our more general Janus solutions have not been captured by the classification  given in  \cite{D'Hoker:2008wc,D'Hoker:2009gg}.  The detailed comparison  and the eleven-dimensional   uplift can be found in Appendix \ref{appendixB} of this paper.

The reason we restrict to the three classes of examples listed above is that all of them can be described in a unified fashion by considering consistent truncations of the maximal $\Neql8$ theory in four dimensions (which has 70 scalars) to a sector with a given global symmetry and only one scalar and one pseudoscalar that can be  combined into a complex scalar parametrizing a $SU(1,1)/U(1)$ scalar manifold. One of the features of all our Janus solutions is that they come in continuous families in which one of the parameters is the asymptotic value of the phase  of the complex supergravity scalar. This parameter is rather simple from the point of view of four-dimensional supergravity but in eleven dimensions and in the dual field theory it makes very significant changes to the physics. In eleven dimensions this phase controls the relative amount of metric deformation versus internal magnetic $3$-form flux and on the M2-brane the phase determines the  combination of fermonic and a bosonic bilinear operators that are turned on and develop a non-trivial profile in the bulk ABJM theory.

In addition to the Janus solutions discussed above we also find a holographic realization of the phenomenon of RG flow domain walls, that is, a codimension-one defect that spatially separates  two distinct superconformal fixed points related by an RG flow. See \cite{Gaiotto:2012np} and \cite{Dimofte:2013lba} for recent work on such configurations in two- and three-dimensional CFT's. The examples we present are interfaces between the maximally supersymmetric ABJM theory with $SO(8)$ global symmetry and one of two distinct $\mathcal{N}=1$ SCFT with $G_2$ global symmetry, which are related to the $SO(8)$ theory by an RG flow \cite{Bobev:2009ms}. The two distinct $\mathcal{N}=1$ SCFT are related by a reversal of the sign of the eleven-dimensional magnetic flux for their dual $AdS_4$ solutions and we also present an Janus solution which interpolates between them. On the interface all of these examples preserves $(0,1)$ superconformal symmetry and the $G_2$ global symmetry. To the best of our knowledge these are  the first examples of a holographic description of RG flow domain walls.  We plan to explore more general examples in the upcoming work \cite{BPWRG}.

Previous efforts to construct supersymmetric Janus solutions were generally made using IIB or eleven-dimensional supergravity \cite{D'Hoker:2006uu,D'Hoker:2007xy,Suh:2011xc,D'Hoker:2009gg}. The advantage of gauged supergravity is that it is extremely efficient in encoding some of the very complicated background fields of the higher-dimensional supergravity theories. As a result, it does not introduce a new level of difficulty if one wants to study superconformal defects that preserve less than half of the maximal supersymmetry since the system of BPS equations always reduces to a coupled system of ODEs for the four-dimensional metric coefficients and the supergravity scalars.  If one were to study these solutions directly in eleven-dimensional supergravity one would typically have the daunting task of solving a system of coupled, non-linear PDE's. Another advantage of the  four-dimensional description is that it should allow for more efficient calculations of correlation functions in the dual field theory in the presence of the Janus defect \cite{Papadimitriou:2004rz}. 

In the next section we first review the holographic dictionary for M2-branes and how the four-dimensional scalars of interest are embedded in eleven-dimensional supergravity. In Section \ref{Sect:JanusSols} we  summarize the basic structure of the class of supergravity truncations that we wish to study and in  Section \ref{Sect:DetailSusy} we perform the detailed analysis of the supersymmetry and derive a universal set of BPS equations for all our Janus solutions. In Section \ref{sec:SO4} we present an analytic supergravity solution corresponding to a $(4,4)$-supersymmetric interface with $SO(4)\times SO(4)$  $\cR$-symmetry. In Section \ref{sec:SU3} we find numerical solutions describing a $(0,2)$ Janus with $SU(3)\times U(1)\times U(1)$ global symmetry. We then find Janus solutions and RG flow domain walls with $(0,1)$ supersymmetry and $G_2$ global symmetry in Section \ref{sec:G2}. We conclude, in Section \ref{sec:Conclusions}, with a discussion of some problems for future study. In Appendix \ref{appendixA} we summarize various technical aspects of  $\Neql8$ supegravity and in Appendix \ref{appendixC} we discuss alternative choices for the supergravity truncations in the $SU(3)\times U(1)^2$ and $G_2$ sector and show that they do not yield Janus solutions.  Appendix~\ref{appendixB} contains details of the eleven-dimensional uplift of our $(4,4)$-supersymmetric Janus solutions and a detailed comparison with the results of  \cite{D'Hoker:2009gg}. We also show that most of our new  $(4,4)$-supersymmetric Janus solutions are not covered by the earlier classification in \cite{D'Hoker:2008wc}.

\section{The holographic dictionary and eleven-dimensional supergravity}
\label{Sect:HolDict}

Before diving into the details of the new Janus solutions it is valuable to review some of the subtleties in the holographic dictionary for the $\Neql8$ supergravity and to recall how the supergravity scalars encode different aspects of the eleven-dimensional theory.  

First, the seventy-dimensional scalar manifold of the $\Neql8$ theory consists of $35$ scalars in the ${\bf 35}_s$ of $SO(8)$ and $35$ pseudoscalars in the $\mathbf{35}_c$ of $SO(8)$.  To linear order in the $S^7$ truncation of eleven-dimensional supergravity, the former correspond to metric perturbations and the latter correspond to modes of the tensor gauge field, $A^{(3)}$.  At higher orders these modes, of course, mix through the non-linear interactions.  

The basic holographic dictionary\footnote{Here we will ignore subtle issues about monopole operators in the ABJM theory and treat them as bosonic/fermionic bilinear operators for simplicity. Alternatively one can view our discussion as applicable to the BLG theory \cite{Bagger:2007jr,Gustavsson:2007vu}.} implies that the scalars are dual to the dimension-one operators which may be thought of as bosonic bilinears of the form
\begin{equation}
O_{b}^{AB} ~=~ \text{Tr}(X^{A}X^{B}) - \frac{1}{8}\delta^{AB} \text{Tr}(X^C X^C)\;, \qquad A,B,C=1,\ldots,8\;, 
\label{Obdef}
\end{equation}
and the pseudoscalars are dual to dimension-two operators, which can be thought of as fermionic bilinears of the form
\begin{equation}
O_{f}^{\dot A \dot B}  ~=~  \text{Tr}(\lambda^{\dot A}\lambda^{\dot B}) - \frac{1}{8}\delta^{\dot A  \dot B} \text{Tr}(\lambda^{ \dot C} \lambda^{ \dot C})\;, \qquad \dot A,\dot B, \dot C=1,\ldots,8\;.
\label{Ofdef}
\end{equation}
However, as discussed in \cite{Klebanov:1999tb}, there are subtleties in this dictionary coming from the choice of how one quantizes the modes.  

The problem is how to distinguish between operator perturbations of the field theory Lagrangian and the development of vevs of the same operator. Usually  non-normalizable supergravity modes  correspond to coupling constants in perturbations of the Lagrangian of the dual theory, while normalizable supergravity modes correspond to states of the field theory, described by vevs. However, as discussed in \cite{Klebanov:1999tb}, this ``standard quantization'' does not necessarily apply in four dimensions if the scalars in the gravitational bulk theory have masses in the range $-9/4<m^2L^2<-5/4$, where $L$ is the scale of the $AdS_4$ fixed point. One can equally well choose  ``alternative quantization,''   which  reverses the standard dictionary with non-normalizable modes describing vevs and normalizable modes representing perturbations of the Lagrangian.  For the  70 scalars of the $\cN=8$ supergravity theory we have $m^2L^2=-2$ around the maximally supersymmetric vacuum dual to the ABJM theory and thus one can choose alternative quantization.  On the other hand, it was shown in  \cite{Breitenlohner:1982jf} that to preserve the supersymmetry in $\cN=4$ supergravity (and therefore to preserve the supersymmetry in $\cN=8$ supergravity) the supergravity pseudoscalars must be quantized in {\it exactly the opposite way}  to the supergravity scalars.  Thus, if the supergravity scalars follow the rules of standard quantization then the supergravity pseudoscalars must undergo alternative quantization,  and vice versa.  

As noted in  \cite{Bobev:2011rv}, there are thus two possible choices of holographic dictionary for the seventy spin-$0$ particles of supergravity but there is only one choice in which the scaling dimensions of the supergravity modes match precisely with the scaling dimensions of the operators or couplings of the dual M2-brane theory. The correct holographic dictionary is thus:
\begin{itemize}
\item The non-normalizable ($\Delta =1$)  modes of the $35$  pseudoscalars  describe fermion masses on the M2-brane  while for the $35$ scalars the $\Delta =1$ modes correspond to vevs of boson bilinears.

\item The normalizable ($\Delta =2$) modes of the $35$  pseudoscalars describe vevs of fermion bilinears on the M2-brane  while for the $35$ scalars the $\Delta =2$ modes correspond to bosonic masses.
\end{itemize}
This is the {\it only} dictionary that is consistent with the following three features of the maximally supersymmetric $AdS_4$ vacuum (where all the supergravity scalars and pseudoscalars vanish) and the Hilbert space erected on it:  a)   $\cN=8$ supersymmetry,   b)  the relationship between supergravity scalars and  bosonic couplings/vevs on the M2-brane and  supergravity pseudoscalars and fermionic couplings/vevs on the M2-brane, and c)  the scaling dimensions of supergravity fields match the scaling dimensions of dual couplings or vevs.

To summarize, suppose that the $AdS_4$ has the Poincar\'e form:
\begin{equation}
ds_{AdS_4} ~=~  \frac{1}{\rho^2}\, (-dt^2~+~ dx^2~+~ dy^2 )~+~ \frac{d\rho^2}{\rho^2}  \,.
\label{AdS4Met}
\end{equation}
Denote the  35 scalars by $\Phi_i$ and the 35 pseudoscalars by $\Psi_i$, then they will generically have the following asymptotic expansion close to the $AdS_4$ boundary at $\rho \to 0$:
\begin{equation}\begin{split}
\Phi_ i \approx \phi_i^{(v)} \rho + \phi_i^{(s)} \rho^2 + \mathcal{O}(\rho^3)~,\\
 \Psi_ i \approx \psi_i^{(s)} \rho + \psi_i^{(v)} \rho^2 + \mathcal{O}(\rho^3)~.
\end{split}
 \label{quantconds}
\end{equation}
The coefficients $\phi_i^{(v)}$ and $\phi_i^{(s)}$ are related to the vev and the source for the bosonic bilinear operator of dimension $\Delta=1$ and $\psi_i^{(v)}$ and $\psi_i^{(s)}$ are related to the vev and the source for the fermionic bilinear operator of dimension $\Delta=2$.  It should, however, be remembered that if a supergravity mode involves a non-zero, non-normalizable part ($\cO(\rho )$) then it can source the normalizable part  ($\cO(\rho^2)$) and so disentangling the independent physical meaning of the  normalizable components can be subtle and one should use holographic renormalization.  There is, of course, no such difficulty if the  non-normalizable part vanishes.

The truncations of four-dimensional supergravity that we consider here consists of a complex scalar,  $z$, in an $SU(1,1)/U(1) = SL(2,\IR)/SO(2)$ coset. The real part of $z$ is a supergravity scalar and the imaginary part of $z$ is a pseudoscalar.   Thus the real part of $z$, at linear order, encodes metric perturbations in eleven dimensions and is dual to operators of the form \eqref{Obdef} and the imaginary part of $z$, at linear order, encodes flux perturbations and is dual to a linear combination of the operators in \eqref{Ofdef}.  The precise holographic dictionary is then governed by  (\ref{quantconds}).  One of the interesting features of all our solutions is the phase of $z$ and the choice of its boundary values.  From the perspective of both eleven-dimensional supergravity and for the field theory on the M2-branes, the families of such solutions represent very different physics.

\section{The BPS defects: The family of Janus solutions}
\label{Sect:JanusSols}

\subsection{The bosonic background}

We are seeking the gravity duals of $(1+1)$-dimensional conformal defects in $(2+1)$-dimensional conformal field theories.
This means that we are looking for solutions with four-dimensional metrics that are sectioned by $AdS_3$:
\begin{equation}
ds^2 \eql e^{2A}ds_{AdS_3(\ell)} ~+~ d\mu^2\,,
\label{JanMet}
\end{equation}
with boundary conditions that produce $AdS_4$ at $\mu= \pm \infty$.  While the radius of the $AdS_3$ sections  can be scaled away, we find it convenient to have this radius appear as an explicit parameter, $\ell$.  In the Poincar\'e patch we therefore have:
\begin{equation}
ds_{AdS_3(\ell)} ~=~ e^{2r/\ell}(-dt^2~+~ dx^2)~+~ dr^2  \,.
\label{AdS3Met}
\end{equation}
Note that the metric, (\ref{JanMet}),  is precisely that of an $AdS_4$ of radius $L$ if one has: 
\begin{equation}
A  ~=~ \log\Big(\frac{L}{\ell}  \cosh \Big(\frac{\mu}{L} \Big)\Big) \,.
\label{Aasymp}
\end{equation}
This will therefore determine the boundary conditions at $\mu = \pm \infty$.

Since we are working in gauged supergravity, the only other non-trivial aspect to the background will be scalar fields in the four-dimensional theory.  Furthermore, we restrict to sectors of gauged $\Neql8$ supergravity that are invariant under some group $G \subset SO(8) \subset E_{7(7)}$ and we choose this invariance group, $G$, so that it only commutes with  an $SL(2, \IR)/SO(2)$ coset in $E_{7(7)}/SU(8)$.   There are three intrinsically different   possibilities for such an embedding and these are described in Section \ref{E7embed}.  Here we simply use the $SL(2, \IR)$ structure and the fact that the embedding is characterized by a positive integer, $k$, known as the embedding index.

Our scalar sub-sector thus always reduces to $SL(2, \IR)/SO(2) =SU(1, 1)/U(1)$, which we can parameterize by
\begin{equation}
g ~=~   \exp\left(
\begin{array}{cc} 
0 & \alpha\, e^{i\zeta} \\  \alpha\, e^{-i\zeta} & 0 
\end{array} \right)  ~=~  \left( 
\begin{array}{cc} 
\cosh \alpha &\sinh \alpha \, e^{i\zeta} \\  \sinh \alpha \, e^{-i\zeta} &  \cosh \alpha
\end{array} \right)   \,,
\label{gmat}
\end{equation}
for some real variables $\alpha$ and  $\zeta$ with $\alpha \ge 0$, $-\pi \le \zeta <  \pi$ .   The kinetic term, $\cA$,  and the  composite  $U(1)$ connection, $\cB$,  are then given by 
\begin{equation}
g^{-1} dg ~=~  \left( 
\begin{array}{cc} 
\cB & \cA \\  \bar  \cA  &  -\cB
\end{array} \right)  ~=~  \left( 
\begin{array}{cc} 
i \sinh^2 \alpha \, d \zeta &(d \alpha  + \frac{i}{2} \,\sinh 2 \alpha \,  d \zeta)\, e^{i\zeta}  \\  (d \alpha  - \frac{i}{2} \,\sinh 2 \alpha\,  d \zeta)\, e^{-i\zeta}  &  -i \sinh^2 \alpha \,  d \zeta 
\end{array} \right)  \,.
\label{kinmat}
\end{equation}
The standard normalized scalar kinetic term in the Lagrangian is then $\frac{1}{2} |\cA|^2$.

In the foregoing discussion we used the $SL(2, \IR)$ group element, $g$, in one copy of the fundamental representation.   However in  $\Neql 8$ supergravity the kinetic term is normalized based upon the fundamental representation of $E_{7(7)}$ and this will generically decompose into larger representations of  $SL(2,\IR) \subset E_{7(7)}$.   The index of the representation\footnote{See, for example, \cite{McKayPatera, Schellekens:1986mb}.} gives the embedding index or winding number, $k$,  that multiplies both the canonically normalized $SL(2,\IR)$ kinetic term, $\cA$, as well as the connection, $\cB$, that arise from the corresponding canonically normalized $E_{7(7)}$ terms.  Thus we will find this (positive) integer consistently arising throughout our discussions of various embeddings.   The complete scalar Lagrangian also involves a scalar potential inherited from the potential of the $\cN=8$ theory and this, of course, depends upon the details of the embedding of $SL(2,\IR)$ in $E_{7(7)}$.  

The Lagrangian can be conveniently described by parametrizing everything in complex variables.   Indeed, the coset $SL(2, \IR)/SO(2)$ is a K\"ahler manifold  with canonical complex coordinate, $z$,  defined by
\begin{equation}
z ~=~ \tanh \alpha\, e^{i\zeta}  \,.
\label{zparam}
\end{equation}
The scalar Lagrangian is then parametrized by a K\"ahler potential, $\cK(z, \bar z)$, and a holomorphic superpotential, $\cV(z)$.  Specifically, the Lagrangian of the theories of interest can be expressed in the form:\footnote{All models we consider arise as truncations of the $\mathcal{N}=8$ supergravity. However, it should be possible to rewrite them as four-dimensional, $\mathcal{N}=2$ gauged supergravity theories. This underpins the holomorphic structure that we are exploiting.}
\begin{equation}
e^{-1}{\cal L}\ ~=~ \coeff{1}{2}\, R  ~-~ g^{\mu \nu} \cK_{z\bar{z}} \partial_\mu z  \, \partial_\nu \bar{z}~-~ g^2\mathcal{P} (z, \bar z )\,,
\label{genLag}
\end{equation}
where $g$ is the coupling constant of the gauged supergravity and
\begin{equation}
\mathcal{K}_{z\bar{z}} ~=~  \partial_{z}\partial_{\bar{z}}\mathcal{K}\;.
\label{derivK}
\end{equation}
We will also  define $\mathcal{K}^{z\bar{z}}$ to be the inverse of $\mathcal{K}_{z\bar{z}}$.    The potential, $\mathcal{P}(z,\bar{z})$, can be obtained from a holomorphic superpotential, $\cV(z)$, via:
\begin{equation}
\mathcal{P} = e^{\mathcal{K}}(\mathcal{K}^{z\bar{z}}\nabla_{z}\mathcal{V}\nabla_{\bar{z}}\overline{\mathcal{V}} -3 \mathcal{V}\overline{\mathcal{V}})\;,
\label{Pform}
\end{equation}
where the covariant derivatives are defined in the usual way:
\begin{equation}
\nabla_{z}\mathcal{V} = \partial_{z}\mathcal{V} + (\partial_{z}\mathcal{K})\mathcal{V}\;, \qquad \nabla_{\bar{z}}\overline{\mathcal{V}} = \partial_{\bar{z}}\overline{\mathcal{V}} + (\partial_{\bar{z}}\mathcal{K})\overline{\mathcal{V}}\;.
\end{equation}
For the  $SL(2, \IR)/SO(2)$ coset we have
\begin{equation}
\mathcal{K} = -k\, \log(1-z\bar{z})\;, 
\label{Kahlerpot}
\end{equation}
where $k \in \ZZ_+$ is the embedding index of the $SL(2, \IR)$ in the $E_{7(7)}$ of $\Neql8$ supergravity.  Thus the scalar kinetic term is given by the canonical sigma-model expression:
\begin{equation}\label{Kform}
\mathcal{K}_{z\bar{z}} ~=~ \dfrac{k}{(1-z\bar{z})^2}\;.
\end{equation}
As we will see, the holomorphic superpotential, $\cV(z)$, is generically a polynomial of degree $k$, or less.

\subsection{The $SL(2, \IR)$  embeddings in $E_{7(7)}$ defined through invariance}
\label{E7embed}

Underlying our Janus solutions are consistent truncations of $\Neql8$ supergravity down to the scalar coset $SL(2, \IR)/SO(2)$.  As we remarked earlier, we will find all such truncations that arise from $G$-invariant sectors of the $\cN=8$ theory where $G \subset SO(8) \subset E_{7(7)}$ and so we require that $G$ only commute with $SL(2,\IR)$ in the  $E_{7(7)}$.  Once one has found the subgroup $G$ it will generically be contained in a larger, possibly non-compact group, $\widehat G$ so that $\widehat G \times SL(2,\IR)$ is a maximal embedding in   $E_{7(7)}$.  Such maximal embeddings are well-known and, for example, a list may be found in \cite{Schellekens:1986mb}.  The complete list with $SL(2,\IR)$ factors is
\begin{itemize} 
\item[(i)]  $ (SO(4) \times SO(4)) \times SL(2,\IR) ~\subset~ SO(6,6)  \times SL(2,\IR)~\subset~ E_{7(7)}$\,,  \  with  $k =1$
\item[(ii)]  $ (SU(3) \times U(1)\times U(1))  \times SL(2,\IR) ~\subset~ F_{4(4)} \times SL(2,\IR)~\subset~ E_{7(7)}$\,,  \  with   $k =3$
\item[(iii)]  $ (G_2) \times SL(2,\IR) ~\subset~ G_2 \times SL(2,\IR)  ~\subset~ E_{7(7)}$\,, \ with   $k =7$
\end{itemize} 
where the first group in parenthesis defines $G \subset SO(8)$ and the second inclusion defines $\widehat G$.  The integer, $k$, is the embedding index of the $SL(2, \IR)$ factor.  

We thus have three distinct classes of models that we discuss systematically in the subsequent sections.  These three consistent truncations have been considered before but not in the context of Janus solutions.  Holographic flows of (i), and their eleven-dimensional uplifts, were extensively analyzed in \cite{Pope:2003jp}.  The $SU(3)$-invariant sector has been studied in many papers  \cite{Bobev:2009ms,Warner:1983vz,Ahn:2000mf,Ahn:2000aq,Ahn:2001by,Ahn:2001kw,Ahn:2002eh,Ahn:2002qga,Bobev:2010ib} and one can obtain  (ii) and (iii) through further truncations of this sector.   However, unlike some of the earlier analysis of such truncations, here  we do not necessarily restrict ourselves to $G$-invariant supersymmetries and consider the more general possibility of supersymmetries  that transform in a non-trivial representation, $\cR_{\epsilon}$, of $G$. 
  
All of these $SL(2, \IR)/SO(2)$ embeddings in $E_{7(7)}/SU(8)$ have a very important feature:  The $SO(2)$ generator lies in the purely imaginary part of $SU(8)$ which means that it is {\it not} generically a symmetry of the gauged theory and that it rotates between the scalar and pseudoscalar sectors of the $\Neql{8}$ supergravity theory.  Thus our complex scalar, $z$, has a real part that is a supergravity scalar and an imaginary part that is a pseudoscalar.  In the UV limit of the holographic dual theory the real part of $z$ therefore encodes details of a boson bilinear and the imaginary part of $z$ encodes a fermion bilinear.  

As described in Section \ref{Sect:HolDict}, the  action of the $SO(2)$ is very interesting from the perspective of the holographic  field theory in that in the UV it interpolates between bosonic and fermionic  bilinears and thus changes the physics underlying the entire flow.  Similarly, in eleven-dimensional supergravity,  the $SO(2)$ action interpolates between metric fields and $3$-form fluxes and so, once again, changing the phase of $z$ makes dramatic changes in the boundary conditions and overall structure of the eleven-dimensional solution.  Indeed, it was this observation that was a major motivation for the analysis in \cite{Pope:2003jp,Bena:2004jw}.

\section{Solving the BPS equations for $G$-invariant Janus solutions}
\label{Sect:DetailSusy}

We now take the general supersymmetry structure of the $\cN=8$ theory and make the detailed reduction to the class of truncations described in Section \ref{E7embed}. 

\subsection{Some supergravity preliminaries}
\label{gensusy}

Our metric has ``mostly plus'' signature and the gamma matrices are defined by $\{ \gamma^a \,, \gamma^b \} = 2\eta^{a b}$ where $\eta = {\rm diag} (- 1,+1,+1,+1)$.  Thus $\gamma^a$, $a=1,2,3$ will be hermitian and $\gamma^0$ is anti-hermitian. We choose an explicit Majorana representation in which the $\gamma^a$, $a=0,1,2,3$ are real and in this representation, the helicity projector, $\gamma_5$, is purely imaginary and anti-symmetric.

Following the standard practice in four dimensions, spinors will be written in terms of the chiral projections of the corresponding Majorana spinors as  described  in \cite{deWit:1978sh}.  For example:
\begin{equation}\label{chirdefs}
\begin{split}
\epsilon^i & ~\equiv~  \coeff 1 2\,(1+\gamma_5){ \epsilon}^i_M\,,\qquad 
\epsilon_i  ~\equiv~  \coeff 1 2\,(1-\gamma_5){  \epsilon}^i_M\,,\\
\bar \epsilon^{\,i} & ~\equiv~ \coeff 1 2\,{  \bar\epsilon}^i_M(1+\gamma_5)\,,\qquad 
\bar \epsilon_{i}   ~\equiv~  \coeff 1 2\, {\bar\epsilon}^{\,i}_M(1-\gamma_5)\,,\\
\end{split}
\end{equation}
where $\epsilon^i_M$, $i=1,\ldots,8$,  are the underlying Majorana spinors.  Since $\gamma_5$ is purely imaginary in this Majorana representation,  complex conjugation raises and lowers the $SU(8)$ indices of the $\Neql8$ theory.

The  supersymmetry variations of the 8 gravitinos and the 56 gauginos in the $\cN=8$ theory are given by \cite{de Wit:1982ig}
\begin{equation}
\delta\psi_\mu{}^i \eql 2 D_\mu\epsilon^i ~+~ \sqrt {2} \, g\,A_1{}^{ij}\gamma_\mu\epsilon_j\,, 
\label{deltagravitino}
\end{equation}
and
\begin{equation}
\delta\chi^{ijk} \eql -\cals A_\mu{}^{ijkl}\,\gamma^\mu\,\epsilon_l  ~-~  2\, g\, A_{2\,l}{}^{ijk}\epsilon^l\,,
\label{deltagaugino}
\end{equation}
respectively. The definitions of the various $E_{7(7)}$ tensors above are summarized in Appendix~\ref{appendixA}.

 Since we are considering backgrounds that are invariant under some subgroup, $G \subset SO(8)$, the unbroken supersymmetries will lie in some representation, $\cR_{\epsilon}$, of $G$.  We will denote the helicity components, $\epsilon^l$ and $\epsilon_l$, in  $\cR_{\epsilon}$ generically by $\epsilon$ and $\epsilon^*$ respectively and since the $SO(8)$ has a real action on $\epsilon^l$ and $\epsilon_l$, both sets of helicity components must  transform in the same $SO(8)$ representation.

Our task will ultimately be to solve the BPS conditions $\delta\psi_\mu{}^i=0$ and $\delta\chi^{ijk} =0$ within the truncations of interest. We will do this in detail below. As often happens with the BPS equations, we find that the solutions also automatically solve the equations of motion.

\subsection{The gaugino variation}

The fields are assumed to be invariant under the $SO(2,2)$ action on the $AdS_3$ and so the scalars can only depend upon the coordinate $\mu$ in (\ref{JanMet}).   This means that the vanishing of the gaugino variation (\ref{deltagaugino}) only involves $\gamma^3$ and can be generically re-written as: 
\begin{equation}
\gamma^3\,\epsilon\eql M\,\epsilon^*\,,\qquad \gamma^3\,\epsilon^*\eql  M^*\,\epsilon\,,
\label{gauginovar}
\end{equation}
where we have used the reality of $\gamma^3$.   In particular, this implies $MM^* = 1$ and hence we have 
\begin{equation}
M \eql e^{i \Lambda}\,,
\label{MLambda}
\end{equation}
for some real phase, $\Lambda$. We can therefore define $\varepsilon$ by 
\begin{equation}
\epsilon   \eql e^{i \Lambda/2} \,\varepsilon  \,,
\label{varespsdefn}
\end{equation}
and then we have
\begin{equation}
\gamma^3\varepsilon\eql\varepsilon^*\,,\qquad \gamma^3\varepsilon^*\eql\varepsilon\,.
\label{varepsproj}
\end{equation}

Explicitly,  multiplying (\ref{deltagaugino}) by $\gamma^3$ we find that the quantity $M$ is given by:  
\begin{equation}
 M  ~=~    \big(g \,\cK^{z\bar{z}} \, e^{\cK/2} \, \nabla_z \cV \big)^{-1}\, \bar{z}'    \,, 
\label{Mexplicit}
\end{equation}
and so (\ref{MLambda}) implies:
\begin{equation}
\label{eqzdermag}
z'\bar{z}' ~=~   g^2 \, (\mathcal{K}^{z\bar{z}})^2\, e^{\mathcal{K}}\, \nabla_{z}\mathcal{V}\, \nabla_{\bar{z}}\overline{\mathcal{V}}\;.
\end{equation}
%

\subsection{The gravitino variation}
\label{Sect:Gravvar}

In looking for the Poincar\'e supersymmetries parallel to the $(1+1)$-dimensional flat sections of the $AdS_3$ metric (\ref{AdS3Met}), we assume that the supersymmetries are independent of $t$ and $x$.  This means that the spin-$\frac{3}{2}$ variations along $t$ and $x$ reduce to
\begin{equation}
\Big(A'\,\gamma^3\,+{1\over\ell}\, e^{-A}\,\gamma^2 \Big)\,\epsilon+g\,\overline {\cals W }\,\epsilon^*\eql 0\,, 
\label{gravvar}
\end{equation}
where ${\cals W}$ is the appropriate eigenvalue of $\frac{1}{\sqrt{2}} A_{1\,ij}$. 
Indeed, ${\cals W }$ is related to the holomorphic superpotential via:
\begin{equation}
{\cals W } \eql  e^{{\cK}/2} \, \cV\,.
\label{WVreln}
\end{equation}

Taking the complex conjugate of (\ref{gravvar}) and iterating, one obtains the quadratic constraint:
\begin{equation}
(A')^2\eql - {1\over\ell^2} \, e^{-2A}+g^2\,|\cals W|^2\,.
\label{quadconstr}
\end{equation}
However, the two projection conditions (\ref{gauginovar}) and (\ref{gravvar}) must be compatible with one another.  In particular, one can use  (\ref{gauginovar}) to eliminate $\gamma^3 \epsilon$ in favor of $\epsilon^*$ and obtain a projection condition solely involving $\gamma^2$, which must have the form:
\begin{equation}
\gamma^2 \epsilon\eql i\,\kappa\,e^{i \Lambda}\, \epsilon^*  \quad \Leftrightarrow \quad \gamma^2 \varepsilon\eql i\,\kappa\,\varepsilon^*\,,\qquad |\kappa | \eql 1\,.
\label{3proj}
\end{equation}
After using (\ref{varepsproj}) in this projection condition one finds that compatibility ($\gamma^2 \gamma^3 = - \gamma^3 \gamma^2$) requires:
\begin{equation}
\kappa^2 \eql 1\,.
\label{projcomp}
\end{equation}
Explicitly,  using (\ref{3proj}) and (\ref{gauginovar})  in (\ref{gravvar})  we find: 
\begin{equation}
\Big(A' ~+~ {i \kappa \over\ell}\, e^{-A} \Big)\,e^{i \Lambda}   \eql  - g\,\overline {\cals W }   \eql - g\, e^{{\cK}/2} \, \overline {\cV}   \,, 
\label{Lambdaexplicit}
\end{equation}
which provides a ``square root'' of (\ref{quadconstr}).  In particular, note that we now know that $\kappa = \pm 1$ and is thus a constant.

The variation along the $AdS_3$ radial direction is
\begin{equation}\label{varr}
2\,\partial_r\epsilon+A'\,e^A\,\gamma^2\gamma^3\,\epsilon+g\, e^A \,\overline{\cals W}\,\gamma^2\epsilon^*\eql 0\,.
\end{equation}
Using \eqref{gravvar}, this reduces to  
\begin{equation}\label{derreps}
2\,\partial_r\epsilon\eql {1\over \ell}\,\epsilon\,,
\end{equation}
and is solved by
\begin{equation}
\label{epsrdep}
\epsilon\eql e^{r/2\ell}\,\tilde \varepsilon\,,
\end{equation}
where $\tilde \varepsilon$ is independent of $r$.

Finally, in general one knows that $\bar \epsilon \gamma^\mu \epsilon$ is a timelike (or null) Killing vector and so consistency with  (\ref{JanMet}),  (\ref{AdS3Met}), (\ref{varespsdefn}) and (\ref{epsrdep}) means that we must have
\begin{equation}\label{varmu}
\epsilon\eql e^{(A(\mu)+ r/\ell + i \Lambda)/2}\,\varepsilon_0\,, 
\end{equation}
where $\varepsilon_0$ could have a phase that depends upon $\mu$.  Explicit calculations in each example show that the phase dependence of $\epsilon$ is determined precisely by $\Lambda$ in (\ref{MLambda}) and (\ref{varespsdefn}) and thus $\varepsilon_0$ is simply a constant spinor satisfying:
\begin{equation}
\gamma^3\varepsilon_0 \eql\varepsilon_0^*\,,\qquad \gamma^2\varepsilon_0 \eql  i\, \kappa\, \varepsilon_0^* \,,
\label{vareps0proj}
\end{equation}
as a consequence of (\ref{varepsproj}) and (\ref{3proj}).

\subsection{The supersymmetries}
\label{Sect:susies}

As we remarked earlier,  the unbroken supersymmetries will lie in some representation, $\cR_{\epsilon}$, of $G$ and $\epsilon$ and $\epsilon^*$  respectively denote the helicity components, $\epsilon^l$ and $\epsilon_l$,  of any spinor in  $\cR_{\epsilon}$.  The elements of $\cR_{\epsilon}$ can be distinguished by comparing (\ref{gravvar}) and (\ref{deltagravitino}):  The supersymmetries are then simply determined by the space of $\epsilon_j$ upon which  $A_1^{ij}$ has the eigenvalue $\frac{1}{\sqrt{2}} \overline{\cW}$.  This determines the number, $\widehat \cN$, of supersymmetries, $\epsilon_j$, that go into the foregoing calculation.  However, it is still possible for $\cR_{\epsilon}$ to be a reducible representation of $G$ and  the phases $e^{i \Lambda}$ and $\kappa$ can differ between  irreducible components of $\cR_{\epsilon}$.  For the present we will assume that are dealing with $\widehat \cN$ supersymmetries in one irreducible component of $\cR_{\epsilon}$ and hence $e^{i \Lambda}$ and $\kappa$ are the same for all  $\widehat \cN$ supersymmetries.  We will return to this issue in Section \ref{sec:SO4} where  $\cR_{\epsilon}$ will have two irreducible components.

The supersymmetric Janus solutions require that we impose the additional conditions (\ref{gauginovar}) and (\ref{3proj}).  These each cut the four independent (real, Majorana) components down by half, leaving a single real component.  In particular, these constraints imply 
\begin{equation}
\gamma^2 \gamma^3 \varepsilon  \eql  -i\, \kappa\,\varepsilon \,. 
\label{helproj}
\end{equation}
However, since  $\epsilon$ represents some set of $\epsilon^l$,  the helicity condition (\ref{chirdefs}) implies that 
\begin{equation}
\gamma_5 \,  \varepsilon  \eql  \varepsilon  \qquad \Rightarrow \qquad \gamma_5 \,  \varepsilon^*  \eql  - \varepsilon^* \,. 
\label{helcond}
\end{equation}
 and since  $\gamma_5 = i\gamma^0 \gamma^2 \gamma^2 \gamma^3$,   (\ref{helproj}) implies that the spinors are projected onto $(1+1)$-dimensional chiral components: 
\begin{equation}
\gamma^0 \gamma^1 \varepsilon  \eql  \kappa\,\varepsilon  \qquad \Rightarrow \qquad \gamma^0 \gamma^1 \varepsilon^*  \eql  \kappa\,\varepsilon^*  \,,
\label{chirproj}
\end{equation}
where we have again used the reality of the $\gamma^a$.  The  conditions (\ref{gauginovar}) and (\ref{3proj}) thus serve to impose the Majorana condition in  $(1+1)$ dimensions and so the four real components of $\epsilon$ are reduced to a single, real component of definite chirality (\ref{chirproj}), determined by $\kappa$,  in $(1+1)$ dimensions.  The theory on the interface thus has $(\widehat \cN,0)$ supersymmetry  for $\kappa= +1$ and $(0,\widehat \cN)$ supersymmetry. for $\kappa= -1$.

As we will see in Section \ref{Sect:JanusEqns}, the choice of $\kappa$ enters directly into the BPS equations underlying the Janus solution and once a choice has been made and a solution has been constructed, the helicity of the supersymmetries of that solution is fixed.  This observation becomes particularly important when $\cR_{\epsilon}$ has more than one irreducible piece.

\subsection{The Janus BPS equations}
\label{Sect:JanusEqns}

Taking the real and imaginary parts  of (\ref{Lambdaexplicit}) one obtains:
\begin{align}
A' \eql& -  \coeff{1}{2}\,g\, e^{{\cK}/2} \, \left( e^{i\Lambda}\, \cV ~+~ e^{-i\Lambda}\, \overline {\cV}  \,\right) \,,
\label{dAsusy1} \\[6 pt]
e^{-A} \eql& - \coeff{1}{2}\,i \kappa \, g\,\ell\, e^{{\cK}/2} \, \left( e^{i\Lambda}\,\cV  ~-~ e^{-i\Lambda}\, \overline {\cV} \,\right)  \,.
\label{eAres1}
\end{align}
We can now use  (\ref{Lambdaexplicit}) to eliminate $M = e^{i \Lambda}$ in (\ref{Mexplicit})  to obtain the BPS equations for the scalars:
\begin{equation} \label{BPSz}
\begin{split}
z' & ~=~  -\mathcal{K}^{z\bar{z}}\,  \big( \overline{\cV}^{-1}  \nabla_{\bar{z}} \overline{\cV} \big)  \,   \left(A' + i \, \kappa \,\dfrac{e^{-A}}{\ell}\right)\,, \\ 
\bar{z}' & ~=~ -\mathcal{K}^{z\bar{z}}\,   \big( \cV^{-1}  \nabla_{z} \cV \big)  \,   \left(A' - i \, \kappa  \, \dfrac{e^{-A}}{\ell}\right) \;.
\end{split}
\end{equation}
These four equations represent a first-order system for the four unknown quantities $z(\mu)$, $\bar z(\mu)$,  $A(\mu)$ and  $\Lambda(\mu)$.  

Note that this shows that the supersymmetric $AdS_4$ critical points are determined by:
\begin{equation} 
\nabla_{z} \cV ~=~0\;.
\label{susycrit}
\end{equation}
Moreover, because $\cV$ is holomorphic and $\cK$ is real,  if $z_0$ satisfies (\ref{susycrit}) then so does  $\bar{z}_0$.  Thus if $z_0$ has a non-trivial imaginary part, then the supersymmetric critical point  comes in a pair related by $z_0 \to {\bar z}_0$.  We will see an example of this in Section \ref{G2truncation}.  In terms of eleven-dimensional supergravity, this complex conjugation corresponds to reversing the sign  of the internal (magnetic) components of the tensor gauge field, $A^{(3)}$. This can be explicitly demonstrated within the $G_2$ truncation as well as for the $SO(4)\times SO(4)$ one, see \eqref{fourflux} and \eqref{intflux}.\footnote{More generally, it is clear at linear order in the consistent truncation and at non-linear order it holds because both the pseudoscalars and the internal components of $A^{(3)}$ are  odd under the parity symmetry that flips all the internal coordinates.} 

One can eliminate $\Lambda$ from  (\ref{dAsusy1}) and  (\ref{eAres1}) and rederive  (\ref{quadconstr}).  One can then view (\ref{BPSz}) and (\ref{quadconstr})  as three equations for the three physical quantities $z(\mu)$, $\bar z(\mu)$ and $A(\mu)$.  
One can easily show that  for any holomorphic superpotential, $\cV$, these  BPS equations imply the equations of motion  derived from the action (\ref{genLag}), or (\ref{simpLag}).

\subsection{The general behavior of the Janus solutions} 
\label{FOsystem}

The first order system, \eqref{BPSz}, can be given a more intuitive form if one writes it  in terms of the real fields, $\alpha(\mu)$ and $\zeta(\mu)$, and the real superpotential, $W$, defined by
\begin{equation}
W^2 ~\equiv~   | \cW|^2 ~=~ e^\cK \, |\cV|^2 \;.
\label{normW} 
\end{equation}
One can then express the potential as
\begin{equation}
\label{Simppot}
\cals P\eql  \frac{1}{k}\,\bigg[ \left({\partial W \over\partial\alpha}\right)^2+ \frac{4}{\sinh^{2}(2\alpha)}\left({\partial W\over\partial\zeta}\right)^2\bigg] ~-~ 3 \, W^2 \,,
\end{equation}
where $k$ is the embedding index that appears in the normalization of the K\"ahler form \eqref{Kform}. The scalar BPS equations may then be written:
\begin{align}
\alpha' & \eql -{1\over k}\,\bigg({A'\over W}\bigg) \,{\partial W\over\partial\alpha}~+~{2 \, \kappa \over k}\, \bigg({e^{-A}\over\ell}\bigg) \,{1\over\sinh(2\alpha) }\,\frac{1}{W}\,{\partial W\over\partial\zeta}\,,\label{simpalphaeqn} \\[10 pt]
\zeta' & \eql-{4\over k}\,\bigg({A'\over W}\bigg) \,{1 \over\sinh^2(2\alpha) }\,{\partial W\over\partial\zeta} ~-~ {2\, \kappa \over k}\, \bigg({e^{-A}\over\ell}\bigg) \,{1\over\sinh(2\alpha)}\,\frac{1}{W}\,{\partial W\over\partial\alpha} \label{simpzetaeqn}  \,.
\end{align}
These scalar equations must be solved together with \eqref{quadconstr}, which in the real notation reads:\footnote{Similar BPS equations for holographic domain walls with curved slices were written down in \cite{Clark:2005te,LopesCardoso:2001rt,LopesCardoso:2002ec}.}
 \begin{equation}
(A')^2\eql g^2\,W^2 ~-~{e^{-2A}\over\ell^2} \,.
\label{Aprimeeqn}
\end{equation}

The fact that \eqref{Aprimeeqn}  is quadratic in $A'$ means that the solution may have a branch cut ambiguity when $A'(\mu) = 0$.  We will see  that the interesting Janus solutions do indeed move across these branches in the ``center'' of the solution:  In particular, we will see that the interesting solutions have $A'(\mu) = \pm c_{\pm}$, where $c_{\pm} >0$, as $\mu  \to  \pm \infty$.

Observe that if one takes the limit $\ell \to \infty$, in which the $AdS_3$ sections (\ref{AdS3Met}) become flat,  then the BPS equations become the familiar steepest descent equations of holographic RG flows (see, for example, \cite{Freedman:1999gp}). Note that in this limit (\ref{Aprimeeqn}) yields $A' = \pm W$ and this sign ambiguity is transmitted to (\ref{simpalphaeqn}) and (\ref{simpzetaeqn}).  This sign choice is then determined in holographic RG flows by boundary conditions. Notice also that in the limit $\ell \to \infty$ there is a simplification in solving the system of equations \eqref{simpalphaeqn}, \eqref{simpzetaeqn}, and \eqref{Aprimeeqn}. The equations for the scalars \eqref{simpalphaeqn} and \eqref{simpzetaeqn} form a closed system which one can integrate and only after that solve \eqref{Aprimeeqn} for $A$.

In Janus solutions we typically want to start from asymptotically $AdS_4$ boundary conditions, which means we start and finish at some critical points of $W$ near which  $e^{A(\mu)}$ is very large and positive.   Since $W$ is manifestly positive this means that we must correlate $A' = \pm W$ as $\mu \to \pm \infty$ and then we have:
\begin{equation}
\alpha'  \eql  \mp {1\over k}  \,{\partial W\over\partial\alpha} \,, \qquad 
\zeta' \eql \mp{ 4 \over k \, \sinh^2(2\alpha) }\,{\partial W\over\partial\zeta} \,.
\end{equation}
This means that near $\mu =- \infty$ the solution starts as a steepest ascent from a critical point and then as $\mu \to +\infty$ the flow changes to steepest descent into another, or possibly the same, critical point.  Indeed, we will typically start and finish at the same critical point and as the solution ascends out of that point the second terms in  (\ref{simpalphaeqn}) and (\ref{simpzetaeqn}) start to play a role and the solution begins to loop around in the $(\alpha, \zeta)$ plane and at some point $A'$ passes through zero onto the other branch of  the $A'$ equation and the evolution starts descending back to the critical point.

In the study of holographic RG flows, it was found that there were flows to ``Hades''  \cite{Freedman:1999gp} in which  either the scalar fields ran off to infinite values of $\cP$, or the metric function $A(\mu)$ diverged at some finite value of $\mu$. It was subsequently shown in \cite{Freedman:1999gk,Gubser:2000nd} that many of these flows to ``Hades'' had a simple physical interpretation in terms of a flow to the Coulomb branch in the dual field theory while others represented unphysical singularities \cite{Gubser:2000nd}.   Here we also find that some of the Janus solutions involve flows to points at which $A(\mu)$ diverges and solutions with similar properties were found in  \cite{Gutperle:2012hy}.  It is possible that these might represent  conformal interfaces between Coulomb branches and other phases of the theory on the M2-branes. This certainly deserves investigation but it will probably require the construction of the eleven-dimensional uplift.  For the present we will confine our attention to regular Janus solutions that start and finish at conformal fixed points, for which the physical interpretation is much clearer.

\section{The $SO(4)\times SO(4)$-invariant Janus}
\label{sec:SO4}

\subsection{The truncation}

The $SO(4)\times SO(4)$ invariant truncation of $\Neql8$ supergravity was discussed extensively in \cite{Pope:2003jp}.  The non-compact generators of the  $SL(2,\IR) \subset  E_{7(7)}$ are defined by:
\begin{equation}
\Sigma_{IJKL} ~\sim~ \big(\, z \,  \delta^{1234}_{[IJKL]} ~+~ \bar z \,\delta^{5678}_{[IJKL]} \,\big)\,,
\label{E7gens1}
\end{equation}
and the embedding index is equal to unity: $k=1$.    The $SO(2)$ or $U(1)$ action is simply the $SU(8)$  transformation:
\begin{equation}
U ~=~  {\rm diag}\,(e^{i \beta}, e^{i \beta}, e^{i \beta}, e^{i \beta}, e^{-i \beta}, e^{-i \beta}, e^{-i \beta}, e^{-i \beta})\,,
\label{SU8a}
\end{equation}
which rotates $z$ by the phase $e^{4i \beta}$.  

The scalar potential is given by 
\begin{equation}
\mathcal{P} \eql -2\, (2 + \cosh 2 \alpha) ~=~ -\frac{2\,(3-  |z|^2)}{1-|z|^2}\;.
\label{PSO4}
\end{equation}
{\it A priori} one does not expect (\ref{SU8a}) to generate a symmetry of the action but in this instance it does and given the consequences in eleven-dimensions this is a very surprising symmetry \cite{Pope:2003jp}.

The effective particle action that encodes all field equations is:
\begin{align}
\cals L ~=~  & e^{3A}\left[3 (A')^2-(\alpha')^2-{1\over 4}\sinh^2(2\alpha)(\zeta')^2+2g^2(2+\cosh(2\alpha))
\right]-{3\over\ell^2}e^A \\
~=~ & e^{3A}\left[3 (A')^2 ~-~ \dfrac{z'\bar{z}'}{1-z\bar{z}} ~+~ 2g^2\,\Big(\dfrac{3-|z|^2}{1-|z|^2}\Big)
\right]-{3\over\ell^2}e^A\,,
\label{EffLag1}
\end{align}
where we have used the K\"ahler potential (\ref{Kahlerpot}) with $k=1$.

The holomorphic superpotential is extremely simple:
\begin{equation}
\mathcal{V} ~=~ \sqrt{2} \qquad \Rightarrow \qquad \cW \eql \sqrt{\frac{2}{1-|z|^2}} \,.
\label{Vres1}
\end{equation}
At the $SO(8)$ critical point one finds
\begin{equation}
\nabla_{z}\mathcal{V} |_{SO(8)} = 0\;.
\end{equation}
There are no other critical points of the potential or the superpotential in this truncation.

In the $\cN=8$ theory,  the eight gravitinos and the supersymmetry parameters, $\epsilon^i$, transform in the $\bfs 8$ of $SO(8)$,\footnote{We have already adopted a convention for the $SO(8)$ representation of the scalars to be $\bfs 35_s$ and pseudoscalars to be $\bfs 35_c$. This implicitly means that the $\epsilon^i$ transform in the $\bfs8_v$.  One can, of course, permute all of this by triality.} which decomposes into $(\bfs 4,\bfs 1)+(\bfs 1,\bfs 4)$ under $SO(4)\times SO(4)$.
As noted in \cite{Pope:2003jp}, the $A_1^{ij}$ tensor is simply $\cosh \alpha \, \delta^{ij}$ so the   spin-3/2 variations are diagonal:
\begin{equation}
\Big(A'\,\gamma^3\,+{1\over\ell}\, e^{-A}\,\gamma^2 \Big)\,\epsilon^j+g\,\overline {\cals W }\,\epsilon_j\eql 0\,, \qquad j=1,\ldots,8\,.
\label{so4gravvar}
\end{equation}
This means that $\cR_\epsilon$ consists of all eight spinors but it is a reducible representation of $G=SO(4)\times SO(4)$.
The  spin-1/2 variations, on the other hand, do distinguish between the irreducible components of $\cR_\epsilon$:
\begin{equation}\label{s3proj}
\gamma^3\epsilon^j \eql M\,\epsilon_j\,,\qquad 
\gamma^3\epsilon^{j+4} \eql M^*\,\epsilon_{j+4}\,,\qquad j=1,\ldots,4\,,
\end{equation}
where (using (\ref{zparam}) and (\ref{Vres1}) in (\ref{Mexplicit}) for $k=1$) we find
\begin{equation}\label{}
M\eql e^{i \Lambda}\eql {1\over\sqrt 2\,g}(\text{csch}\,\alpha\,\alpha'-i\cosh\alpha\,\zeta')\,.
\end{equation}

Following the analysis of Section \ref{Sect:Gravvar}, we can now use either one of the $\gamma^3$-projection conditions in equation (\ref{so4gravvar}) to obtain the $\gamma^2$-projection conditions: 
\begin{equation}\label{s2proj}
\gamma^2\epsilon^j \eql i\,\kappa\,e^{i \Lambda}\, \epsilon_j\,,\qquad 
\gamma^2\epsilon^{j+4}  \eql  - i\,\kappa\,e^{-i \Lambda}\,\epsilon_{j+4}\,,\qquad j=1,\ldots,4\,.
\end{equation}
Since $\cW$ is real, the $\gamma^2$-projections on $\epsilon^{j+4}$ can be obtained from those of $\epsilon^{j}$ by complex conjugating (\ref{so4gravvar}).  We therefore see that the effective sign of $\kappa$ changes between the two irreducible pieces of $\cR_\epsilon$ and, in particular:
\begin{equation}
\gamma^0 \gamma^1 \, \epsilon^j   \eql  \kappa\,\epsilon^j  \,,  \qquad \gamma^0 \gamma^1 \, \epsilon^{j+4}  \eql  -  \kappa\, \epsilon^{j+4}\,,\qquad j=1,\ldots,4    \,.
\label{so4chirproj}
\end{equation}
Thus the $(\bfs 4,\bfs 1)$ and $(\bfs 1,\bfs 4)$ correspond to supersymmetries with opposite $(1+1)$-dimensional helicity and hence we have an interface theory with  $\widehat \cN_L =\widehat \cN_R\eql 4$, or $(4,4)$ supersymmetry. This is consistent with the unbroken supersymmetries of the corresponding eleven-dimensional lift discussed in Appendix~\ref{appendixB}. 

Writing the symmetry action in terms of $SU(2)^4$, the action of the $\cR$-symmetry on the supersymmetries, the bosons, $X^A$, and the fermions, $\lambda^{\dot{A}}$, decomposes as: 
\begin{equation}
\begin{split}
\epsilon^i: \qquad & \bfs8_v ~=~  (\bfs2,\bfs2,\bfs1,\bfs1) \ \oplus \ (\bfs1,\bfs1,\bfs2,\bfs2)\,, \\ 
X^A: \qquad & \bfs8_s ~=~  (\bfs2,\bfs1,\bfs2,\bfs1) \ \oplus \ (\bfs1,\bfs2,\bfs1,\bfs2) \,,  \\ 
\lambda^{\dot{A}}: \qquad & \bfs8_c ~=~ (\bfs2,\bfs1,\bfs1,\bfs2) \ \oplus \ (\bfs1,\bfs2,\bfs2,\bfs1)  \,. 
\end{split}
\end{equation}
The group theory implies that the $(\bfs2,\bfs2,\bfs1,\bfs1)$ supersymmetries must relate the $(\bfs2,\bfs1,\bfs2,\bfs1)$ bosons to the $(\bfs1,\bfs2,\bfs2,\bfs1)$ fermions and the $(\bfs1,\bfs2,\bfs1,\bfs2)$ bosons to the $(\bfs2,\bfs1,\bfs1,\bfs2)$ fermions. On the other hand, the $(\bfs1,\bfs1,\bfs2,\bfs2)$ supersymmetries must relate the $(\bfs2,\bfs1,\bfs2,\bfs1)$ bosons to the $(\bfs2,\bfs1,\bfs1,\bfs2)$ fermions and the $(\bfs1,\bfs2,\bfs1,\bfs2)$ bosons to the $(\bfs1,\bfs2,\bfs2,\bfs1)$ fermions.  Thus each set of four symmetries naturally decomposes the bosons and fermions into two copies of a standard $\Neql 4$ representation, however the two different sets of four supersymmetries pair the boson and fermion decompositions differently.

\subsection{The BPS solutions}
\label{SO4sols}

As noted above, we have:
\begin{equation}
M\eql e^{i \Lambda} \eql {1\over\sqrt 2\,g}(\text{csch}\,\alpha\,\alpha'-i\cosh\alpha\,\zeta')\,.
\label{SO4M}
\end{equation}
One then finds that (\ref{dAsusy1}) simplifies to 
\begin{equation}
\tanh\alpha \, A' ~+~  \alpha'  ~=~ 0\;,
\label{simpcomp1}
\end{equation}
which can be integrated to yield
\begin{equation}
A = - \log (\sinh\alpha) + c_A\;,
\label{Asolnalpha}
\end{equation}
where $c_A$ is an integration constant.  

Reality of this solution naturally requires that  one has $\alpha >0$ and that $c_A$ is real.  Alternatively, one could allow  $\alpha < 0$ by making a purely imaginary shift in  $c_A$.  However, once $c_A$ is chosen, this option disappears and so we will require: 
\begin{equation}
\alpha  ~>~ 0 \,, \qquad c_A \in \IR \;.
\label{constr1}
\end{equation}

The fact that (\ref{EffLag1}) is independent of $\zeta$ means that there is a conserved Noether charge:
\begin{equation}
e^{3A}\, \sinh^2 2 \alpha  \,  \, \zeta'    ~=~ \text{const.}\;.  
\label{Nother1}
\end{equation}
Using  (\ref{Asolnalpha})  in (\ref{BPSz})  leads to a trivial identity in $\alpha'$ and it fixes the constant in (\ref{Nother1}):
\begin{equation}
\zeta' = - \dfrac{\kappa e^{-c_A}}{\ell} \dfrac{\sinh\alpha}{\cosh^2\alpha}\;.
\label{zetaeqn1}
\end{equation}
The last of the BPS equations, (\ref{quadconstr}), is simply
\begin{equation}\label{theAqss}
(A')^2=-\dfrac{e^{-2A}}{\ell^2} +2g^2\cosh^2\alpha\;,
\end{equation}
and using (\ref{Asolnalpha}) one obtains:
\begin{equation}
 (\alpha')^2 ~=~ - \frac{e^{-2c_A}}{\ell^2}\,\frac{\sinh^4\alpha}{\cosh^2\alpha} ~+~ 2g^2\sinh^2\alpha\;. 
 \label{alphaeqn1}
\end{equation}
Define the parameter
\begin{equation}
a ~\equiv~ \sqrt{2}\,g\,\ell\, e^{c_A} \,,  \label{adefn1}
\end{equation}
then (\ref{alphaeqn1}) is easily integrated to obtain, for $a<1$:
\begin{equation}
\sinh (\alpha(\mu)) ~=~ \kappa_{\alpha} \, \frac{a}{\sqrt{1- a^2}} \, \frac{1}{\cosh\big(\sqrt{2}\, g (\mu-\mu_0)\big)}\,,
\label{alphasol1}
\end{equation}
or, for $a>1$:
\begin{equation}
\sinh (\alpha(\mu)) = \kappa_{\alpha} \, \frac{a}{\sqrt{a^2 - 1}} \, \frac{1}{\sinh\big(\sqrt{2}\, g (\mu-\mu_0)\big)}\,,
\label{alphasol2}
\end{equation}
where  $\kappa_{\alpha}^2 =1$.

The requirement (\ref{constr1}) that  $\alpha > 0$ means that for the solutions (\ref{alphasol1}) we must take:
\begin{equation}
\kappa_\alpha ~=~ +1 \,,
\label{constr2}
\end{equation}
while for the solutions (\ref{alphasol2}) we must take either $\kappa_{\alpha} =+1$ and $\mu > \mu_0$ or $\kappa_{\alpha} =-1$ and $\mu < \mu_0$.  Without loss of generality we will take the former choice and hence always choose (\ref{constr2}).  The parameter $\mu_0$ is an integration constant and without loss of generality one can also take $\mu_0=0$.  As we noted earlier, the parameter $\ell$ is spurious and, if it is finite, we can scale the metric so that  $\ell=1$.

One can now solve (\ref{zetaeqn1}) and the result is:
\begin{align}
\tan (\zeta(\mu)-\zeta_0) &~=~  - \kappa \,  \kappa_{\alpha}   \sqrt{1- a^2}\, \sinh\big(\sqrt{2} \, g (\mu-\mu_0)\big) \,, \qquad a <1 \,;  \label{zetasol1} \\ 
\tan (\zeta(\mu)-\zeta_0) &~=~  - \kappa \,  \kappa_{\alpha}   \sqrt{a^2-1}\, \cosh\big(\sqrt{2} \, g (\mu-\mu_0)\big) \,, \qquad a > 1  \,.\label{zetasol2}
\end{align}

Finally the solution for the metric function $A(\mu)$ is obtained from (\ref{Asolnalpha})
\begin{align}
e^{A(\mu)} &~=~ \kappa_{\alpha}\,  \frac{ \sqrt{1- a^2}}{\sqrt{2}\, g \, \ell} \, \cosh \big(\sqrt{2}\, g (\mu-\mu_0)\big) \,, \qquad a <1 \,; \label{Asol1} \\
e^{A(\mu)} &~=~ \kappa_{\alpha}\,  \frac{ \sqrt{a^2-1}}{\sqrt{2}\, g \, \ell} \, \sinh \big(\sqrt{2}\, g (\mu-\mu_0)\big) \,, \qquad a > 1 \,. \label{Asol2} 
\end{align}

Scaling out $\ell$ by absorbing it in $c_A$, and then replacing this  $c_A$ via  (\ref{adefn1}) means that the  free parameters in the solution are: 
\begin{equation}\label{parameters}
a \;, \qquad \zeta_0\;, \qquad g\;,
\end{equation}
and there is also the sign choice, $\kappa$ ($\kappa_\alpha$ was fixed in (\ref{constr2})).

\begin{figure}[t]
\centering
\includegraphics[width=5.25cm]{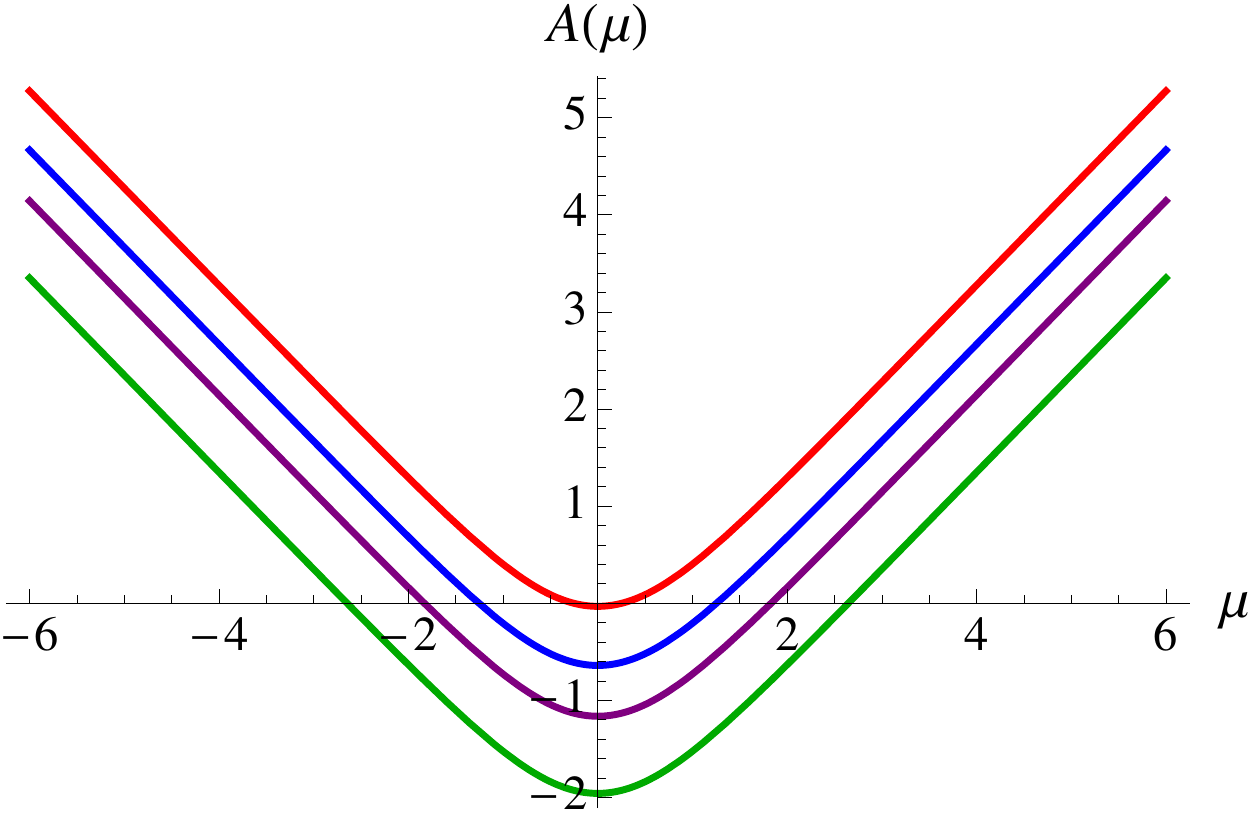}
\hfill
\includegraphics[width=5.25cm]{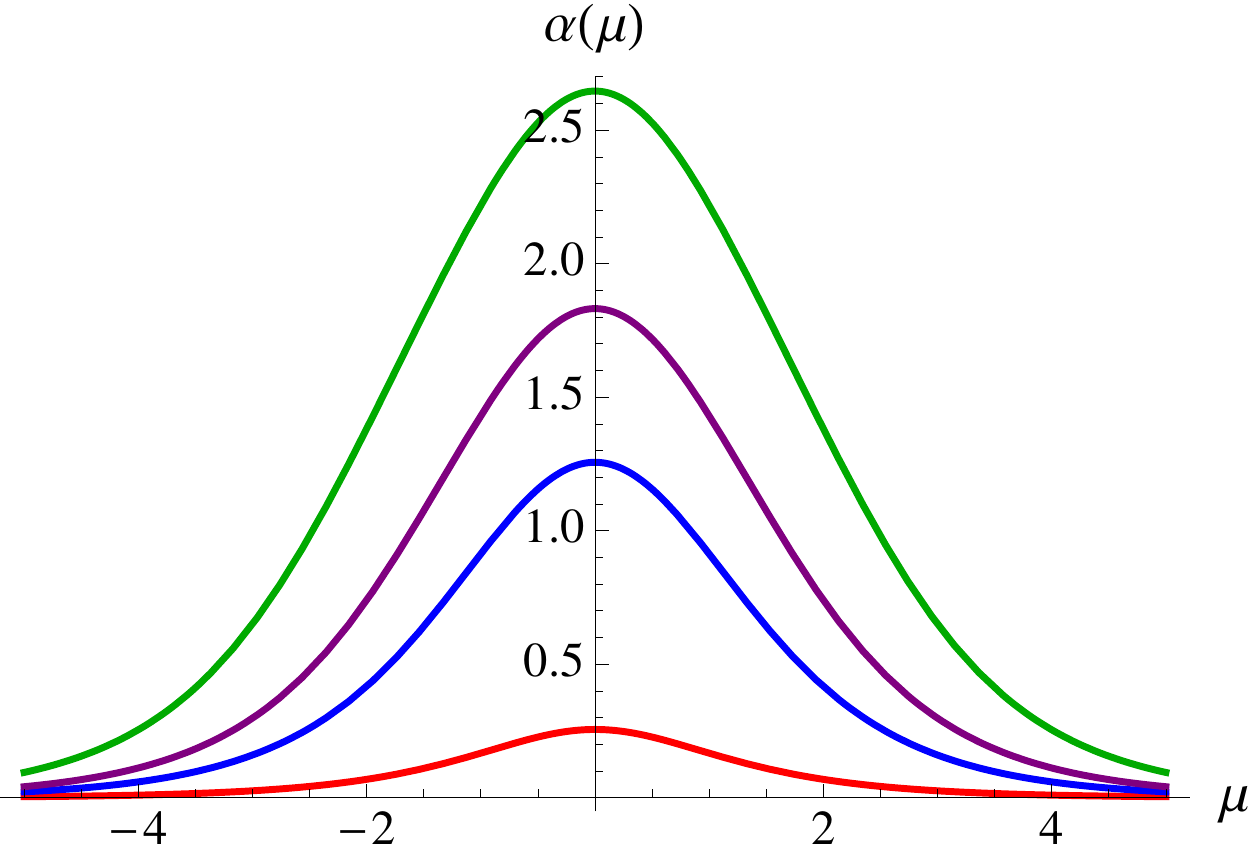}
\hfill
\includegraphics[width=5.25cm]{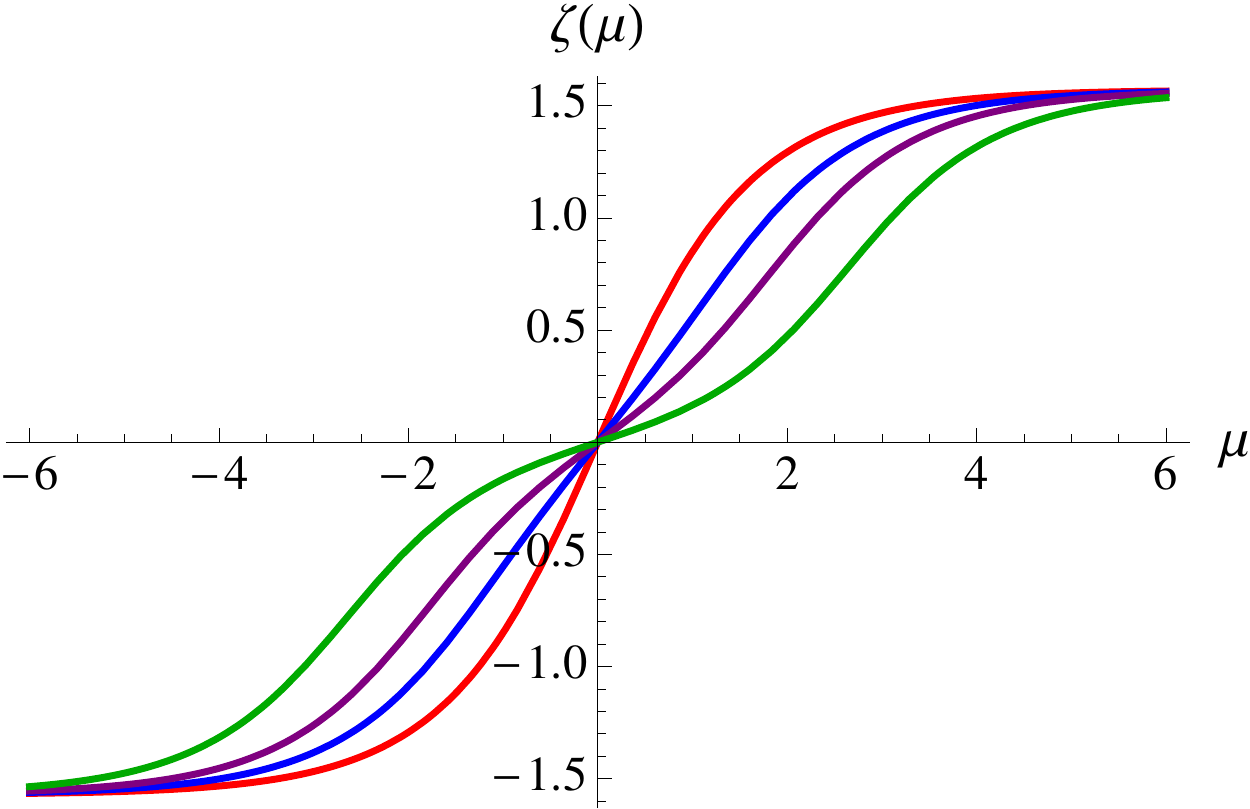}
\caption{Typical profiles for the $SO(4)^2$ Janus solutions. We have set $\mu_0=\zeta_0=0$, $\ell=1$, $g=1/\sqrt{2}$, and $\kappa = -1$. The curves are for $a=0.25$ (red), $a=0.85$ (blue), $a=0.95$ (purple), and $a=0.99$ (green).}
\label{plotsAalphazeta}
\end{figure}

For $a  < 1$  we get Janus solutions that are smooth for $-\infty < \mu < \infty$.  The profiles of these  solutions are all fairly similar in appearance.  From (\ref{alphasol1}) it is evident that the scalar field, $\alpha$, is globally positive, vanishing at $\mu  = \pm \infty$ and with a peak value of $ \frac{a}{\sqrt{1- a^2}}$ at $\mu  = \mu_0$.  From (\ref{zetasol1}) we see that the phase, $\zeta - \zeta_0$, goes between $ \frac{\kappa \pi}{2}$ and $- \frac{\kappa \pi}{2}$ as $\mu$ goes from $-\infty$  to $+\infty$.  Similarly, (\ref{Asol1}) shows that $A(\mu) \sim \pm \sqrt{2} g \mu$ as $\mu \to \pm \infty$ and reaches a minimum value at $\mu  = \mu_0$. Typical profiles are shown in Figure \ref{plotsAalphazeta}. The meaning of the parameters for this family of Janus solutions is as follows. The parameter $a<1$, controls the ``height of the bump" in the scalar $\alpha$. In field theory this parameter should map to the strength of the coupling between the $(1+1)$-dimensional defect and the $(1+2)$-dimensional bulk field theory.  The parameter, $\zeta_0$, determines which linear combination of the fermionic bilinear and bosonic bilinear operators in field theory we turn on.  Finally the parameter $g$ is the usual scale of $AdS_4$ which maps to the rank of the two CS gauge groups in the ABJM theory, that is, to the number of M2-branes.

For $a  > 1$ and taking $\kappa_\alpha = +1$, $\mu > \mu_0$ in (\ref{alphasol2})  we get solutions in which $\alpha$ vanishes at $\mu  = + \infty$ and runs off to $+\infty$ at $\mu  = \mu_0$.   From (\ref{Asol2})  we see that the metric function diverges:  $A(\mu) \to -\infty$ at  $\mu  = \mu_0$ and the geometry becomes singular.  It is also interesting to note that  $A'(\mu)$ never vanishes. From (\ref{zetasol2}) we see that the phase, $\zeta - \zeta_0$, asymptotes to  $- \frac{\kappa \pi}{2}$ as $\mu$ goes  $+\infty$ and at $\mu  = \mu_0$ this phase limits to some finite value whose sign is that of $-\kappa$.  Thus the phase swings through a finite range of less than $\frac{\pi}{2}$. These singular ``flows to Hades" may have an interesting physical interpretation but we will refrain from discussing them further here.

\subsection{Holographic analysis and interpretation}
\label{subsec:HoloSO4}

While the singular solutions that run off to Hades (flows with $a>1$) might ultimately admit some interpretation involving domain walls between the $SO(8)$ invariant conformal phase and a Coulomb phase, we will focus here on the smooth flows with $a<1$ that evidently represent domain walls separating two $SO(8)$ invariant conformal fixed points. 
 
We will therefore take $a<1$ and fix $\mu_0=0$ and $\kappa_\alpha=1$. To expand around $\mu \to \pm \infty$ it is convenient to define a new radial variable
\begin{equation}
\mu = \ds\frac{\mp1}{\sqrt{2}g} \log\left(\ds\frac{\sqrt{1-a^2}}{2a}\rho\right)~,
\end{equation}
and it is clear that for $\rho \to 0$ one has $\mu\to \pm \infty$. The scalars and the metric function have the following expansions for $\mu\to \pm \infty$ (the signs below are correlated)
\begin{equation}
\begin{split}
\alpha (\rho) &\approx \rho + \frac{1}{4}\left(\frac{1}{3} - \frac{1}{a^2}\right) \rho^3 + \mathcal{O}(\rho^5)~,\\
\zeta (\rho) &\approx \left(\zeta_0 \mp \kappa\ds\frac{\pi}{2}\right) \pm \frac{\kappa}{a} \rho \mp \kappa\frac{(1+3a^2)}{12a^3} \rho^3 + \mathcal{O}(\rho^5)~,\\
A (\rho) &\approx -\log \rho + \log\frac{a}{\sqrt{2}g\ell} + \frac{1}{4}\left( \frac{1}{a^2}-1\right) \rho^2 + \mathcal{O}(\rho^4)~.\\
\end{split}
\end{equation}
For holographic purposes and for comparison with the eleven-dimensional solution of \cite{D'Hoker:2009gg} it is convenient to work with the scalars
\begin{equation}
x=\text{Re}(z)=\tanh\alpha \,  \cos\zeta~, \qquad\qquad y = \text{Im}(z)=\tanh\alpha \, \sin\zeta~.
\end{equation}
One can expand the scalars $x(\mu)$ and $y(\mu)$ as
\begin{equation}\label{xydef}
\begin{split}
x (\rho) &\approx \cos\left(\zeta_0\mp \kappa \frac{\pi}{2}\right) \rho \mp \frac{\kappa}{a} \sin\left(\zeta_0\mp \kappa \frac{\pi}{2}\right)\rho^2  + \mathcal{O}(\rho^3)~,\\
y (\rho) &\approx \sin\left(\zeta_0\mp \kappa \frac{\pi}{2}\right) \rho \pm \frac{\kappa}{a} \cos\left(\zeta_0\mp \kappa \frac{\pi}{2}\right)\rho^2  + \mathcal{O}(\rho^3)~.
\end{split}
\end{equation}

Recalling the  holographic dictionary from Section \ref{Sect:HolDict},  our general Janus solution  is somewhat non-standard since the phase $\zeta_0$ ``rotates" scalars into pseudoscalars ({\it i.e.} bosonic bilinears into fermionic bilinears). For the solution at hand the scalar $x(\mu)$ in \eqref{xydef} is dual to a bosonic bilinear operator ${\cO}_1$ of dimension 1  and the scalar $y(\mu)$ in \eqref{xydef} is dual to a fermionic bilinear operator ${\cO}_2$ of dimension 2.  These may be written as:
\begin{equation}
\label{O12}
\begin{split}
{\cO}_1 & ~=~ \text{Tr}\big( (X^{1})^2+(X^{2})^2+(X^{3})^2+(X^{4})^2 -((X^{5})^2+(X^{6})^2+(X^{7})^2+(X^{8})^2)\big)  \,,\\
{\cO}_2  & ~=~ \text{Tr}\big((\lambda^{1})^2+(\lambda^{2})^2+(\lambda^{3})^2+(\lambda^{4})^2-((\lambda^{5})^2+(\lambda^{6})^2+(\lambda^{7})^2+(\lambda^{8})^2)\big) \,.
\end{split}
\end{equation}
By tuning the initial value of the phase $\zeta_0$ we obtain a famly of Janus solutions that are sourced in the boundary field theory  by a linear combination of ${\cO}_1$ and ${\cO}_2$. 

\begin{figure}[t]
\centering
\includegraphics[width=7.5cm]{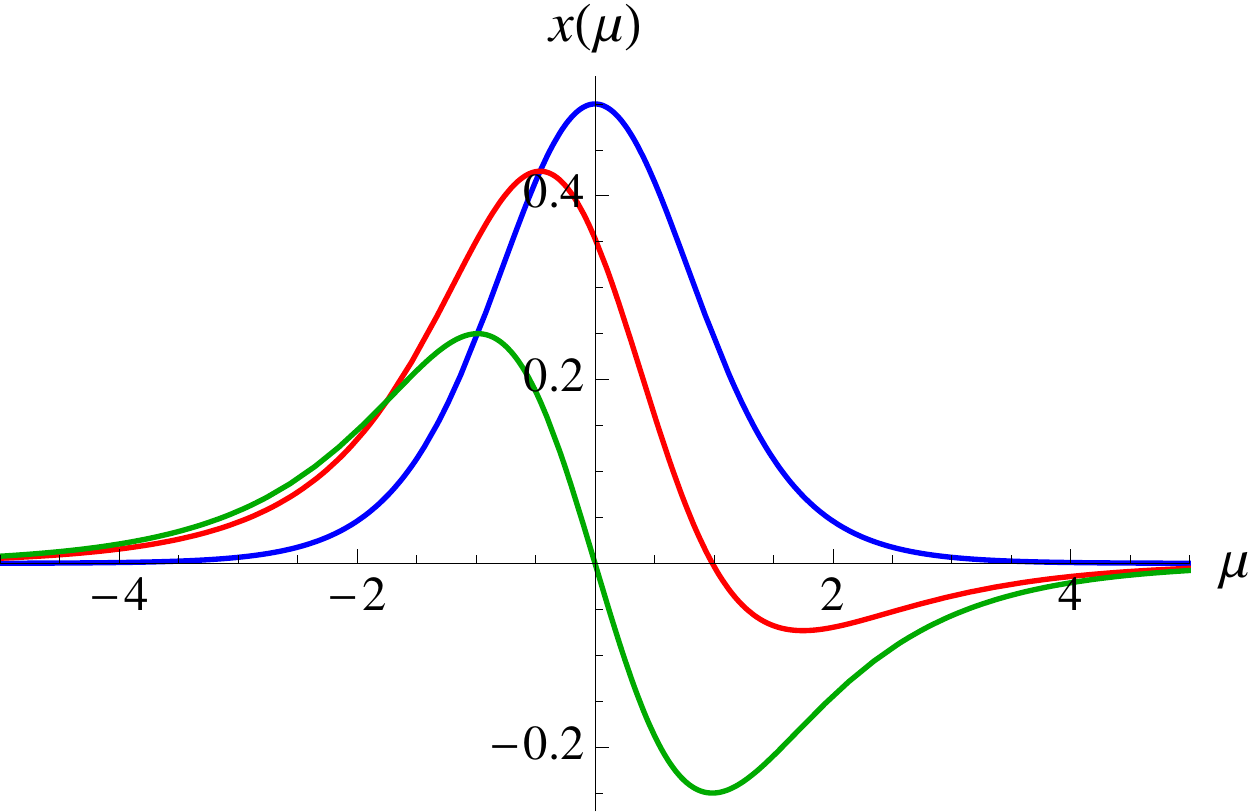}
\qquad
\includegraphics[width=7.5cm]{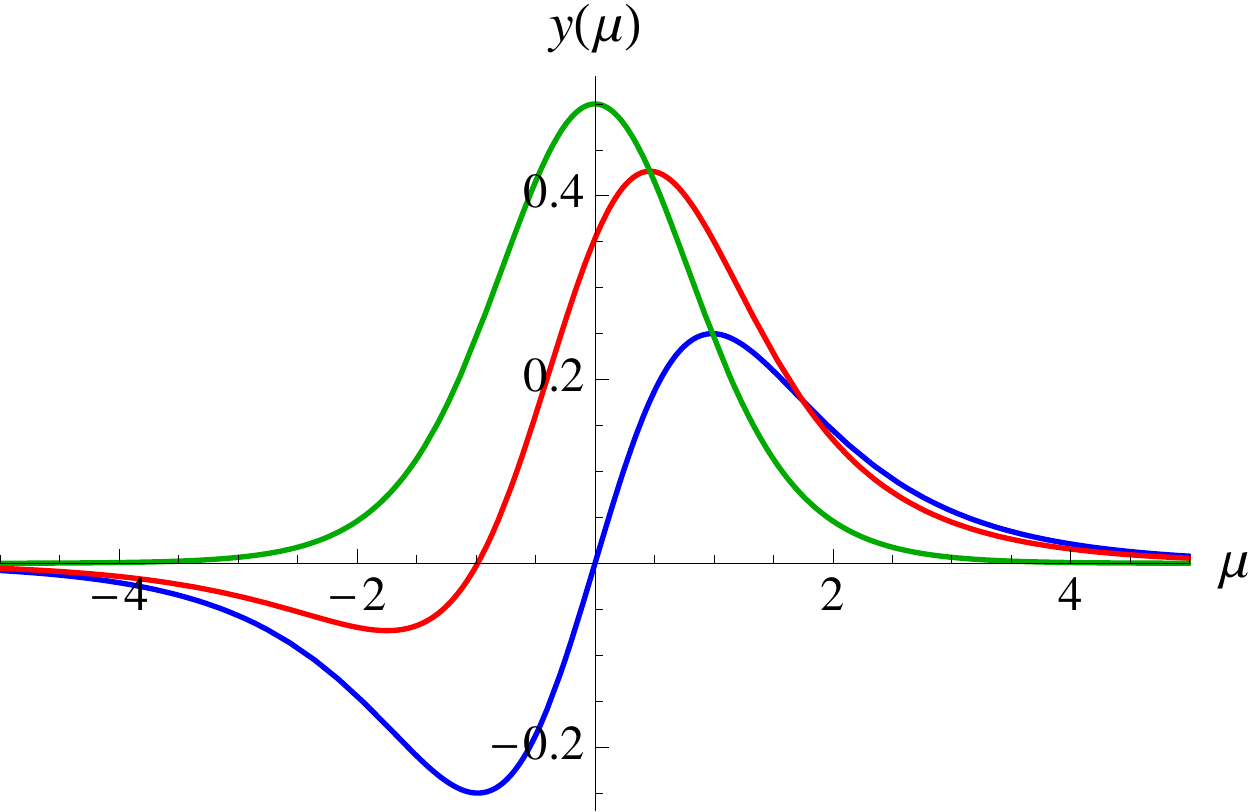}
\hfill
\caption{Plots for $x(\mu)$ and $y(\mu)$ for $\mu_0=0$, $g=\frac{1}{\sqrt{2}}$, $\kappa=-1$, and $a =\frac{1}{2}$. The parameter $\zeta_0$ takes the values $\zeta_0=0$ (blue), $\zeta_0=\pi/4$ (red), and $\zeta_0=\pi/2$ (green).}
\label{plotsxySO4}
\end{figure}

The four-dimensional reduction of the eleven-dimensional Janus solution discussed in \cite{D'Hoker:2009gg} was argued to have a normalizable mode for the pseudoscalar and the text suggests that the metric corrections were of the same, or lower, order.  As we describe in detail in Appendix \ref{appendixB}, the solution of \cite{D'Hoker:2009gg} corresponds to our solution with $\zeta_0= \kappa \pi/{2}$.  On the other hand, it is evident from our analysis in (\ref{xydef}) that if the pseudoscalar mode ($y (\rho)$)  is normalizable then the scalar mode ($x (\rho)$) must be non-normalizable, or {\it vice versa}.  Moreover, whatever the value of $\zeta_0$,  both the scalar $x(\mu)$ and the pseduoscalar $y(\mu)$ always develop a non-trivial profile and therefore we have both operators ${\cO}_1$ and ${\cO}_2$ turned on in the dual field theory.  To illustrate the importance of the parameter $\zeta_0$ we have presented plots of $x(\mu)$ and $y(\mu)$ for different values of $\zeta_0$ in Figure \ref{plotsxySO4}.

The apparent conflict with the asymptotic analysis of  \cite{D'Hoker:2009gg} could stem from the difficulty of correctly identifying the internal metric perturbations from the eleven-dimensional perspective because of the warp factors.  It is evident in  \cite{D'Hoker:2009gg} that they have a non-trivial warp factors  in front of the $AdS_3$ and $S^7$ metric  in a manner that closely parallels ours.  This shows that metric perturbations and hence the scalars are indeed playing a role in the Janus solution of  \cite{D'Hoker:2009gg} and perhaps the expansion of these modes proved rather subtle. 

Returning to our flows, note  that, for generic choices of $\zeta_0$, we have both a source and a vev for the operators in the dual field theory. Naively one might think that inserting a codimension-one defect in the field theory should not induce a deformation of the Lagrangian of the parent theory far away from the defect and thus the only deformation of the parent theory should be by a vev. However it is clear that in our solutions the situation is more general and one has both a source and a vev deformation of the ABJM theory at asymptotic infinity. This implies that in the dual field theory one has relevant couplings turned on which are function of the distance to the interface. Such position dependent couplings may change the nature of relevant and marginal operators as discussed recently in \cite{Dong:2012ena,Dong:2012ua} (see also \cite{Gutperle:2012hy} for a discussion in the present context). It would be very interesting to understand the physics of such position dependent relevant deformations from the point of view of the dual strongly coupled field theory.

It is also curious to note that the ``oblique'' mixtures of scalars and dual operators defined by:
\begin{equation}\label{xyoblique}
\begin{split}
\tilde x (\rho) ~\equiv~ \cos\left(\zeta_0\mp\kappa\frac{\pi}{2}\right) \, x (\rho) + \sin\left(\zeta_0\mp\kappa\frac{\pi}{2}\right)  y (\rho)  &~\approx~  \rho   + \mathcal{O}(\rho^3)\,, \\
\tilde y (\rho) ~\equiv~ \cos\left(\zeta_0\mp\kappa\frac{\pi}{2}\right) \, y (\rho) -  \sin\left(\zeta_0\mp\kappa\frac{\pi}{2}\right)  x (\rho)   &~\approx~   \pm \frac{\kappa}{a}  \rho^2  + \mathcal{O}(\rho^3)\,,
\end{split}
\end{equation}
suggests a simpler holographic interpretation in terms of a pure vev.  However, the standard holographic dictionary discussed in Section \ref{Sect:HolDict} does not seem to admit a simple interpretation of the dual of such mixtures of scalars and pseudoscalars.

Note that in the holographic RG flows studied in \cite{Pope:2003jp} (see also \cite{Bena:2004jw}) the phase $\zeta$ was a constant throughout the flow. For the Janus interfaces we study here $\zeta$ is necessarily a non-trivial function of $\mu$.  This will probably complicate the analysis if one tries to find the Janus-like generalization of the large family of solutions in \cite{Bena:2004jw}.

\section{The $SU(3) \times U(1) \times U(1)$-invariant Janus}
\label{sec:SU3}

\subsection{The truncation}

The $SU(3) \times U(1) \times U(1)$-invariant truncation  is easily extracted from the $SU(3)$ invariant truncation that has been widely studied.  In particular, it can be obtained from \cite{Bobev:2009ms,Warner:1983vz,Bobev:2010ib}.  The non-compact generators, $\Sigma_{IJKL}$, of $E_{7(7)}$ can be associated with differential forms on $\IR^8$:
\begin{equation}
\Sigma ~\equiv~\frac{1}{24}\,  \Sigma_{IJKL}  \, dx^I \wedge dx^J \wedge dx^K\wedge dx^L \,.
\label{E7forms}
\end{equation}
Define the complex variables $z_1=x_1+ix_2\,,\ldots,z_4\eql x_7+ix_8$ and introduce the $2$-forms
 \begin{equation}
 \label{complstr}
J^\pm\eql {i\over 2}\left(\sum_{j=1}^3 dz_j\wedge d \bar  z_j\right)\pm {i\over 2}\,dz_4\wedge d\bar z_4\,,
\end{equation}
The non-compact generators of the  $SU(1,1) \subset  E_{7(7)}$ are then defined by:
\begin{equation}\label{Fone}
F^+\eql \frac{1}{4}\, (J^+ + J^-)\wedge (J^+ + J^-) \,,\qquad F^-\eql   \frac{1}{4}\,  (J^+ + J^-)\wedge  (J^+ - J^-)  \,,
\end{equation}
and the real-form generators of $SL(2,\IR)$ are obtained by taking real and imaginary parts.  The embedding index, $k$, of this $SL(2,\IR)$ in $E_{7(7)}$
is $3$.

The $SO(2)$ or $U(1)$ action is simply the $SU(8)$ transformation acting on the real variables, $(x_1,\dots, x_8)$ by:
\begin{equation}
U ~=~  {\rm diag}\,(e^{i \beta}, e^{i \beta}, e^{i \beta}, e^{i \beta}, e^{i \beta}, e^{i \beta}, e^{-3i \beta}, e^{-3i \beta})\,,
\label{SU8b}
\end{equation}
which rotates $F^+$ by the phase $e^{4i \beta}$ and $F^-$ by the phase $e^{-4i \beta}$.  

These forms are manifestly invariant under the $U(3)$ that acts on $(z_1,z_2, z_3)$ and the $U(1)$ acting on $z_4$.  This $U(3) \times U(1)$ is also manifestly a subgroup of the $SO(8)$ symmetry acting on the $\IR^8$ and hence is a subgroup of the gauge symmetry.

The scalar potential is given by 
\begin{equation}
\mathcal{P} \eql -6\,  \cosh 2 \alpha  ~=~ -\frac{6\,(1 + |z|^2)}{1-|z|^2}\;.
\label{PSU3}
\end{equation}
Once again, one does not expect (\ref{SU8b}) to generate a symmetry of the action but here we find that it does.  This means that there may well be new interesting classes of holographic RG flows along the lines of \cite{Pope:2003jp,Bena:2004jw} in which metric structure can be rotated into internal fluxes. 

The effective particle action that encodes all field equations is:
\begin{equation}\label{su3lag}
\begin{split}
\cals L ~=~  & e^{3A}\left[3 (A')^2  ~-~ 3\, (\alpha')^2  ~-~ {3\over 4}\sinh^2(2\alpha)(\zeta')^2~+~ 6g^2 \, \cosh(2\alpha)
\right]-{3\over\ell^2}e^A \\
~=~ & 3\, e^{3A}\left[(A')^2 ~-~ \dfrac{z'\bar{z}'}{1-|z|^2} ~+~ 2 g^2\,\Big(\dfrac{1+|z|^2}{1-|z|^2}\Big)~-~ {1\over\ell^2}e^{-2A}
\right]  \,,
\end{split}
\end{equation}
where we have used the K\"ahler potential (\ref{Kahlerpot}) with $k=3$.  Once again the unexpected symmetry of the action makes it independent of $\zeta$ and so there is a conserved Noether charge:
\begin{equation}
e^{3A}\, \sinh^2 2 \alpha  \,  \, \zeta'    ~=~ {\rm const.  }
\label{Nother2}
\end{equation}

The tensor $A_1^{ij}$ of the $\Neql8$ theory is, once again, diagonal but there  are only two equal eigenvalues, $\cW$, that can be written in terms of a holomorphic superpotential, $\cV$, as in (\ref{WVreln}).  (We discuss the other six eigenvalues in Appendix \ref{appendixC}.)  This means that the number of supersymmetries, as discussed in Section \ref{Sect:susies}, is $\widehat \cN =2$ and the theory on the $(1+1)$-dimensional defect has $(0,2)$ supersymmetry.   The residual $\cR$-symmetry is the $U(1)$ symmetry that acts on $z_4 = x_7 + i x_8$ (as defined above) and lies outside the global $U(3)$ symmetry.

The holomorphic superpotential is a cubic:
\begin{equation}
\mathcal{V} ~=~ \sqrt{2} (z^3 +1)  \qquad \Rightarrow \qquad \cW \eql \frac{\sqrt{2}\, (z^3 +1)}{(1-|z|^2)^{3/2}}\,.
\label{Vres2}
\end{equation}
Apart from the $SO(8)$ critical point there are no other critical points of the potential or the superpotential within this truncation.

\subsection{Janus solutions}
\label{SU3numerics}

\begin{figure}[t]
\centering
\includegraphics[width=8cm]{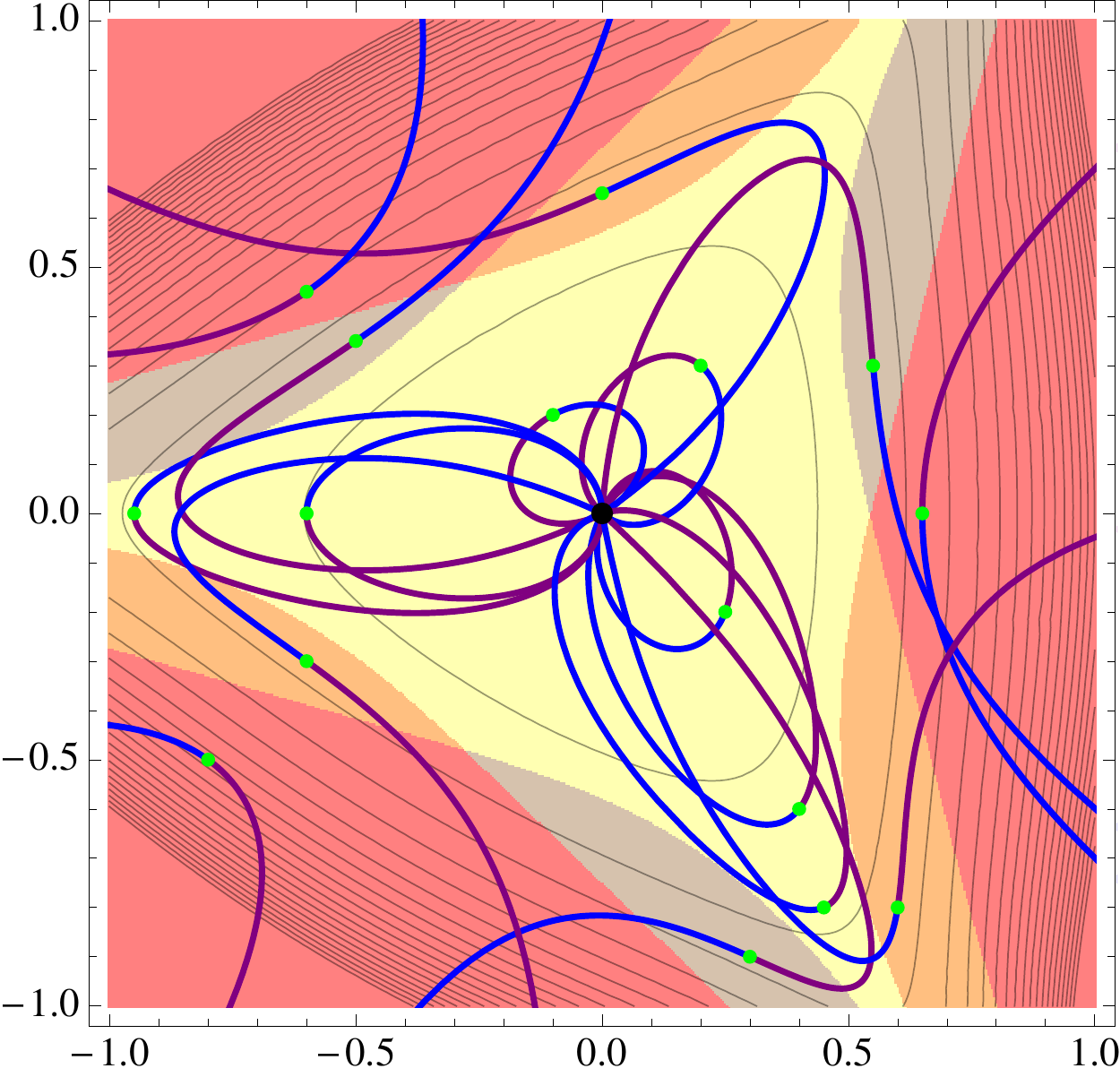}
\caption{The $SU(3)\times U(1)\times U(1)$ space of solutions. The horizontal	and	vertical	axes	are	$ \alpha\cos\zeta$	and	$\alpha\sin\zeta $ and the contour lines are of the real superpotential, $W$.  The maximally supersymmetric $AdS_4$ solution is the black dot in the middle.  The green dots denote turning points of $A(\mu)$ in the solutions. If a turning  point  lies in the yellow region, the solution is a regular Janus solution. Other colored regions   correspond to different types of singular solutions.}
\label{SU3WPplots}
\end{figure}

Unfortunately, unlike in the  $SO(4) \times SO(4)$ sector, one cannot solve analytically the BPS equations \eqref{simpalphaeqn}--\eqref{Aprimeeqn} for  flows based on the superpotential \eqref{Vres2}. In this section we    use numerical methods to explore the space of solutions and identify those solutions  that describe  domain walls between conformal phases. Such solutions  are   asymptotic to $AdS_4$ as $\mu \to \pm \infty$ and have a turning point    $A'(\mu_0) =0$ at some finite $\mu_0$.  This means that in our analysis we may miss some flows to Hades, like those  found in Section~\ref{sec:SO4}.

\begin{figure}[t]
\centering
\includegraphics[width=5.3cm]{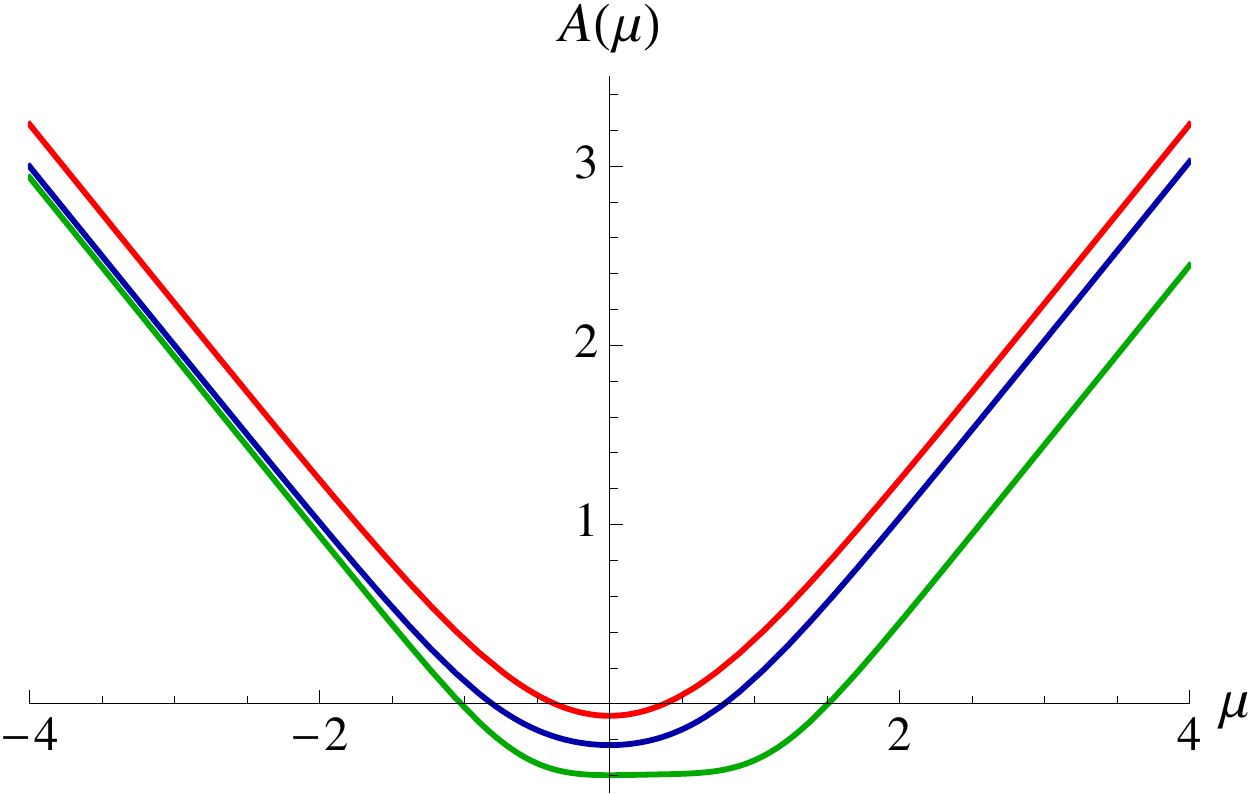}
~
\includegraphics[width=5.3cm]{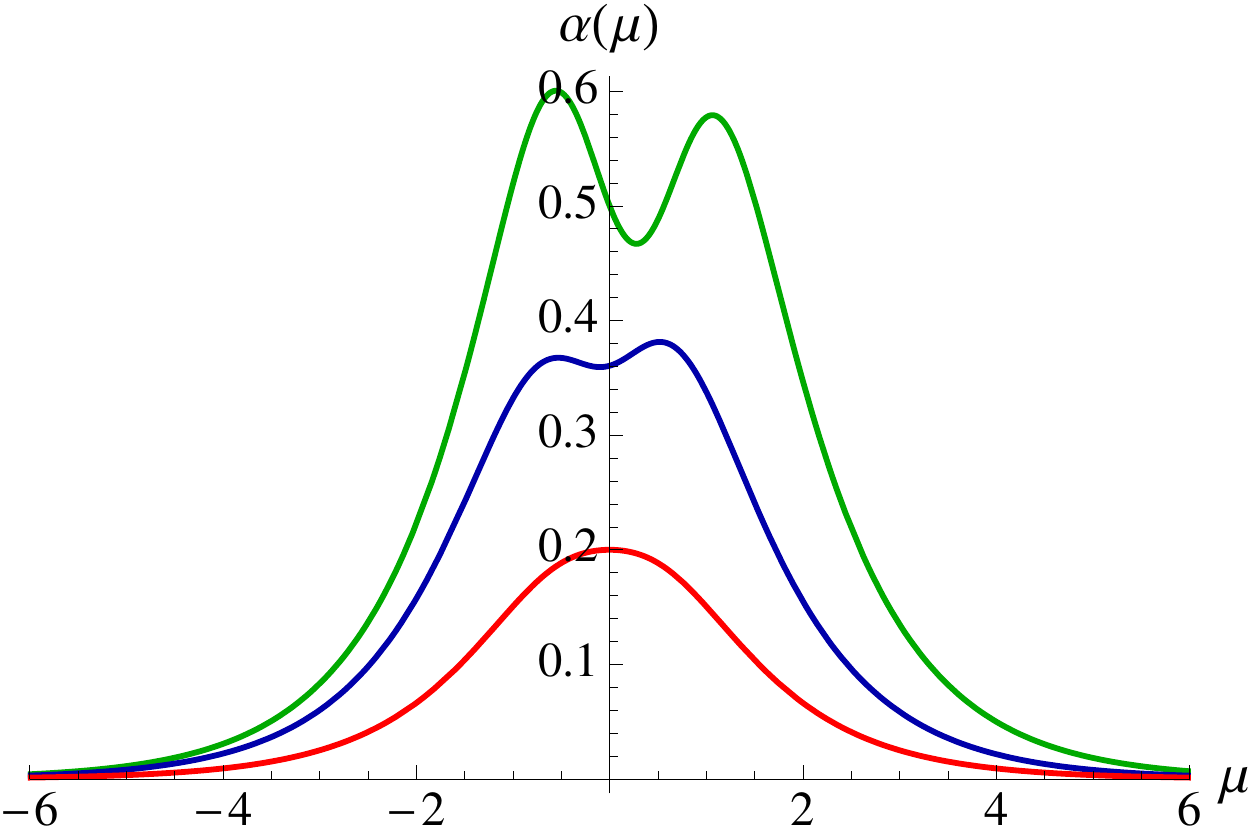}
~
\includegraphics[width=5.3cm]{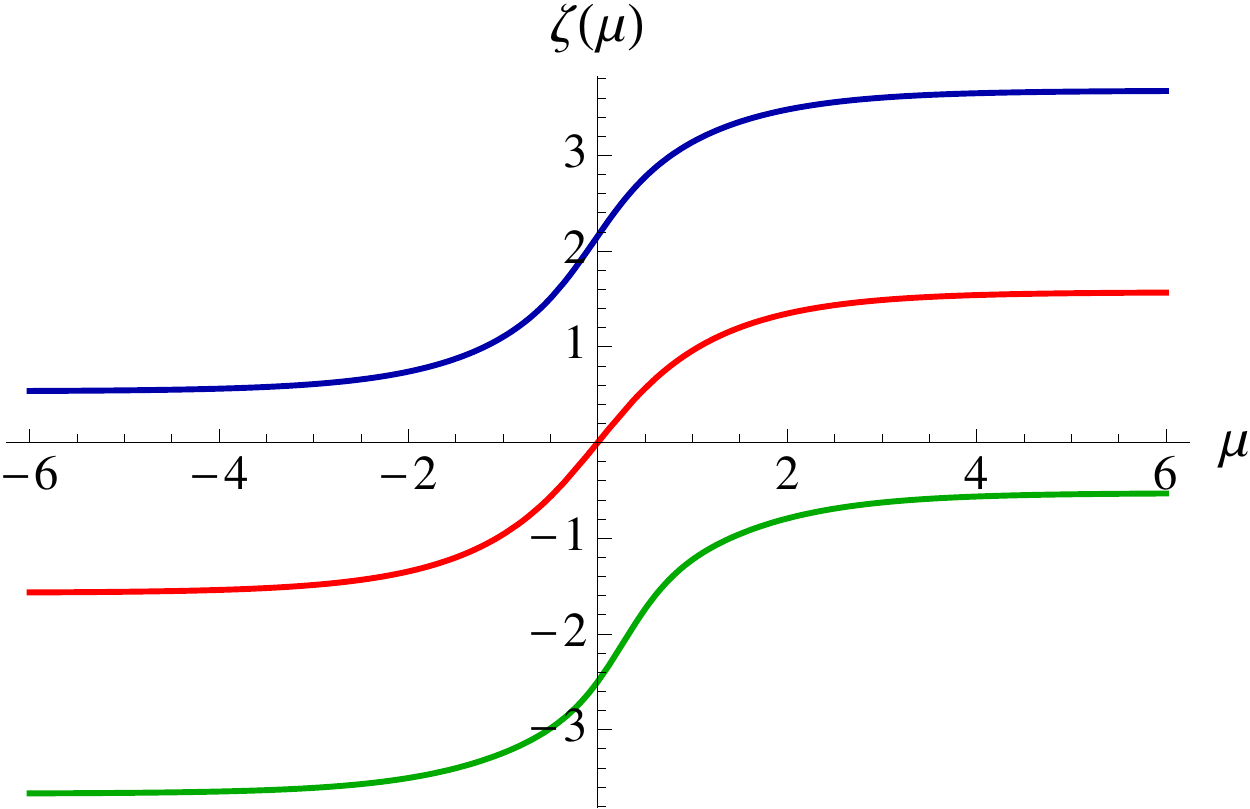}
\caption{ Typical profiles of $A(\mu)$, $\alpha(\mu)$ and $\zeta(\mu)$ for   $SU(3)\times U(1)^2$ Janus solutions. }\label{SU3functPlots}
\end{figure}

The representative numerical solutions presented in Figures~\ref{SU3WPplots} and \ref{SU3functPlots} are obtained as follows: We start by imposing the turning point\footnote{In our discussion, the ``turning point'' will mean the minimum of $A(\mu)$, which generically does not coincide with a turning point of $\alpha(\mu)$.}  condition, $A'(0)=0$ at $\mu=\mu_0=0$, for some finite values  $\alpha_*=\alpha(0)$ and $\zeta_*=\zeta(0)$. Next, for a fixed sign $\kappa=- 1$, we solve the BPS equations  \eqref{simpalphaeqn}, \eqref{simpzetaeqn}, and \eqref{Aprimeeqn}, for $A(0)$, $\alpha'(0)$ and $\zeta'(0)$. This determines a complete set of initial conditions for the second order equations that follow from the Lagrangian \eqref{su3lag}.  Setting $g=1/\sqrt 2$ and $\ell=1$, we then integrate numerically those equations to large positive and negative values of $\mu$.  Finally, we check that the resulting numerical solutions solve the BPS equations. The advantage of numerically integrating   the second order  equations is that they do not contain any branch cuts. The choice of the branch cut in  \eqref{Aprimeeqn} for a particular side of a solution   is controlled  at the outset by the initial conditions and the numerical integration can be carried out smoothly through the entire range of positive and negative values of the radial variable, $\mu$. 

The space of numerical solutions to the BPS equations in the $(\alpha\cos\zeta,\alpha\sin\zeta)$-plane is illustrated in Figure~\ref{SU3WPplots}. The turning  point   is always denoted by a green dot and the blue and purple parts of curves correspond to negative and positive values of $\mu$, respectively. Since there is a clear symmetry of the BPS equations under $\mu\to -\mu$ and $\kappa\to -\kappa$, each ``blue-purple'' curve also has the same ``purple-blue'' counterpart obtained by switching the signs in the initial conditions. 

Representative profiles for the scalars $\alpha$ and $\zeta$ and the metric function $A$ for some of the Janus solutions are shown in Figure~\ref{SU3functPlots}.  They illustrate more precisely the dependence of the solutions on the radial variable $\mu$.

We find four classes of solutions. There are regular Janus solutions that asymptote to the  maximally supersymmetric $AdS_4$ vacuum as $\mu \to \pm \infty$. These solutions exist when the turning point represented by the green dot lies in the yellow region in Figure \ref{SU3WPplots}. There are solutions which asymptote to $AdS_4$ as $\mu \to \infty$ or $\mu \to -\infty$, but are singular at a finite negative or positive value of $\mu$. The turning  point for these solutions lies in the grey or orange regions,  respectively. Finally, the solutions for which the turning point is in the pink region in Figure \ref{SU3WPplots} become singular on both sides of the defect at finite positive and negative values of $\mu$.

It is also clear from the plots  that as in the $SO(4)\times SO(4)$ invariant regular Janus solutions we always find $\lim_{\mu\to \infty}(\zeta(\mu)-\zeta(-\mu))=\pi$. In the dual field theory this implies that on both sides of the Janus interface we turn on the same linear combination of a bosonic and a fermonic bilinear in the ABJM theory. 

The asymptotic expansion of  the Janus solutions here for $\mu \to \pm\infty$ is similar to the one discussed in Section \ref{subsec:HoloSO4}. Depending on the value of $\zeta_0$, we again have a different linear combination of the bosonic and fermonic bilinear operators $\mathcal{O}_1$ and $\mathcal{O}_2$:
\begin{equation}
\mathcal{O}_1 = \mathcal{O}_b^{77} + \mathcal{O}_b^{88}\;, \qquad\qquad \mathcal{O}_2 = \mathcal{O}_f^{77} + \mathcal{O}_f^{88}\;,
\end{equation}
where the bilinears on the right hand sides are defined in \eqref{Obdef} and \eqref{Ofdef}. 

\section{The $G_2$-invariant Janus}
\label{sec:G2}

This sector of $\Neql8$ supergravity has a richer structure than the sectors considered above because there are  two $G_2$-invariant,  supersymmetric critical points, denoted by  $G_2^\pm$, that differ by the   sign of the pseudoscalar.   In eleven dimensions, the sign of the four-dimensional pseudoscalar determines the sign of the internal, or magnetic, components of the three-form flux. Thus the $G_2^+$ and $G_2^-$ critical points represent supergravity phases with opposite magnetic fields.

The families of Janus solutions are also correspondingly much richer and  include, in addition to solutions  representing domain walls between two copies of the $SO(8)$-invariant phase, solutions that involve, or are dominantly controlled by, any combination of the three supersymmetric critical points.  Indeed, we will find classes of solutions that start out in the $SO(8)$ phase at $\mu = \pm \infty$ but are perturbed by relevant operators  that drive the solution, via the standard holographic RG flow \cite{Bobev:2009ms}, to either one or both $G_2$ phases. We will argue that in a certain limit they should give rise to  new  families of $SO(8)/G_2^\pm$ domain walls and a special $G_2^+/G_2^-$ domain wall.


\subsection{The truncation}
\label{G2truncation}

The $G_2$-invariant truncation  can also be extracted from the $SU(3)$ invariant truncation.  Indeed, the $SO(7)$-invariant self-dual tensor is given by \cite{Bobev:2009ms,Warner:1983vz,Godazgar:2013nma}:
\begin{equation}
\begin{split}
\Sigma^+_{IJKL} ~=~  & \Big( \delta^{1234}_{IJKL}  +
\delta^{5678}_{IJKL} + \delta^{1256}_{IJKL} + \delta^{3478}_{IJKL}
+ \delta^{3456}_{IJKL} + \delta^{1278}_{IJKL} \\ 
& -(\delta^{1357}_{IJKL} + \delta^{2468}_{IJKL} )+
(\delta^{2457}_{IJKL} + \delta^{1368}_{IJKL}) +
(\delta^{1458}_{IJKL} +\delta^{2367}_{IJKL})  + (
\delta^{1467}_{IJKL} + \delta^{2358}_{IJKL}) \Big) \,.
\end{split}
\end{equation}
and the  $SO(7)$-invariant anti-self-dual tensor, $\Sigma^-_{IJKL}$, can be obtained from this by making the reflection $x_8 \to -x_8$.

The $SO(2)$ or $U(1)$ action is simply the $SU(8)$ transformation acting on the real variables, $(x_1,\dots, x_8)$ by:
\begin{equation}
U ~=~  {\rm diag}\,(e^{i \beta}, e^{i \beta}, e^{i \beta}, e^{i \beta}, e^{i \beta}, e^{i \beta}, e^{i \beta}, e^{-7i \beta})\,,
\label{SU8c}
\end{equation}
which rotates $\Sigma^+_{IJKL} \pm   \Sigma^-_{IJKL}$ by a phase $e^{\pm 4i \beta}$.  These $E_{7(7)}$ Lie algebra elements generate the $SL(2,\IR)$, with embedding index, $k=7$. 

The detailed structure of this supergravity sector can be read-off from \cite{Bobev:2009ms,Bobev:2010ib}. In particular, the tensor $A_1^{ij}$ of the $\Neql8$ theory is diagonal and has two distinct sets of eigenvalues according to the branching ${\bfs 8}_v\rightarrow \bfs 7 +\bfs 1$ of the gravitino representation under $G_2$. However, only one eigenvalue 
\begin{equation}
\label{G2superpot}
\cW\eql \sqrt{2} \left[\, \cosh ^7 \alpha    + 7   \cosh ^3 \alpha \sinh ^4 \alpha \, e^{4 i \zeta }+   7  \cosh^4 \alpha  \sinh ^3  \alpha  \, e^{3 i \zeta }   +  \sinh ^7 \alpha \, e^{7 i \zeta }  \,\right] \,, 
\end{equation}
corresponding to the singlet of $G_2$,  can be written in terms of a holomorphic superpotential,\footnote{One can read-off this superpotential from Eqs.\ (2.34) and (2.35) in   \cite{Bobev:2010ib} by setting $z=0$ and identifying $\zeta_{12}$ in \cite{Bobev:2010ib} with the $z$ below.}
\begin{equation}
\cV ~=~  \sqrt{2}(z^7+7z^4+7z^3+1)\;,
\label{G2V}
\end{equation}
as in \eqref{WVreln}. This means that the number of supersymmetries, as discussed in Section~\ref{Sect:susies}, is $\widehat \cN =1$ and the theory on the $(1+1)$-dimensional defect has $(0,1)$ supersymmetry. 

The effective, one-dimensional Lagrangian is:
\begin{equation}\label{G2Lag}
\cL ~=~   e^{3A}\Big[\, 3 (A')^2 -7\,\Big[(\alpha')^2 + \frac{1}{4}\sinh^2(2\alpha)(\zeta')^2\,\Big]  ~-~ g^2\,\cP \, \Big]~-~ {3\over\ell^2}e^A \,.
\end{equation}
where the supergravity potential, $\cP$, can be obtained from \eqref{normW}  and \eqref{Simppot}, or, equivalently from \eqref{Pform}  with \eqref{Kahlerpot} and $k=7$. 

\begin{figure}[t]
\centering
\includegraphics[width=7cm]{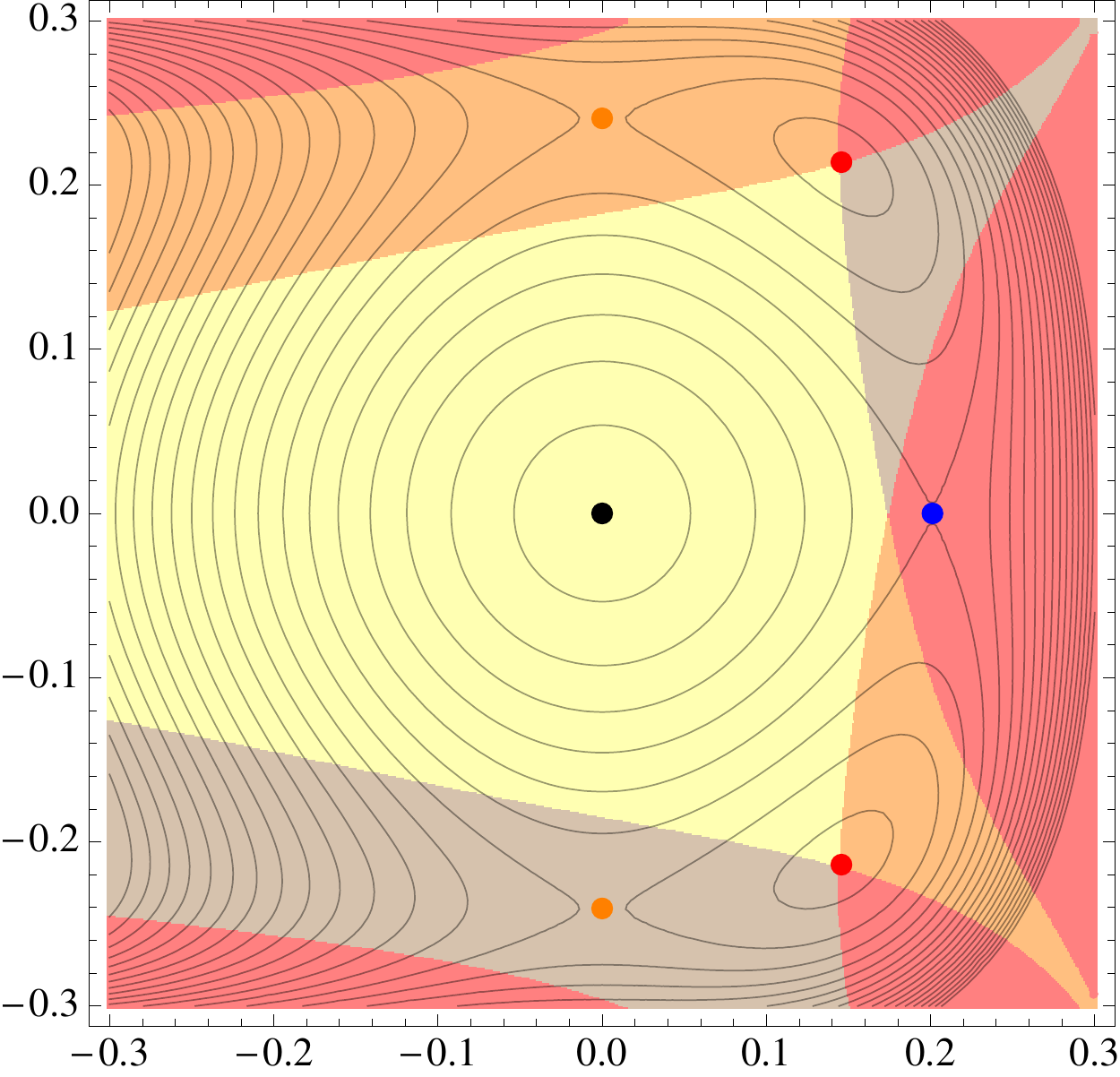}
\qquad
\includegraphics[width=7cm]{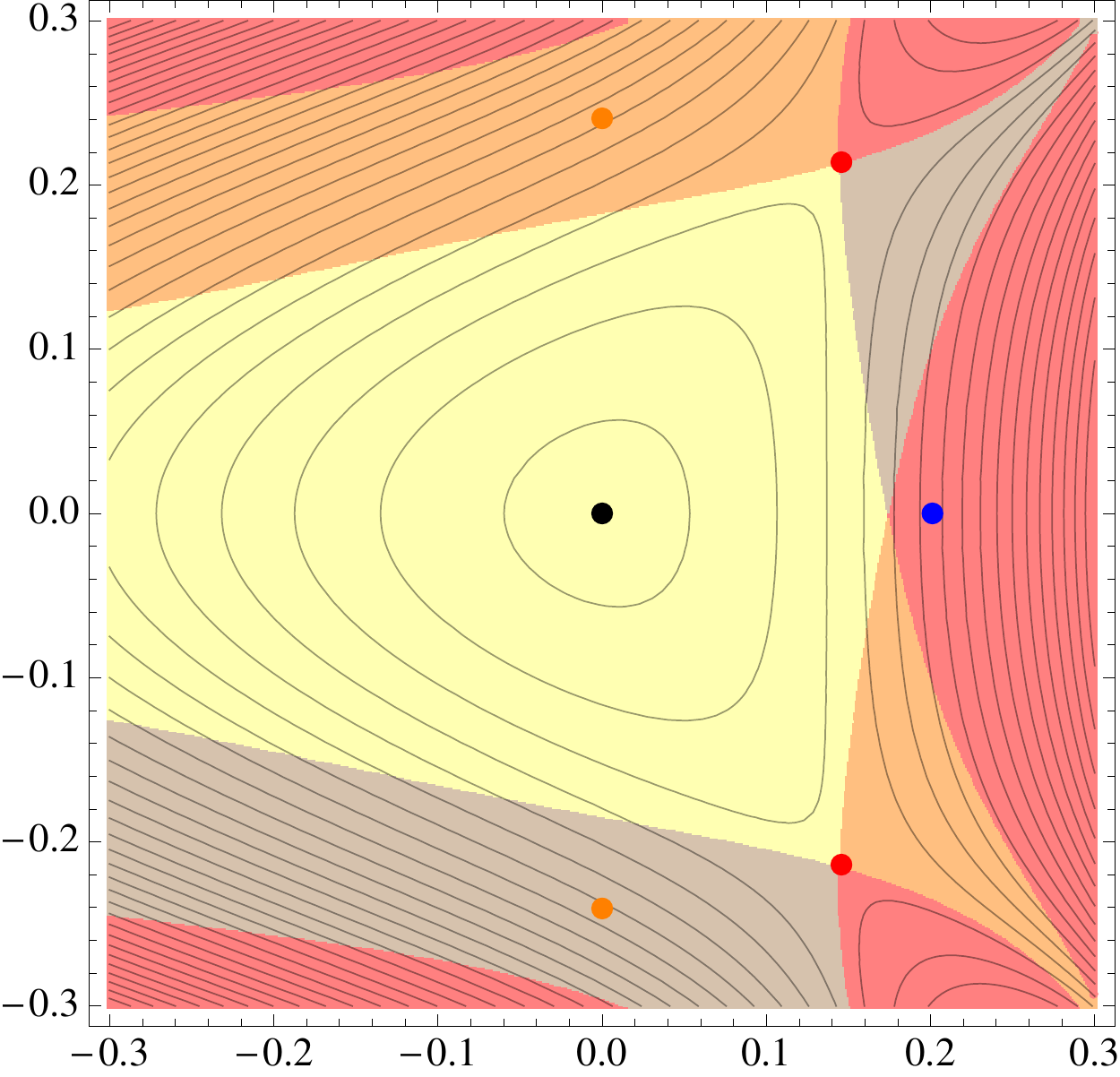}
\caption{Contour plots of the potential $\cP$ (left) and the real superpotential, $W = |\cW|,$ (right). The horizontal and vertical axes are $\alpha\cos\zeta$ and $\alpha\sin\zeta$. The $SO(8)$, $SO(7)^{+}$, $SO(7)^{-}$ and $G_2$ invariant critical points are denoted by black, blue, orange and red dots, respectively. The  shading of various domains is described in Section \ref{G2Janusflows}.}
\label{G2WPplots}
\end{figure}

The scalar potential, $\cP$, has a number of critical points \cite{Warner:1983vz} shown in Figure~\ref{G2WPplots}: 
\begin{itemize}[noitemsep,topsep=0pt]
\item [(i)]  the maximally supersymmetric point (black dot) at $z=0$; 
\item [(ii)] the non-supersymmetric point  with $SO(7)^+$ symmetry (blue dot) at $\alpha= {1\over 8}\log 5$ and $\zeta=0$; 
\item [(iii)] two non-supersymmetric points with $SO(7)^-$ symmetry (orange dots) at $\alpha= {1\over 2}\mathop{\rm arccsch} 2$ and $ \zeta=  \pm {\pi\over 2}$;
\item [(iv)] two  supersymmetric $G_2$-invariant points, $G^\pm_2$, (red dots) at 
\begin{equation}
\alpha\eql \frac{1}{2} \mathop{\rm arcsinh}\left(\sqrt{\frac{2\sqrt{3}-2}{5}}\right)\approx 0.2588  \,,\quad \zeta\eql \pm \mathop{\rm arccos}{1\over 2}\sqrt{3-\sqrt 3} \approx \pm 0.9727\,.
\label{G2pts}
\end{equation}
\end{itemize}
The $SO(8)$ and $G_2^\pm$ supersymmetric points are also critical points of the superpotential $\cals W$. For future reference we note that the slope of the function $A$ for the two supersymmetric critical points (where we fix $g=1/\sqrt{2}$) is given by:
\begin{equation}
\lim_{\mu\to\pm\infty}A'(\mu)|_{SO(8)} = \pm 1\;, \qquad\qquad \lim_{\mu\to\pm\infty}A'(\mu)|_{G_2} = \pm\left(\frac{3^92^{10}}{5^{10}}\right)^{\frac{1}{8}}\approx \pm1.0948 \;.
\end{equation}
This determines the $AdS$ radius of the corresponding vacua. 

The non-supersymmetric points  are perturbatively unstable \cite{Bobev:2010ib} and they do not give rise to any supersymmetric Janus sulutions.\footnote{By solving numerically the second order equations for \eqref{G2Lag}, we  have, in fact,  found some non-supersymmetric Janus solutions and RG flow domain walls in those  sectors. However, it is very likely that these solutions are unstable and we refrain from discussing them here.}  Similarly, there are no supersymmetric Janus solutions with $\cals R_\epsilon=\bfs 7$ of $G_2$. See  Appendix~\ref{appendixC} for some additional details.

\subsection{Janus solutions}
\label{G2Janusflows}

\begin{figure}[t]
\centering
\includegraphics[width=8cm]{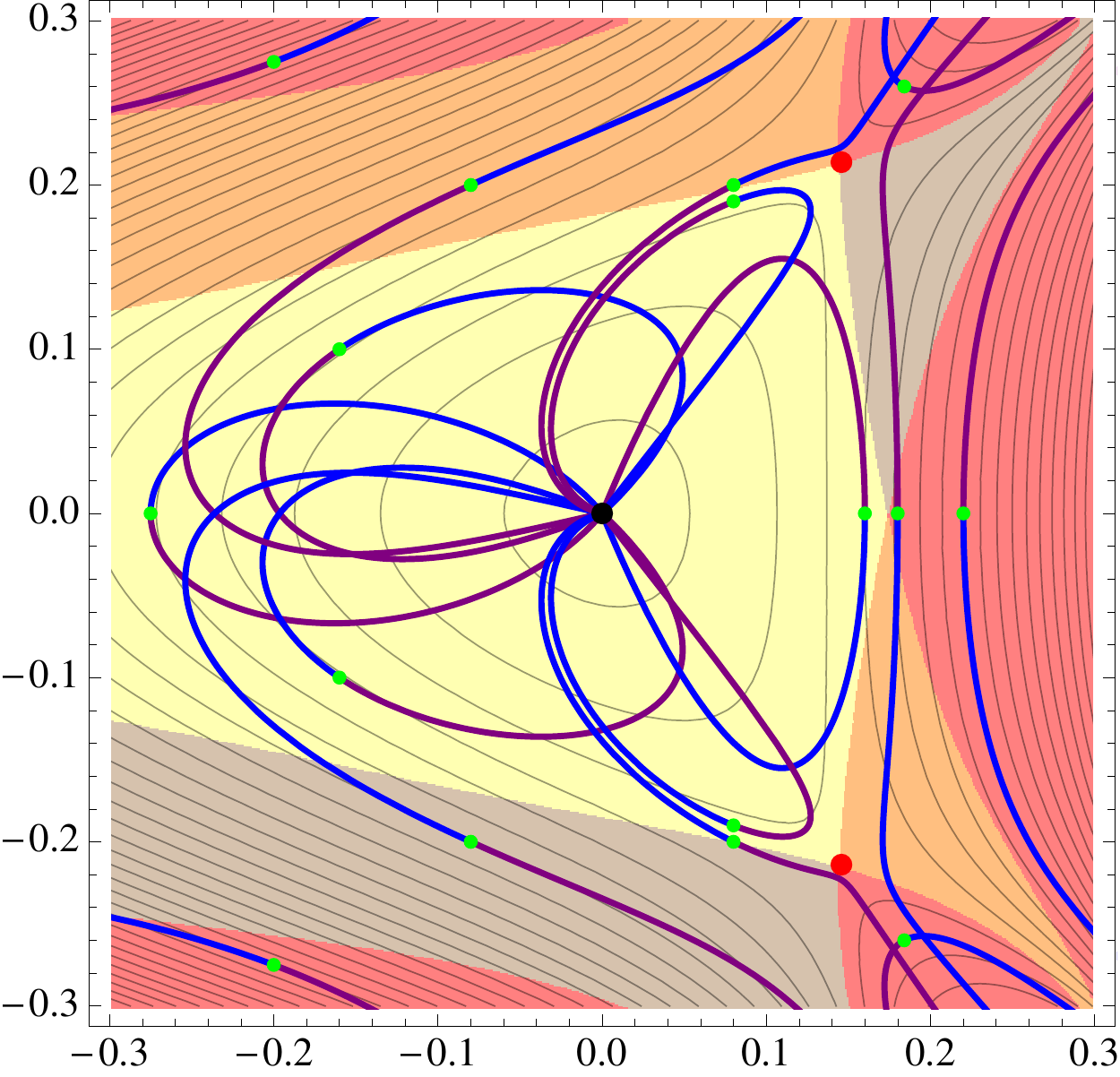}
\caption{ The ``phase diagram" of $AdS_3$ sliced domain wall solutions in the $G_2$ invariant sector of $\cN=8$ gauged supergravity.}
\label{AllTogether}
\end{figure}

Not surprisingly,  the BPS equations \eqref{simpalphaeqn}--\eqref{Aprimeeqn} in the $G_2$ sector can only be   solved   numerically. Using the same method as in Section~\ref{SU3numerics}, we have carried out an extensive search for different classes of solutions shown in Figure~\ref{AllTogether} and these will be discussed in more detail below. 
Just as in the $SU(3)\times U(1)^2$ sector, we  find that there is a ``basin of attraction'' around the maximally supersymmetric $SO(8)$ critical point where the solutions typically start and/or finish. We also find good numerical evidence for classes of solutions that start and/or finish at the $G_2^\pm$  points. 

The details of the solutions are primarily controlled by the location, $(\alpha_*, \zeta_*)$, in the scalar manifold  of the turning point of $A(\mu)$  in the Janus solution ({\it i.e.} by the values of $(\alpha, \zeta)$ at which $A'(\mu)$ momentarily vanishes).  As usual, this point will be marked by a green dot in all the contour plots.

\subsubsection{Symmetric solutions}
\label{SymmJanus}

\begin{figure}[t]
\centering
\includegraphics[width=7cm]{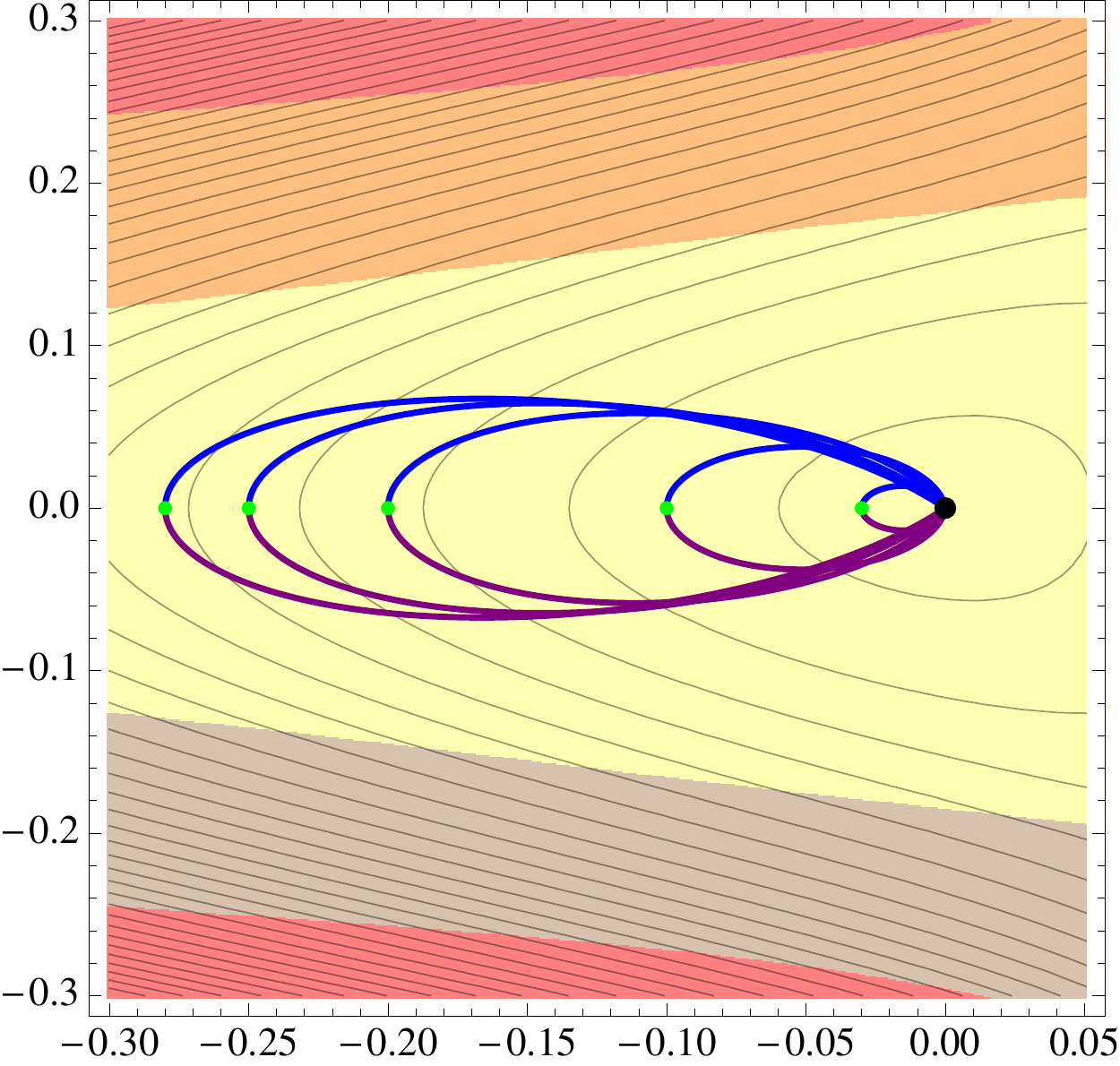}
\qquad
\includegraphics[width=7cm]{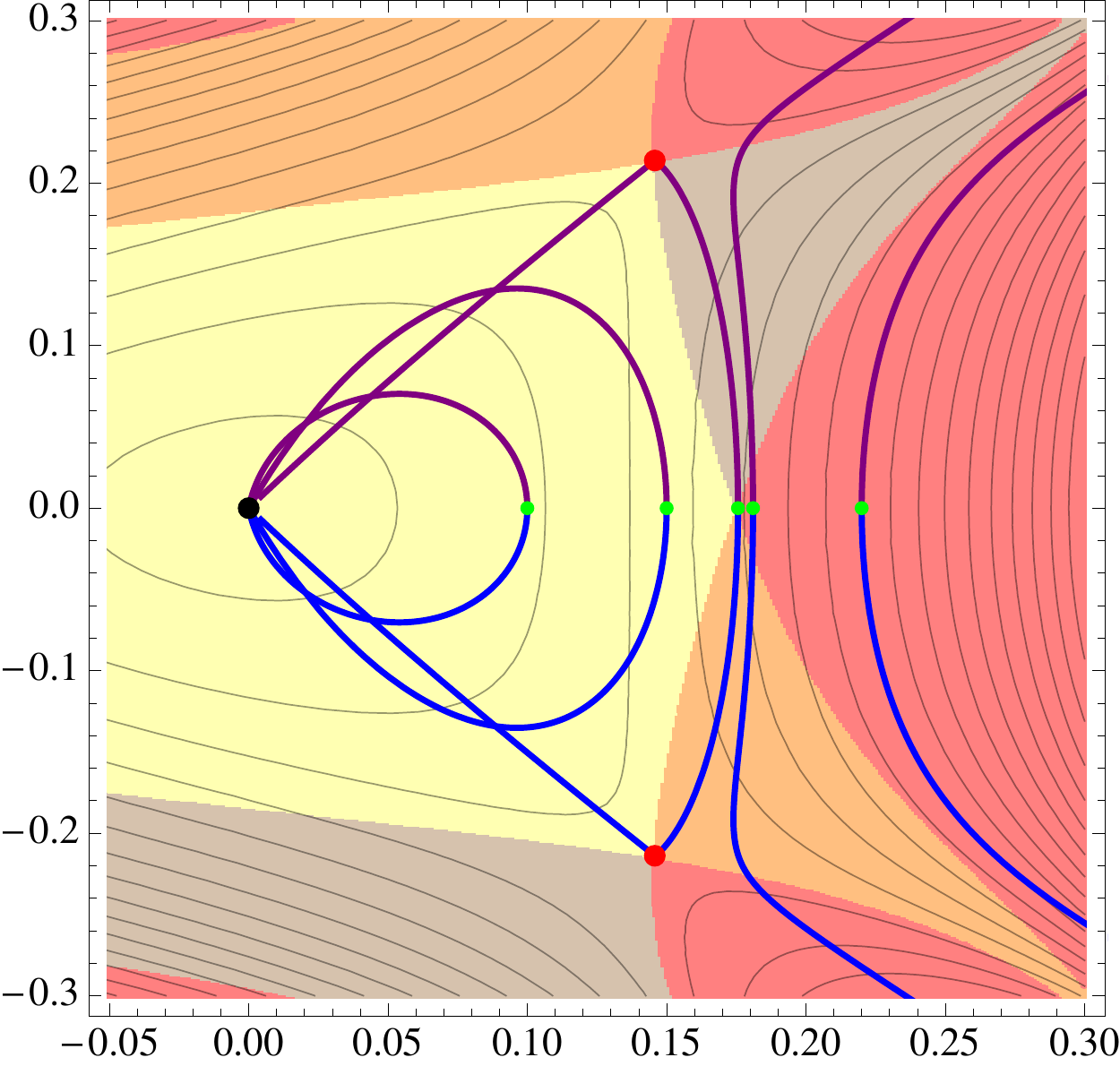}
\caption{The symmetric solutions  with (a) $\zeta_0=\pi$ and (b) $\zeta_0=0$. As the turning point  approaches the point $\alpha_0=\alpha_{cr}$, $\zeta_0=0$, the Janus solution asymptotes to the $G_2^+/G_2^-$   Janus solution.}
\label{symmetricG2plots}
\end{figure}

The simplest class of solutions have the  turning point of $A(\mu)$ on the real axis of the scalar manifold: $\zeta_*=0$ or $\zeta_*=\pi$, and thus are invariant under the  $\ZZ_2$ symmetry generated by $\zeta\rightarrow-\zeta$.  

Representative solutions with the turning point on the negative real axis, $\zeta_*=\pi$, are shown in the first plot in 
Figure~\ref{symmetricG2plots}.  We find only closed loops of $SO(8)$/$SO(8)$ Janus solutions that are similar  to those in the previous two sections, but  there is one significant difference.  In the previous Janus solutions, the net change of the phase, $\Delta\zeta=\zeta(+\infty)-\zeta(-\infty)$, between the two sides was always equal to $\pi$, but here the net change of phase for solutions in Figure \ref{symmetricG2plots}, measured by  the opening angle of the loops, is less than $\pi$ and depends on the initial data. We attribute this to a non-trivial dependence of the potential, $\cals P$, on the phase, $\zeta$, and hence the absence of a conserved quantity such as \eqref{Nother1}
or \eqref{Nother2}.

\begin{figure}[t]
\centering
\includegraphics[width=7cm]{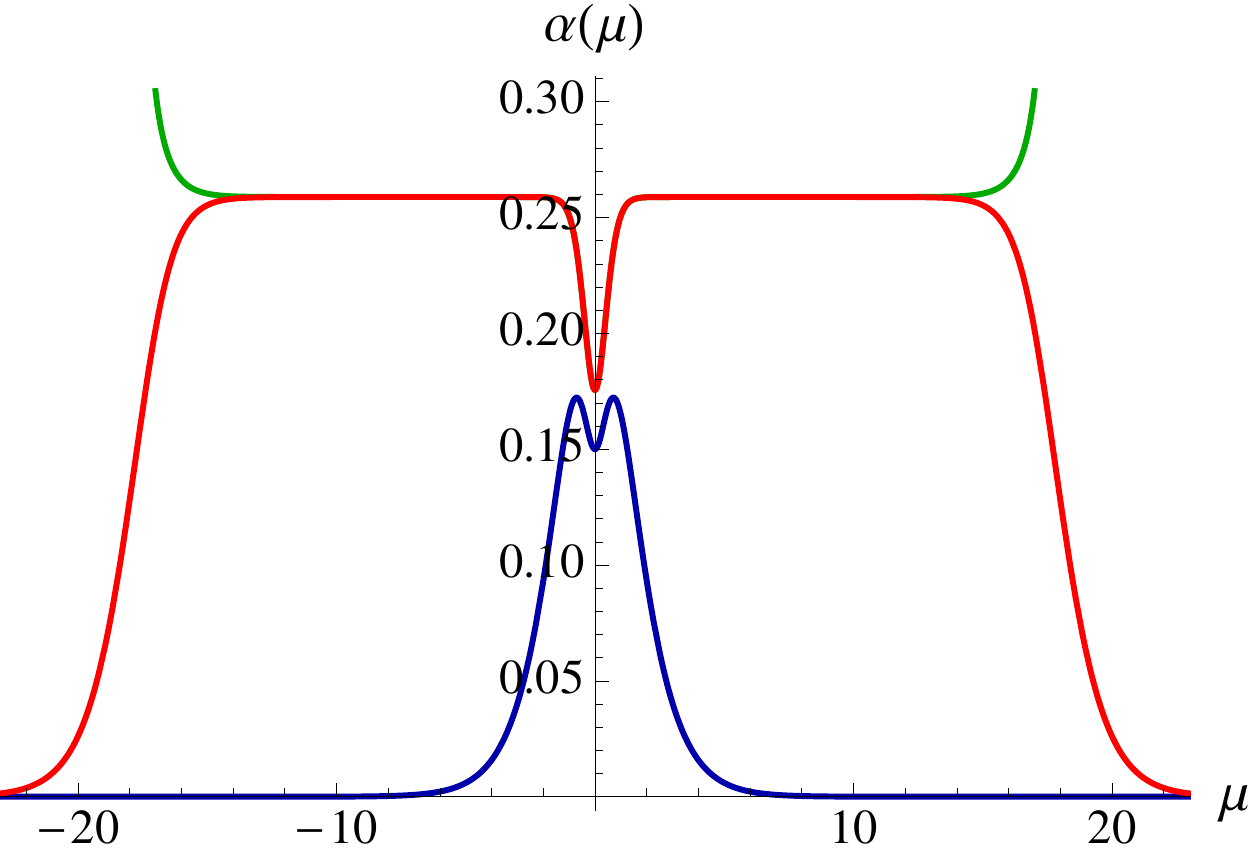}
\qquad
\includegraphics[width=7cm]{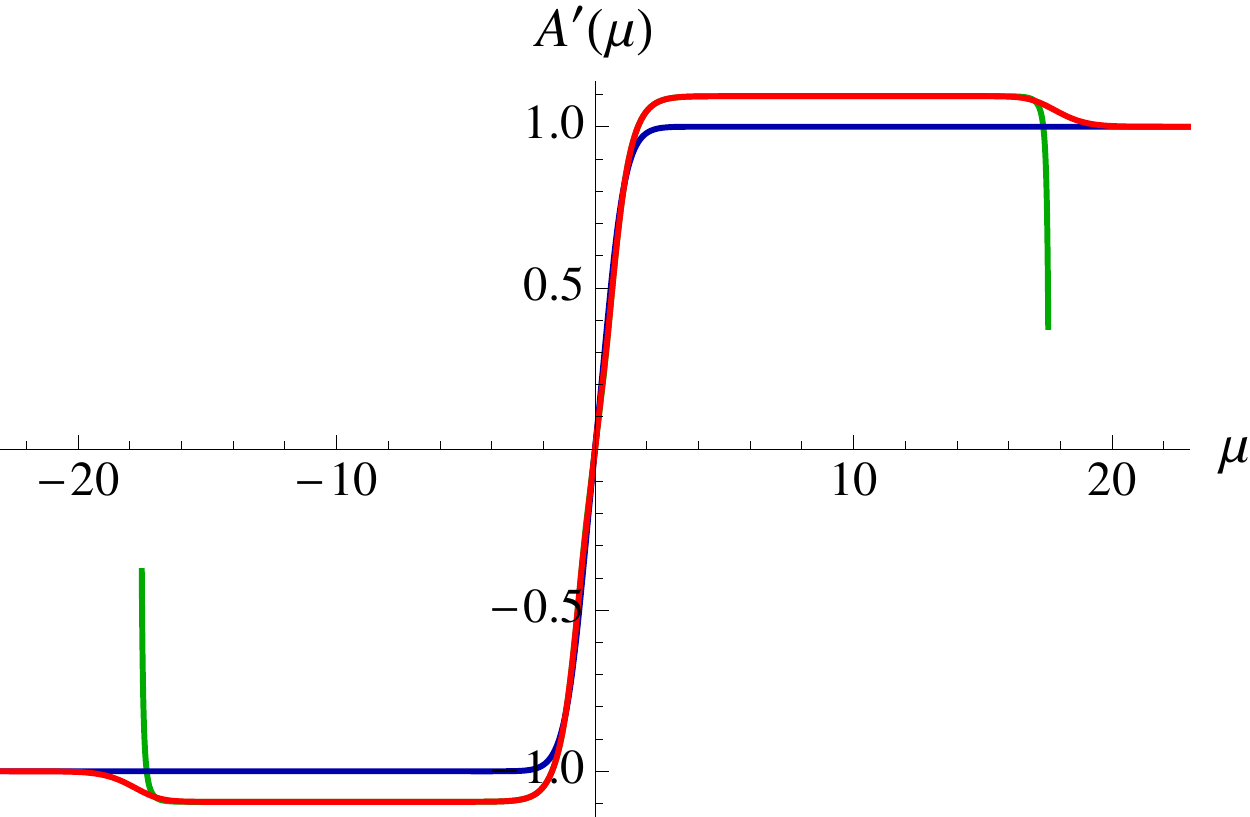}
\caption{Plots of $\alpha(\mu)$ and $A'(\mu)$ for three solutions with $\zeta_0=0$ and  $\alpha_0=0.15$ (blue), $\alpha_0=0.1756087990472$  (red) and $\alpha_0= 0.1756087990474$ (green).}
\label{aldAz0}
\end{figure}

The more interesting class of solutions arises when the turning point lies on the positive real axis, $\zeta_*=0$. This is evident from the  second plot in Figure~\ref{symmetricG2plots}.  Once again, for small values of $\alpha_*$, we find closed loops of $SO(8)$/$SO(8)$ Janus solutions with different values of  $\Delta\zeta<\pi$. However, as the  turning point approaches the point at the intersection of the four colored regions at $\alpha_*=\alpha_{cr}$, where 
\begin{equation}\label{}
0.1756087990472\ldots \leq \alpha_{cr}\leq 0.1756087990473\ldots\,,
\end{equation}
the solution also begins to swing close to the $G_2$ critical points. In particular, for $\alpha_*$ very close, but smaller than $\alpha_{cr}$, one obtains what looks like a ``limiting loop:'' At each end it is almost exactly a steepest ascent from the $SO(8)$ to the $G_2^\pm$ points along the ridges of the real superpotential, $W$, and then it swings between the two $G_2$ points. If one examines the plot of $\alpha(\mu)$ and $A'(\mu)$ in Figure~\ref{aldAz0}, one sees that such a  solution (plotted in red)   involves a  rapid evolution from the $SO(8)$  to  $G_2^\pm$ critical points, where it spends a long period  before it  swings between the two $G_2$ points relatively rapidly.  Numerical results suggest  that by fine tuning $\alpha_*$ to $\alpha_{cr}$ the solution can be made to approach the $G_2^\pm$ points arbitrarily close and stay there arbitrarily long. 

On the other side of the special point, where $\alpha_*>\alpha_{cr}$, we find solutions that become singular on both sides at finite values of $\mu$. Once more, as $\alpha_*$ approaches $\alpha_{cr}$, those solutions approach the $G_2^{\pm}$ points arbitrarily close and run off to infinity afterwards along the ridge of $W$, see Figure~\ref{symmetricG2plots} and the green plots in Figure~\ref{aldAz0}.

Since the two families, $\alpha_*<\alpha_{cr}$ and $\alpha_*>\alpha_{cr}$,  of solutions depend continuously on $\alpha_*$, and given the behavior of those solutions close to  the $G_2$ points, we expect that there exists a unique separating solution for $\alpha_*=\alpha_{cr}$ that describes a  $G_2^-/G_2^+$ interface. 

It appears that such a solution might be rather special in that it stays close to the $G_2^-$ and $G_2^+$ points infinitely long and then makes a quick transition between the two points close to $\mu=0$. Given the limited numerical accuracy and very slow convergence, we cannot predict whether that transition will be smooth, as for the approximating solution in  Figure~\ref{aldAz0}, or whether it will become a discrete jump. In other words, looking at the plots in Figure~\ref{aldAz0}, the question is whether in the limit $\alpha_*\rightarrow\alpha_{cr}$,  as the two sides of the plots asymptote the $G_2$ values over an increasing range of $\mu$, the transition around $\mu=0$ shrinks to zero width. 

On the other side, there is a compelling physical argument for the existence of a $G_2^-/G_2^+$ interface solution.  First, the loops to the left of the $SO(8)$ point and the smaller loops to the right represent Janus interfaces between $SO(8)$ phases. As  $\alpha_*$ approaches $\alpha_{cr}$,  the solution gets more and more controlled by the $G_2$ points.  The limiting loops describe solutions in which the theory is initially perturbed so that it undergoes a rapid and standard holographic RG flow, as in \cite{Bobev:2009ms}, to settle in a $G_2$ phase on each side of the defect, where it remains for a significant interval in $\mu$.  The  limiting solution is thus a $G_2^-$ to $G_2^+$ Janus and the only role of the $SO(8)$ point is to provide a way to generate the $G_2^\pm$  phases on either side of the defect.

What makes this solution especially interesting is the fact that the two $G_2$ phases on either side of the defect are physically distinct:  They have different signs for $\zeta$, which means that they have different signs for the pseudoscalar. In eleven dimensions this means that the two phases have opposite signs for the components of the $A^{(3)}$ gauge field on the $S^7$.\footnote{ 
Indeed, given that   the complete set of uplift formulae for the $G_2$ invariant $AdS_4$ critical point is now known \cite{Godazgar:2013nma}, one can demonstrate this explicitly: Our phase parameter $\zeta$ is a called $\alpha$ in  \cite{Godazgar:2013nma} and from formulae (73)--(76) and (96) of  \cite{Godazgar:2013nma} one can see that $A^{(3)}$ changes sign if one changes the sign of the phase.}  This is thus the M-theory analog of a conformal domain wall between two opposing magnetic fields.

\subsubsection{Asymmetric solutions}
\label{AsymmJanus}

\begin{figure}[t]
\centering
\includegraphics[width=7cm]{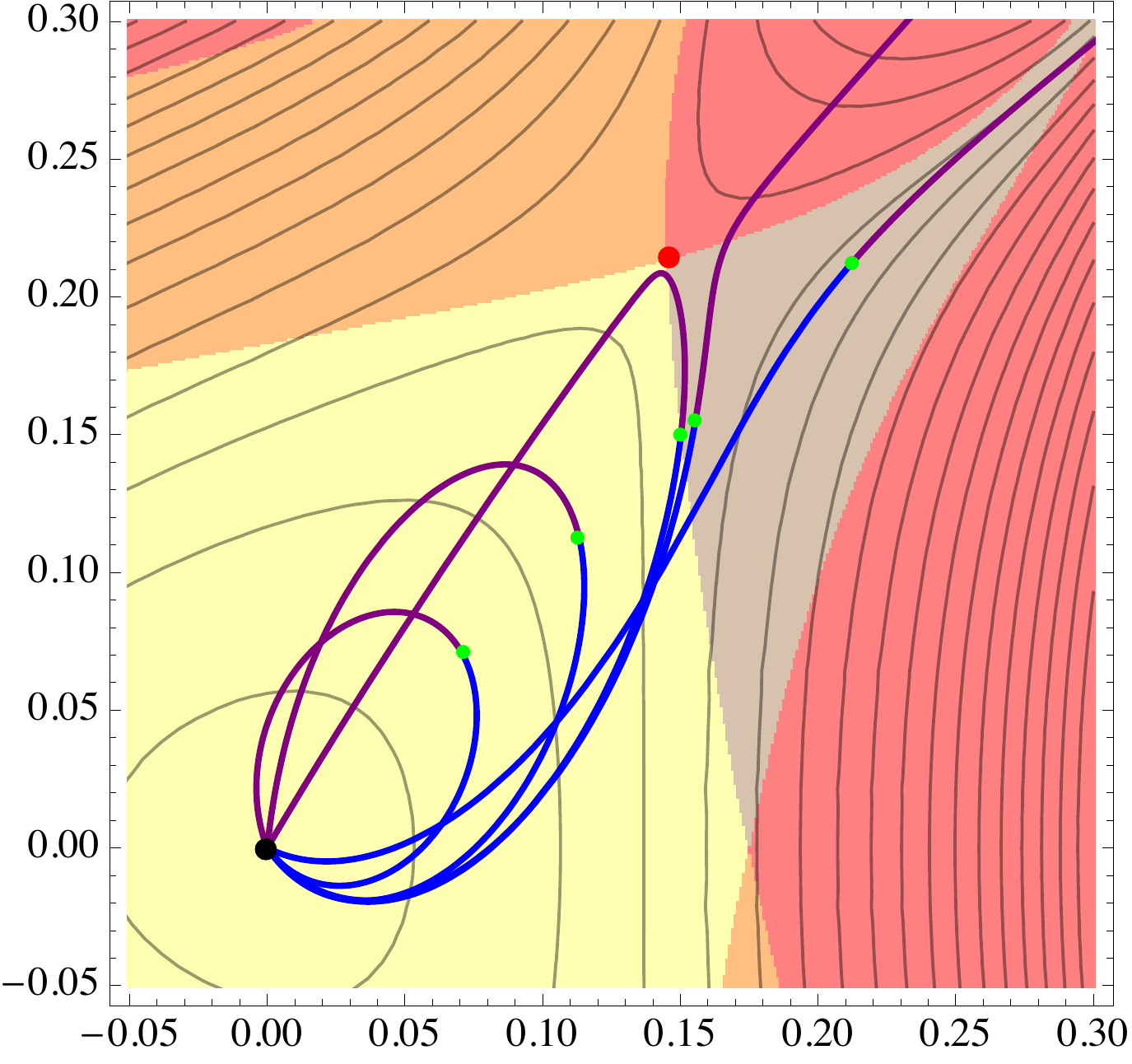}
\qquad
\includegraphics[width=7cm]{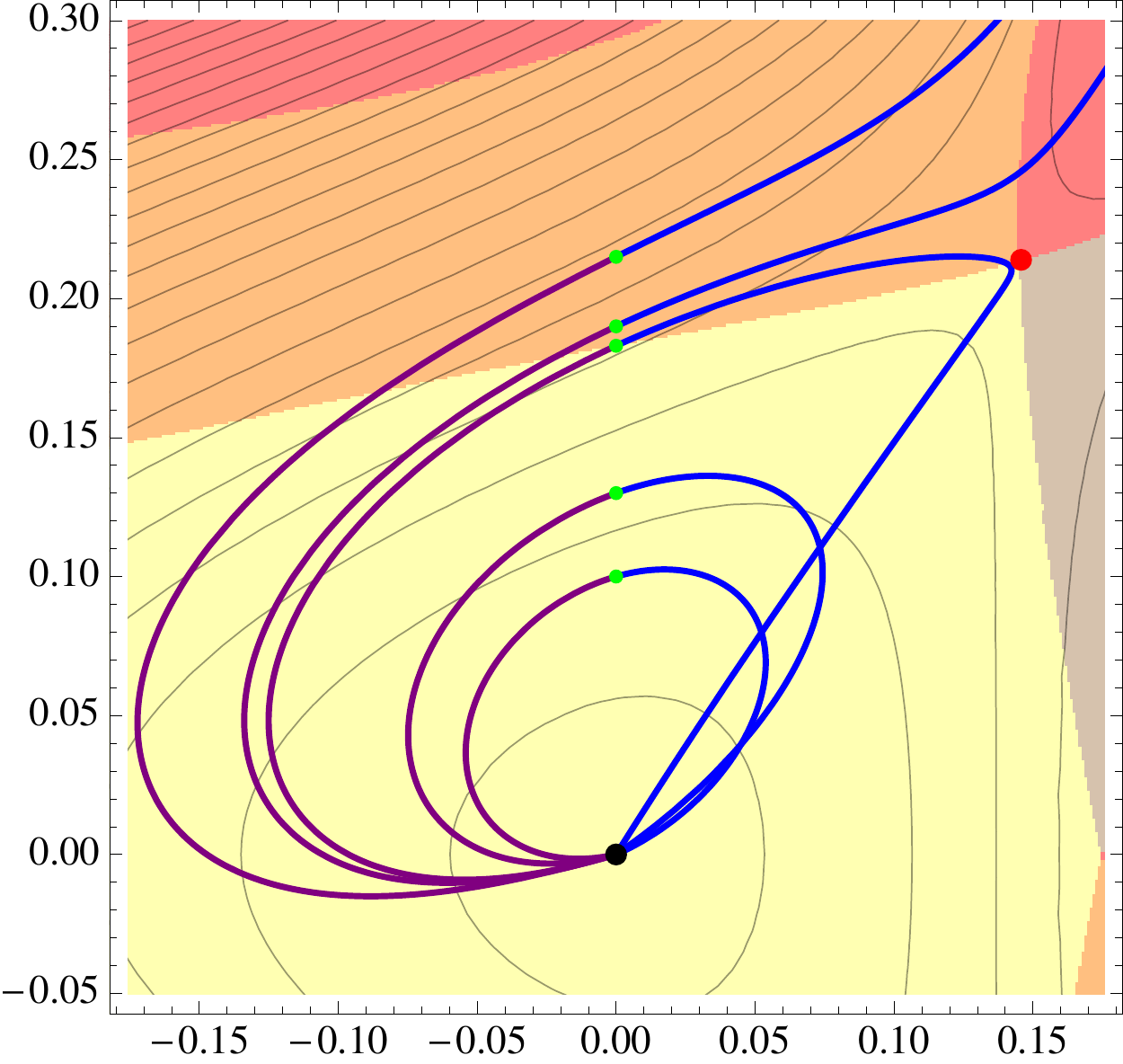}
\caption{A representative set of solutions for $\zeta_0=\pi/4$ and $\pi/2$. As the turning point  approaches the orange  or grey   boundary the  $SO(8)/SO(8)$ Janus solutions asymptote  to a $G_2/SO(8)$ or  $SO(8)/G_2$  Janus solution, respectively. }
\label{OrgBdryApproach}
\end{figure}

\begin{figure}[t]
\centering
\includegraphics[width=7cm]{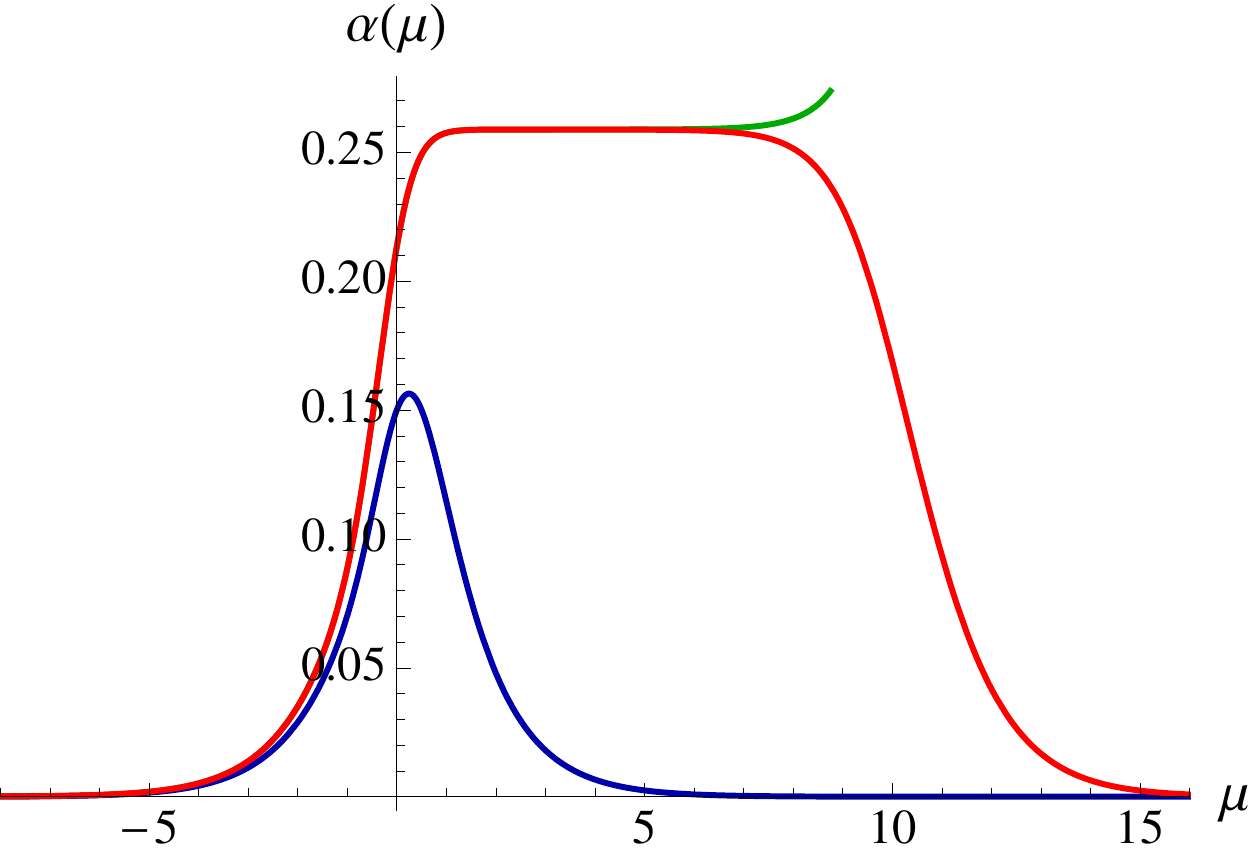}
\qquad
\includegraphics[width=7cm]{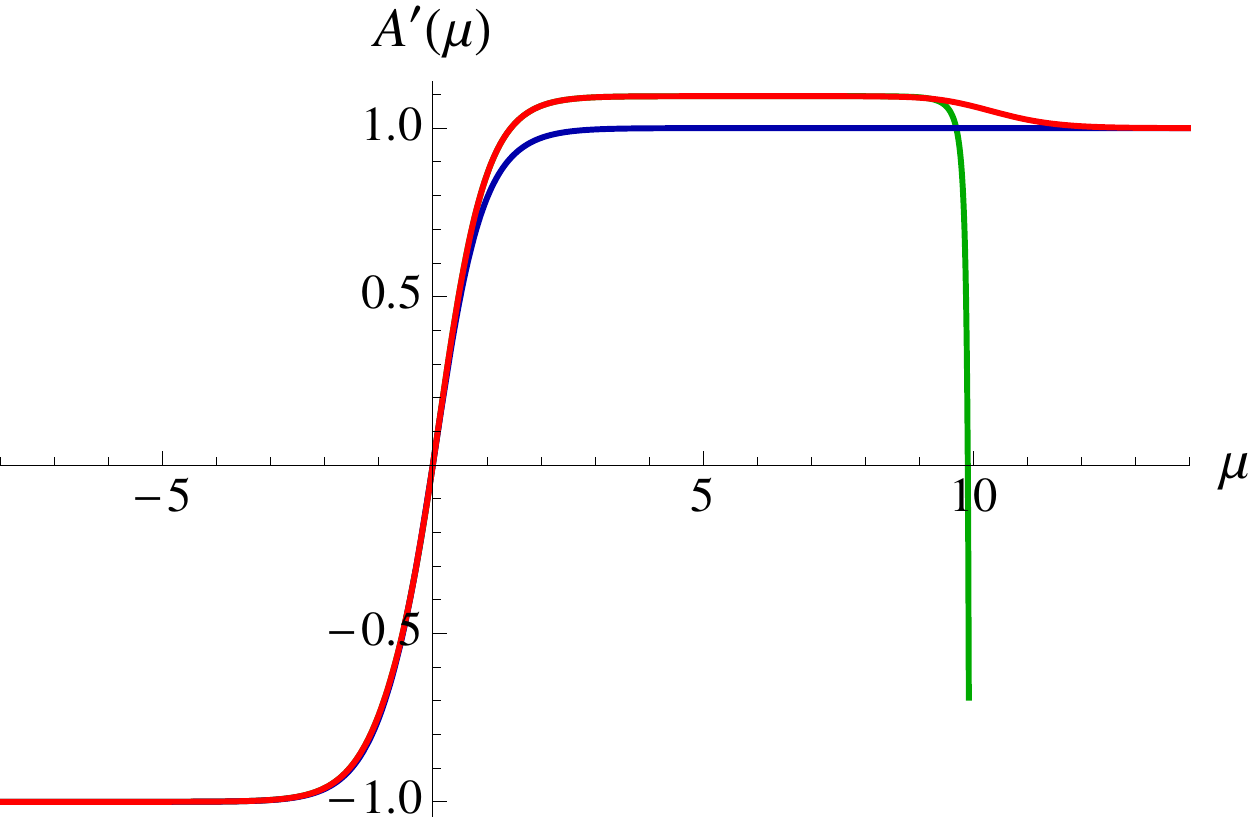}
\caption{Plots of $\alpha(\mu)$ and $A'(\mu)$ for three solutions with $\zeta_0=\pi/4$ and  $\alpha_0=0.15$ (blue), $\alpha_0=0.21332461$  (red) and $\alpha_0= 0.21332464$ (green).}
\label{aldAzetapio4}
\end{figure}

\begin{figure}[t]
\centering
\includegraphics[width=7cm]{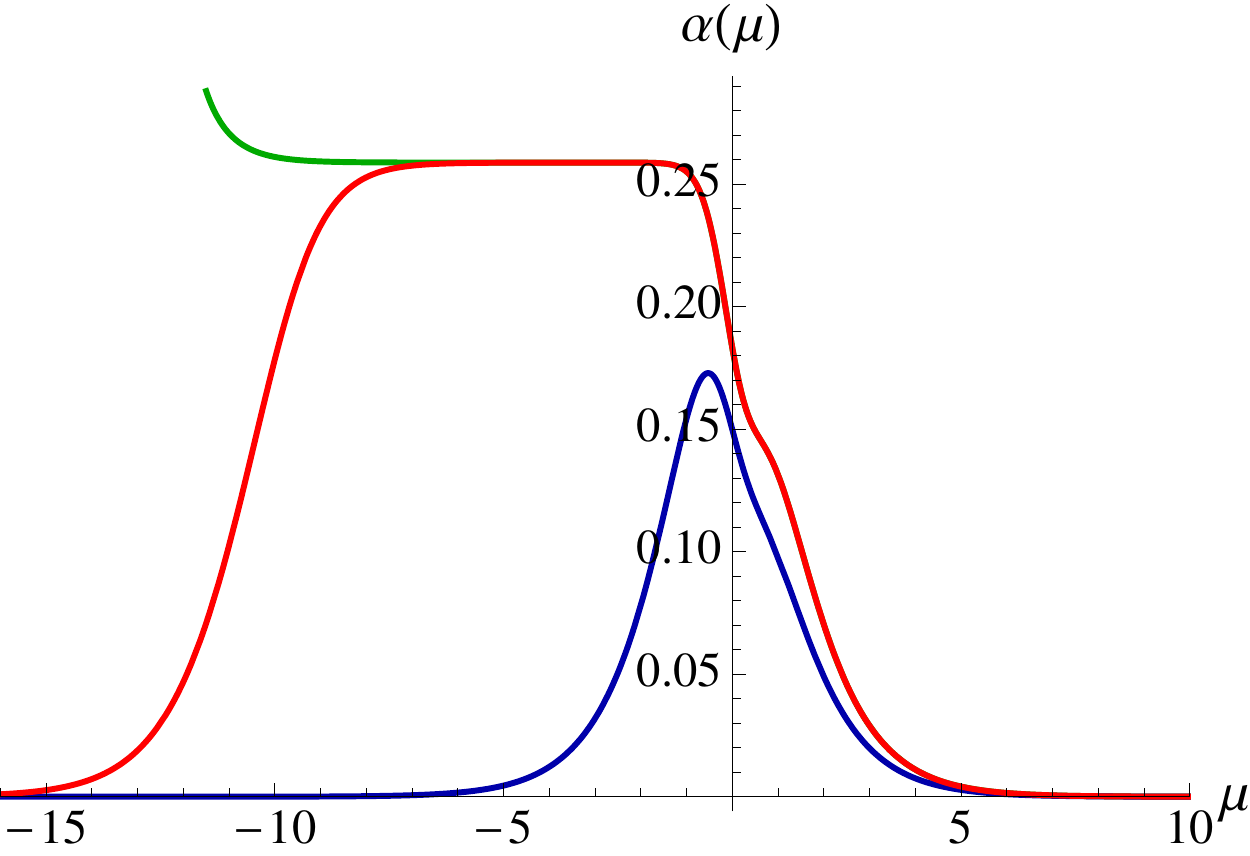}
\qquad
\includegraphics[width=7cm]{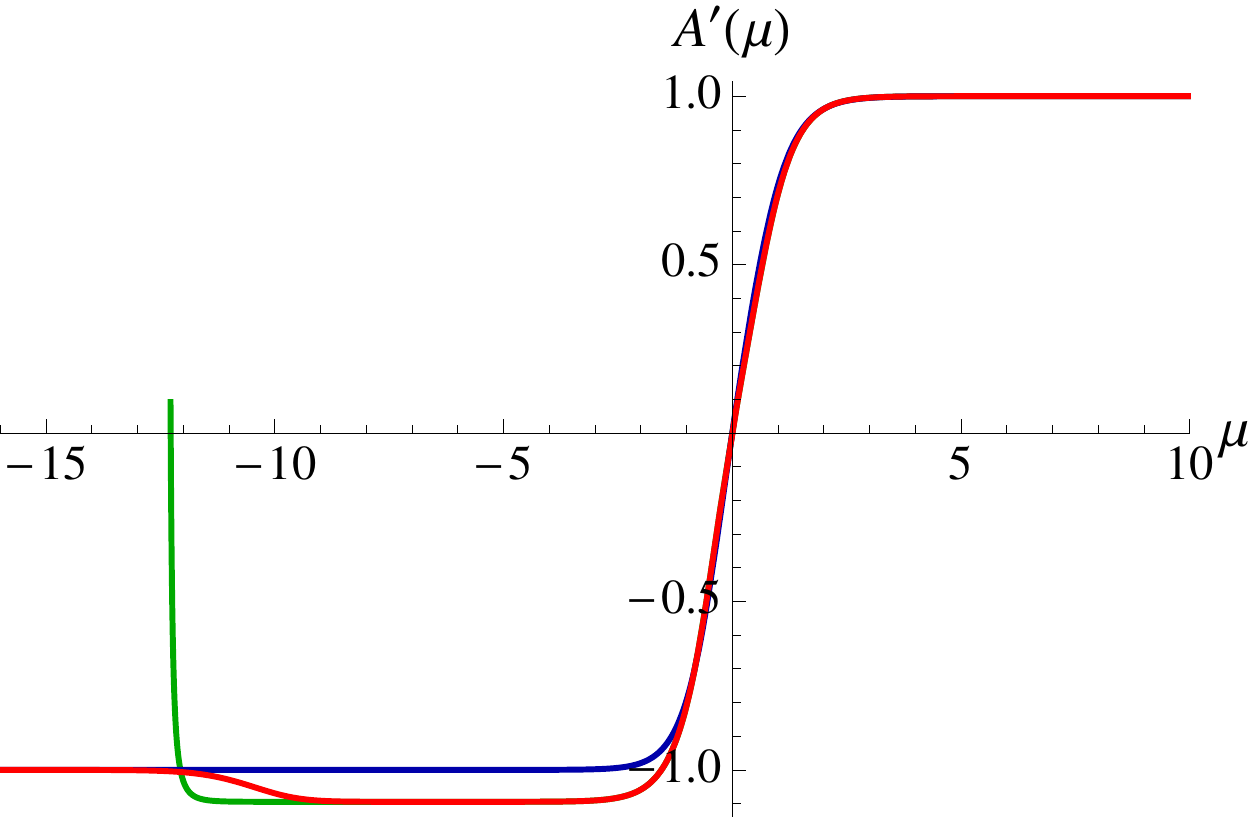}
\caption{Plots of $\alpha(\mu)$ and $A'(\mu)$ for three solutions with $\zeta_0=\pi/2$ and  $\alpha_0=0.15$ (blue), $\alpha_0=0.18337147$  (red) and $\alpha_0= 0.18337149$ (green).}
\label{aldAzetapio2}
\end{figure}

One can obviously move the turning point  of $A(\mu)$ for the Janus solution into the upper or lower half-plane of the scalar manifold.  These classes of solutions are related to each other by complex conjugation and so we focus on solutions with $\zeta_* >0$.    Once again, if the turning point lies within the yellow region,   see Figure~\ref{AllTogether}, the solutions are loops that start and finish at the $SO(8)$ point. As above, we interpret them as  Janus solutions between two copies of the $SO(8)$ phase, where,   depending on the asymptotic value of the angle, $\zeta$, at infinity, different mixtures of dual operators have been added to the field theory Lagrangian or are developing vevs within the phase on each side of the domain wall.

There are two interesting boundaries of the yellow region:  the orange boundary and the grey boundary. 
As the turning point approaches the grey boundary, see   Figure~\ref{OrgBdryApproach} and Figure~\ref{aldAzetapio4}, the purple side of the solution, $\mu>0$, becomes more and more controlled by the $G_2^+$ point.  At the grey boundary, the $SO(8)$ phase on the $\mu>0$ side rapidly undergoes an RG flow to establish a $G_2$ phase.  The solution  then loops back to the $SO(8)$ point via the $A(\mu)$-turning point.  Thus the right-hand side of the interface ($\mu>0$) is in the $G_2$ phase while the left-hand side ($\mu<0$) is controlled by the $SO(8)$ point.  This therefore represents a Janus interface with the $G_2$ phase on the right and the  $SO(8)$ phase on the left.  This description is also evident from the values of $A'(\mu)$ on either side of the interface in Figure \ref{aldAzetapio4}.

As the turning point approaches the orange boundary, see   Figure~\ref{OrgBdryApproach} and Figure~\ref{aldAzetapio2}, 
the solution for $\mu<0$ becomes increasingly controlled by the $G_2^+$ point.  At the  boundary, the $SO(8)$ phase described by that side of the solution   rapidly undergoes an RG flow to establish a $G_2^+$ phase for $\mu\rightarrow-\infty$  while the   phase for $\mu\rightarrow+\infty$ is controlled by the $SO(8)$ point.  This therefore represents an interface with the $SO(8)$ phase on the left and the $G_2^+$ phase on the right.

If the $A(\mu)$-turning point, $(\alpha_*,\zeta_{*})$, crosses into an orange or grey region then one end of the solution runs to Hades and the other end goes back to the the $SO(8)$ point, and if the turning point moves into a pink region then both ends of the solution run to Hades. Figure \ref{AllTogether} displays the features of  the various domains we have described here.

For the $SO(8)$/$SO(8)$ Janus solutions the asymptotic analysis at $\mu \to \pm \infty$ is again similar to that of  Section \ref{subsec:HoloSO4}. The operators $\cO_{1,2}$ in \eqref{Obdef}--\eqref{Ofdef} are given by:
\begin{equation}\label{O1O2G2}
\cO_1 = \cO_b^{88}\;, \qquad\qquad \cO_2 = \cO_f^{88}\;.
\end{equation}
The value of $\zeta$ at $\mu \to \pm \infty$ controls the linear combination of the operators $\mathcal{O}_1$ and $\mathcal{O}_2$ that is being turned on. The new feature however is that for a generic Janus solution in the $G_2$ truncation we have $\lim_{\mu\to \infty}(\zeta(\mu)-\zeta(-\mu))\neq\pi$. This means that different linear combination of the bosonic and fermonic bilinear operators $\mathcal{O}_1$ and $\mathcal{O}_2$ are driving the flow on each side of the interface. 

The three-dimensional field theory dual to the $G_2$ critical point is poorly understood since it is strongly coupled and has only $\cN=1$ superconformal symmetry \cite{Bobev:2009ms}. The limited information we have about this theory comes from holography and therefore it is hard to identify the field theory deformations that trigger the $SO(8)$/$G_2^\pm$ and $G_2^+$/$G_2^-$ Janus solutions.

\section{Conclusions}
\label{sec:Conclusions}

We have seen, once again, that gauged supergravity can be an immensely powerful tool for constructing interesting holographic solutions. While the truncation to gauged $\Neql8$ supergravity limits one to the holographic duals of essentially bilinear operators and thereby limits the classes of flows that can be studied, the fact that the higher-dimensional fields are relatively simply and highly efficiently encoded in the four-dimensional theory means that one can find many solutions that would represent a formidable, if not impossible, task from the perspective of the higher-dimensional supergravities. Even with the four-dimensional solutions that we constructed at hand it is generally not a simple task to construct their eleven-dimensional uplift. Due to the large global symmetry and the previous results in the literature on consistent truncations it is possible to uplift the $SO(4)\times SO(4)$ Janus solutions to eleven dimensions with little effort (see Appendix \ref{appendixB}). The uplift of arbitrary solutions in the $SU(3)\times U(1)\times U(1)$ and $G_2$ truncations is generically not known. The uplift of the metric is relatively straightforward to perform using the uplift formula of \cite{de Wit:1986iy}. It is much more subtle to obtain the fluxes of the eleven-dimensional solution and the recent results on consistent truncations of eleven-dimensional supergravity \cite{Nicolai:2011cy,deWit:2013ija,Godazgar:2013nma} may provide useful methods for attempting such a construction.

It would be nice to have a better field-theory understanding of the interface defects we have constructed holographically. We provided evidence that, in addition to vevs, the presence of the defect introduces a deformation of the Lagrangian and it is important to clarify how this happens in the dual field theory. The analysis for the field theory duals to the $SO(8)$/$SO(8)$ Janus solutions should proceed along the lines of the calculations performed in \cite{D'Hoker:2006uv} for $\cN=4$ SYM. It will be much more challenging to understand the $SO(8)$/$G_2^{\pm}$ and $G_2^{+}$/$G_2^{-}$ solutions in field theory due to the minimal amount of supersymmetry and the limited field theory information about the $G_2$ fixed points. More generally it will be nice to have a field theory classification of superconformal defects in the ABJM theory. There has been recent interesting work on boundary conditions in $\cN=2$ theories in three dimensions \cite{Okazaki:2013kaa} and one should be able to use similar techniques to systematically classify at least the 1/2-BPS defects as was done in \cite{Gaiotto:2008sd,Gaiotto:2008sa,Gaiotto:2008ak} for $\cN=4$ SYM.

Even within the extremely simple class of $SU(1,1)/U(1)$ coset models studied here we have found a plethora of new Janus solutions.  Of particular interest are the interfaces between different superconformal fixed points and especially the $G_2^+$/$G_2^-$ interface between two domains of opposite magnetic fields.  This leads to the obvious question of possible generalizations.  We have done some calculations within the larger $SU(3)$-invariant sector that has been much studied in ordinary holographic RG flows \cite{Bobev:2009ms,Warner:1983vz,Ahn:2000mf,Ahn:2000aq,Ahn:2001by,Ahn:2001kw,Ahn:2002eh,Ahn:2002qga,Bobev:2010ib}. It is evident that there are indeed Janus solutions that involve not only the $SO(8)$ and $G_2^\pm$ phases but incorporate the $\Neql2$ supersymmetric $SU(3) \times U(1)^\pm$ critical point as well.  We are continuing to investigate these flows \cite{BPWRG} and because of the $U(1)$ $\cR$-symmetry at the $\Neql2$ points,  the holographic field theory phase is better understood \cite{Benna:2008zy} and perhaps can lead to some non-trivial tests within the  theory. Then there are the flows to Hades:  From the field theory perspective it seems difficult for there to be a conformal interface between a superconformal phase and a Coulomb phase.  However, it would certainly be interesting to see if such an interface is predicted by holography.

Although we have concentrated on examples of four-dimensional gravitational actions that arise as a consistent truncation of the $\Neql8$ gauged supergravity it should be emphasized that our construction works for any holomorphic superpotential, $\cal{V}$, and any real number, $k$. Therefore any four-dimensional supergravity theory with a $SU(1,1)/U(1)$ scalar manifold and a holomorphic superpotential will admit Janus solutions of the type discussed here. If $\cal{V}$ has any non-trivial critical points there will also be RG flow domain walls analogous to those that we found in the $G_2$ truncation.

Going beyond ABJM theory and $\Neql8$ supergravity in four dimensions there are obvious questions about the extent to which our results can be generalized to gauged supergravity theories in higher dimensions. Starting at the top, it is relatively easy to see that there are no supersymmetric Janus solutions in seven-dimensional maximal gauged supergravity. We have explicitly looked for such solutions and have shown that they do not exist. If there were Janus solutions they would be dual to codimension-one superconformal defects in the six-dimensional $(2,0)$ theory.  The reason for this negative result is that the five-dimensional superconformal group $F(4)$, which should be the symmetry group of the defect, is not a subgroup (see \cite{D'Hoker:2008ix} for a proof) of the $OSp(8|4)$ superconformal symmetry group of the six-dimensional $(2,0)$ theory. This implies that there are no superconformal codimension one defects in the $(2,0)$ theory and its $(1,0)$ orbifold generalizations.

In five-dimensional, gauged $\Neql8$ supergravity the possibilities are much richer and Janus solutions are already known \cite{Clark:2005te,Suh:2011xc}.  Here we are, of course, dealing with  a consistent truncation of IIB supergravity and the holographic dual of $\Neql4$ Yang Mills theory.  The interfaces are $(2+1)$-dimensional and the superconformal ones, for which the theory living on the two sides of the defect is $\cN=4$ SYM, were classified in \cite{Clark:2004sb,D'Hoker:2006uv}. There are $1/2$, $1/4$ and $1/8$-BPS superconformal interfaces and some of their gravity duals are known. The $1/2$-BPS Janus was found in IIB supergravity in \cite{D'Hoker:2007xy} and the $1/8$-BPS Janus  was found first in five-dimensional supergravity in \cite{Clark:2005te} and then uplifted to ten dimensions in \cite{D'Hoker:2006uu,Suh:2011xc}. The five-dimensional supergravity dual of the $1/4$-BPS Janus will be presented in \cite{BPW5d}.

It is therefore evident that there is still much to be learned about Janus solutions by using gauged supergravity theory in four and five dimensions and that this paper represents a fraction of the interesting results that are within reach.

\bigskip
\bigskip
\leftline{\bf Acknowledgements}
\smallskip
We would like to thank Costas Bachas, Chris Beem, Eric D'Hoker, John Estes, Davide Gaiotto, Jaume Gomis, Michael Gutperle, Murat G\"unaydin, Darya Krym, and Balt van Rees for helpful discussions. Most of this work was done while NB was a postdoc at the Simons Center for Geometry and Physics and he would like to thank this institution for the great working atmosphere. The work of NB is supported by Perimeter Institute for
Theoretical Physics. Research at Perimeter Institute is supported by the Government of Canada through Industry Canada and by the Province of Ontario through the Ministry of Research and Innovation. The work of KP and NPW is supported in part by DOE grant DE-FG03-84ER-40168. NPW is grateful to the IPhT, CEA-Saclay, the Institut des Hautes Etudes Scientifiques (IHES), Bures-sur-Yvette, and Perimeter Institute for Theoretical Physics for hospitality while this work was completed. NPW would also like to thank the Simons Foundation for their support through a Simons Fellowship in Theoretical Physics.  We would all like to thank the KITP, Santa Barbara for warm hospitality during the initial stages of this project.



\begin{appendices}

\appendix
\section{Four-dimensional $\Neql{8}$ supergravity}

\renewcommand{\theequation}{A.\arabic{equation}}
\renewcommand{\thetable}{A.\arabic{table}}
\setcounter{equation}{0}
\label{appendixA}

The scalars of the $\Neql8$ theory lie in the coset $E_{7(7)}/SU(8)$ whose non-compact generators can be represented by a complex, self-dual four-form, considered as a $28 \times 28$ matrix:   $\Sigma_{IJKL}=\Sigma_{[IJ][KL]}$.  That is, one defines a non-compact generator, $G$,  in the 56-dimensional representation of $E_{7(7)}$ (see, Appendix A in \cite{de Wit:1982ig}),
\begin{equation}
\label{esevgen}
G\eql \left(\begin{matrix}
0 & \Sigma_{IJKL} \\[6 pt] \Sigma^{MNKL} & 0
\end{matrix}\right)\,.
\end{equation} 
The components $\Sigma^{IJKL}$ are complex conjugate of $\Sigma_{IJKL}$ and the self-duality constraint is
\begin{equation}
\label{selfdual}
\Sigma_{IJKL}\eql {1\over 24}\epsilon_{IJKLMNPR}\Sigma^{MNPR}\,.
\end{equation}
The exponential map,  $G~\rightarrow~V\equiv \exp(G)$, defines coset representatives and determines the scalar vielbein and its inverse, 
\begin{equation}
\label{scalviel}
V\eql\left( \begin{matrix}
u_{ij}{}^{IJ} & v_{ijKL}\\[6 pt] v^{klIJ} & u^{kl}{}_{KL}\end{matrix}\right)\,,\qquad V^{-1} \eql\left( \begin{matrix}
u^{ij}{}_{IJ}  & - v_{klIJ}\\[6 pt] - v^{ijKL} &   u_{kl}{}^{KL} \end{matrix}\right)\,,
\end{equation}
in terms of which the supergravity action is constructed.\footnote{In this subsection capital Latin indices, $I,J,K,
\ldots$, transform under $SO(8)$ and small Latin indices, $i,j,k,\ldots$, transform under $ SU(8)$.}

One then defines a composite $SU(8)$ connection acting on the $ SU(8)$ indices according to
\begin{equation}
\label{compconn}
\cD_\mu \varphi^i ~\equiv~ \partial_ \mu\varphi^i  + \coeff{1}{2}  \cB^{\,i}_{\mu \, j} \, \varphi^j\,, 
\end{equation}
and introduces the minimal couplings of the $ SO(8)$ gauge fields with coupling constant $g$.  For example
\begin{equation}
\label{covderu}
\cD_\mu u_{ij}{}^{IJ}  \eql \partial _\mu u_{ij}{}^{IJ} -   \coeff{1}{2}  \cB^{\,k}_{\mu \, i} u_{kj}{}^{IJ} - \coeff{1}{2}  \cB^{\,k}_{\mu \, j} u_{ik}{}^{IJ}    - g \left(A_\mu^{KI} u_{ij}{}^{JK} - A_\mu^{KJ}u_{ij}{}^{IK}  \right)\,.
\end{equation}
The composite connections are then defined by requiring that
\begin{equation}
\label{conndefn}
(\cD_\mu V)\, V^{-1}  ~=~ - \ds\frac{\sqrt{2}}{4}\left( \begin{matrix}
0  & \cA_\mu{}^{ijkl}   \\[6 pt]
  \cA_\mu{}_{\,mnpq} & 0  \end{matrix}\right)   \,.
\end{equation}
More directly, 
\begin{equation}\label{atens}
\cA_\mu{}^{ijkl}\eql -2\sqrt 2\,\left(u^{ij}{}_{IJ}\nabla_\mu v^{klIJ}-v^{ijIJ}\nabla_\mu u^{kl}{}_{IJ}\right)\,,
\end{equation}
where the covariant derivative in \eqref{atens} is only with respect to the $SO(8)$ indices of the scalar vielbeins, that is
\begin{equation}\label{covderv}
\nabla_\mu v^{ijIJ}\eql \partial _\mu v^{ijIJ}-g \left(A_\mu^{KI}v^{ijJK}-A_\mu^{KJ}v^{ijIK}\right)\,,
\end{equation}
and similarly for other fields.

The supergravity action involving only the graviton and scalar fields of $\Neql8$ supergravity \cite{de Wit:1982ig} is given by:
\begin{equation}
e^{-1}{\cal L}\ ~=~ \coeff{1}{2}\, R  ~-~ \coeff{1}{96}\, \cA_\mu{}^{ijkl}\cA^\mu{}_{ijkl}  ~-~ g^2 \,\cP(\phi )\,,
\label{scalLag}
\end{equation}
where $g$ is the gauge coupling and the potential, $\cP$, is given by 
\begin{equation}
\label{scalpot}
\cP\eql   \coeff{3}{4}\left|A_1{}^{ij}\right|^2-\coeff{1}{24}\left|A_{2i}{}^{jkl}\right|^2 \,.
\end{equation}
The tensors $A_1{}^{ij}$ and $A_{2i}{}^{jkl}$ that appear in the scalar potential above and in the supersymmetry variations in Section~\ref{Sect:DetailSusy} are defined by \cite{de Wit:1982ig}
\begin{equation}
\label{Atensors}
A_1{}^{ij}\eql \coeff{4}{21} T_k{}^{ikj}\,,\qquad A_{2i}{}^{jkl}\eql -\coeff{4}{3}T_i{}^{[jkl]}\,,
\end{equation}
where
\begin{equation}\label{ttensor}
T_i{}^{jkl}~\equiv~\left(u^{kl}_{IJ}+v^{klIJ}\right)\left(u_{im}{}^{JK}u^{jm}{}_{KI}-v_{imJK}v^{jmKI}\right)\,,
\end{equation}
is the  $T$-tensor.

As discussed in Section~\ref{Sect:JanusSols}, for our $SL(2,\IR)$ coset theories the Lagrangian \eqref{scalpot} reduces to
\begin{equation}
e^{-1}{\cal L}\ ~=~ \coeff{1}{2}\, R  ~-~ k\, \dfrac{ g^{\mu \nu} \, \partial_\mu z  \, \partial_\nu \bar{z}}{(1-z\bar{z})^2}  \, ~-~ g^2e^{\mathcal{K}}(\mathcal{K}^{z\bar{z}}\nabla_{z}\mathcal{V}\nabla_{\bar{z}}\overline{\mathcal{V}} -3 \mathcal{V}\overline{\mathcal{V}}) \,,
\label{simpLag}
\end{equation}
where $\mathcal{K}$ is given by (\ref{Kahlerpot}).
In particular, the $E_{7(7)}$ tensor, $A_\mu{}^{ijkl}$,  gives rise to the scalar kinetic term, $\cA$, in (\ref{kinmat}) and certain eigenvalues of ${A_1}^{ij}$ are proportional to $e^{\cK/2} \cV(z)$, from which one obtains the holomorphic superpotential.  In the examples in which we find Janus solutions,  the tensors $A_{2\,l}{}^{ijk}$  can be written in terms of $\cK$, $\cV$ and their derivatives.

\section{$SO(4)\times SO(4)$ Janus in eleven dimensions} 
\renewcommand{\theequation}{B.\arabic{equation}}
\renewcommand{\thetable}{B.\arabic{table}}
\setcounter{equation}{0}
\label{appendixB}

In this appendix we present an   uplift of the $SO(4)\times SO(4)$ Janus solutions in Section~\ref{sec:SO4} to M-theory using standard uplift formulae from  \cite{Cvetic:1999au} and  verify  that  the resulting solutions in eleven dimensions have the same supersymmetry as in four dimensions. We then compare our solutions with the general form of eleven-dimensional solutions with half-maximal supersymmetry obtained in \cite{D'Hoker:2009gg,D'Hoker:2008wc} and, more recently, in \cite{Estes:2012vm,BDEK}.

\subsection{The uplift}
\label{wehavealift}

A complete uplift of the $SO(4)\times SO(4)$ invariant sector of the $\cN=8$ theory in four dimensions to M-theory was derived in \cite{Cvetic:1999au}. Subsequently, the explicit formulae in \cite{Cvetic:1999au} were used in \cite{Pope:2003jp} to uplift to M-theory the half-BPS holographic RG flows in this sector of four-dimensional supergravity and to prove that the uplift preserved  all supersymmetries of the solutions as expected.

Since the uplift formulae are valid for any solution of the four-dimensional theory, we may use them  to  obtain readily the eleven-dimensional counterparts of the  Janus solutions in Section~\ref{sec:SO4}. Using the same notation  as in \cite{Pope:2003jp},  the metric is
\begin{equation}\label{ourmetr}
ds_{11}^2\eql \Omega^2\,\Big(e^{2A}\,ds_{AdS_3}^2+\,d\mu^2\Big)+{2\over  g^2}\, \Omega^2\,\Big(d\theta^2+{\cos^2\theta\over Y}\,ds^2_{\sigma}+{\sin^2\theta\over \widetilde Y}\,ds^2_{\tilde\sigma}\Big)\,,
\end{equation}
where 
\begin{equation}\label{}
\Omega\eql \big(Y\widetilde Y\big)^{1/6}\,,
\end{equation}
is the warp factor and 
\begin{equation}\label{}
ds_\sigma^2\eql {1\over 4}(\sigma_1^2+\sigma_2^2+\sigma_3^2)\,,\qquad ds_{\tilde \sigma}^2\eql {1\over 4}(\tilde\sigma_1^2+\tilde\sigma_2^2+\tilde\sigma_3^2)\,,
\end{equation}
are the $SO(4)$ invariant metrics on the two unit radius $S^3$'s that are fibered over the interval $0\leq\theta\leq\pi/2$. The  two ubiquitous functions $Y$ and $\widetilde Y$  are defined as
\begin{equation}\label{defYs}
\begin{split}
Y(\mu,\theta) & \eql \cos ^2 \theta \, \left (\cosh (2 \alpha )+\sinh (2 \alpha ) \cos  \zeta \right)+\sin ^2 \theta  \,,\\[6 pt]
\widetilde Y(\mu,\theta) & \eql \sin ^2 \theta \, \left (\cosh (2 \alpha )-\sinh (2 \alpha ) \cos  \zeta  \right)+\cos ^2 \theta \,,
\end{split}
\end{equation}
and depend on both  the scalar fields, $\alpha(\mu)$ and $\zeta(\mu)$,  and the coordinate, $\theta$, on $S^7$. It may be worth noting that this formula for the uplifted metric is valid off-shell and follows from the general embedding of   $\cN=8$ supergravity into M-theory \cite{deWit:1984nz}.

At this point one may verify that the equations of motion and the Bianchi identity completely determine the four-form flux in eleven dimensions in terms of the metric functions in \eqref{ourmetr}. This is manifest in the  original formulae in \cite{Cvetic:1999au} and \cite{Pope:2003jp}. For completeness we quote here the full result in a more convenient form:\footnote{The overall sign of $F_{(4)}$ is opposite to that in  \cite{Pope:2003jp}. This is consistent with the  conventions in eleven dimensions that we are using, see  Appendix ~A in \cite{Bena:2004de}. We also note that there is a typo in the supersymmetry variation (4.1) in \cite{Pope:2003jp}, which  on the right hand side should have the opposite  sign of the flux term. }
\begin{equation}\label{fourflux}
\begin{split}
F_{(4)}\eql {\rm vol}_{AdS_3}\wedge\omega_{(1)}+d A_{(3)}^{\rm{sph}}\,,
\end{split}
\end{equation}
where ${\rm vol}_{AdS_3}$ is the volume form on $AdS_3$ with metric given in \eqref{AdS3Met} and
\begin{equation}\label{}
\omega_{(1)}\eql  e^{3A}(\sqrt 2\,g\, U\, d\mu +V\,d\theta)\,,
\end{equation}
is a one-form, with the functions $U$ and $V$   given by: 
\begin{equation}\label{}
\begin{split}
U(\mu,\theta) & \eql \cos (2 \theta ) \sinh (2 \alpha ) \cos  \zeta+\cosh (2 \alpha )+2\,,\\
V(\mu,\theta) & \eql {1\over 2}\sin (2\theta )  \left(4 \,\alpha ' \cos  \zeta-\zeta ' \sinh (4 \alpha
   )  \sin  \zeta\right)\,.
\end{split}
\end{equation}
Finally $A_{(3)}^{\rm{sph}}$ is a three-form potential along the two $S^3$'s:
\begin{equation}\label{}
A_{(3)}^{\rm{sph}}\eql f\,\sigma^1\wedge\sigma^2\wedge\sigma^3+\tilde f\,\tilde\sigma^1\wedge\tilde\sigma^2\wedge\tilde\sigma^3\,,
\end{equation}
with the functions
\begin{equation}\label{intflux}
\begin{split}
f(\mu,\theta) & \eql -{1\over2\sqrt 2\, g^3}{\cos^4\theta\over Y}\,{\sinh(2\alpha)\,\sin\zeta}  \,,\\[6 pt]\tilde f
(\mu,\theta) & \eql {1\over2\sqrt 2\, g^3}{\sin^4\theta\over\tilde  Y}\,{\sinh(2\alpha)\,\sin\zeta}\,.
\end{split}
\end{equation}
Notice that for $\zeta=0$ the components of the flux \eqref{fourflux} with legs along the internal manifold vanish. This is in agreement with the fact that for $\zeta=0$ the four-dimensional complex scalar $z$ corresponds to a scalar (as opposed to a pseudoscalar) and thus, to linear order, the deformation of the internal $S^7$ should be purely a metric mode.

One can verify using the equations in Section~\ref{SO4sols} that $\omega_{(1)}$ is closed and thus the four-form flux \eqref{fourflux} satisfies the Bianchi identity in eleven dimensions. Similarly, one verifies  directly that the metric \eqref{ourmetr} and the flux \eqref{fourflux} satisfy the field equations of eleven-dimensional supergravity (see, Appendix A in \cite{Bena:2004de} for our conventions) for any four-dimensional solution.\footnote{In fact, it is sufficient to use the equations satisfied by $A$, $\alpha$ and $\zeta$ in four dimensions. 
This is  guaranteed to work by the construction of the lift in \cite{Cvetic:1999au}.}  
 
 \def\kz{{\kappa_\zeta}}
The uplifted solutions at $\mu\rightarrow\pm\infty$ are asymptotic to $AdS_4\times S^7$. To check that explicitly, we substitute solutions  \eqref{alphasol1}, \eqref{zetasol1} and \eqref{Asol1} into the metric  functions above, where we take $\mu_0=0$ and fix the overall scale of the eleven-dimensional solution by setting 
\begin{equation}\label{fixggg}
g\eql {1\over\sqrt 2}\,.
\end{equation}
Then we have 
\begin{equation}\label{detYs}
\begin{split}
Y(\mu,\theta) & \eql 1+{2a(a+\kz\cos\zeta_0)\over 1-a^2}{\cos^2\theta\over\cosh^2\mu}+
{2a\,\kappa\kz \sin\zeta_0\over \sqrt{1-a^2}}\,{\cos^2\theta \tanh\mu\over \cosh \mu}\,,\\[6 pt]
\widetilde Y(\mu,\theta) & \eql 1+{2a(a-\kz\cos\zeta_0)\over 1-a^2}{\sin^2\theta\over\cosh^2\mu}-
{2a\,\kappa\kz \sin\zeta_0\over \sqrt{1-a^2}}\,{\sin^2\theta \tanh\mu\over \cosh\mu}\,,
\end{split}
\end{equation}
where $\kz=\pm 1$ gives the two branches of  the  solution for $\cos\zeta(\mu)$ and $\sin\zeta(\mu)$ in \eqref{zetasol1}. The asymptotic behavior of the solution is now manifest.

\subsection{Supersymmetry} 
 
Since the four-dimensional theory we are starting with is a consistent truncation of   $\cN=8$ gauged supergravity, which in turn, over the course of the past three decades, has been shown to be a consistent truncation of eleven-dimensional supergravity on $S^7$ \cite{deWit:1984nz,deWit:1986mz,de Wit:1986iy, Nicolai:2011cy}, one expects all supersymmetries to be preserved. 
To see that this is indeed the case, we briefly outline a direct calculation of   supersymmetries of our solutions in eleven dimensions following a similar calculation for the RG flows in  \cite{Pope:2003jp}.
Just as in Section~\ref{Sect:DetailSusy}, we find that the $AdS_3$ slicing introduces additional terms into the supersymmetry variations  which modify the usual analysis.
 
We choose the vielbeins, $e^M$, $M=1,\ldots,11$, for the metric \eqref{ourmetr} to be the same as in (3.10) of \cite{Pope:2003jp}, modulo the obvious difference in the $AdS_3$ vs $\mathbb{R}^{1,2}$ part of the metric and the choice of   letters for the coordinates,  $(t,x,r,\mu)$ instead of $(t,x,y,r)$, respectively.
Let us define the  operators, $\cals M_M$,  given by the algebraic part of the supersymmetry 
variations,
\begin{equation}\label{}
\cM_M \epsilon ~\equiv~ \Gamma^M(\delta\psi_M-\partial_M\epsilon)\qquad \text{(no sum on $M$) }\,.
\end{equation}
It follows from the symmetry of the solution that
\begin{equation}\label{}
 \cM_1\eql \cM_2\,,\qquad \cM_6\eql \cM_7\eql \cM_8\,,\qquad   \cM_9=\cM_{10}=\cM_{11}\,.
\end{equation}
   
As in Section~\ref{Sect:DetailSusy}, we look for  Poincar\'e supersymmetries with  $\epsilon$  constant 
along $t$ and $x$. Those  must satisfy the algebraic equation
\begin{equation}\label{mastproj}
\cM_1\epsilon\eql { e^{-A}\over 2\ell}\Omega^{-1}\,\Gamma^3\,\epsilon+\cM^\infty_1\epsilon\eql 0\,,
\end{equation}
where $\cM^\infty_M=\lim_{\ell\rightarrow\infty}\cM_M$ is the corresponding operator for the RG flow. It was was shown in \cite{Pope:2003jp} that all Poincar\'e supersymmetries are given by the solutions to this equation. Here, we find that the same result holds, except that with the additional $1/\ell$-term in \eqref{mastproj} there are only 8 instead of 16 Killing spinors that are constant along $t$ and $x$. 

To exhibit the explicit structure of those eight Killing spinors, let us define\footnote{Using $\Gamma^{12\ldots 11}=1$, one can rewrite the last product of $\Gamma$-matrices as $\Gamma^{123}$. }
\begin{equation}\label{}
\oPi\eql {1\over 4}\,\Big[\,1+\Gamma^{34}\Gamma^{678}+\Gamma^{35}\Gamma^{9\,10\,11}-\Gamma^{45}\Gamma^{67\ldots 11}\Big]\,,
\end{equation}
which is a projector onto an eight-dimensional subspace in the thirty two-dimensional spinor space. Let
\begin{equation}\label{su2elem}
\cals O(a_0,\vec a)\eql a_0+a_1\,\Gamma^{34}+a_2\,\Gamma^{35}+a_3\,\Gamma^{45}\,, 
\end{equation}
where
\begin{equation}\label{}
a_0^2+a_1^2+a_2^2+a_3^2\eql 1\,,
\end{equation}
be  an $SU(2)$ group element, with the inverse element given by $\cals O(a_0,-\vec a)$. It is now a matter of straightforward, albeit tedious, algebra to verify that the matrix equation for $\cals O(a_0,\vec a)$:
\begin{equation}\label{}
\cals M_1\,\cals O(a_0,\vec a)\,\oPi\eql 0\,,
\end{equation}
has a unique solution, up to an overall sign. The resulting expressions  for $a_0$ and $\vec a$ in terms of the flux components and the metric functions of the solution are quite complicated and we will omit  them here.

The Killing spinors that solve  {\it all} supersymmetry variations are now simply given by
\begin{equation}\label{the11ks}
\epsilon\eql \big(Y\widetilde Y\big)^{1/12}\,e^{A/2+r/2\ell}\,\cals O(a_0,\vec a)\,\epsilon_0\,,
\end{equation}
where $\epsilon_0$ is an arbitrary spinor in the eight-dimensional subspace
\begin{equation}\label{}
\qquad \oPi\,\epsilon_0\eql\epsilon_0\,,
\end{equation}
 constant along $t$, $x$, $r$, $\mu$ and $\theta$. The dependence of $\epsilon_0$   on  the coordinates of the two three spheres is given by the transitive action of two $SU(2)$'s as in \cite{Pope:2003jp}. 

One recognizes 
\eqref{the11ks} as a generalization of the corresponding solution \eqref{varmu} for the Killing spinor in four dimensions.  In comparison with various  RG flows and $AdS_4$ solutions,\footnote{See for example \cite{Halmagyi:2012ic} and the references therein.} a novel feature is the presence of an $SU(2)$ rotation \eqref{su2elem} in the $(r\mu\theta)$-subspace as opposed to a simpler   $U(1)$ rotation in the $(\mu\theta)$-subspace. 

Finally, we note that the chirality operator,  $\Gamma^{12}$, commutes with the supersymmetry variations and the eight solutions \eqref{the11ks} for Poincar\'e supersymmetries split into four with the positive and four with the negative chirality, in agreement with the analysis in Section~\ref{sec:SO4}. The negative chirality spinors are also constant along the three spheres. 

To summarize, we have shown that the uplift of the four-dimensional  $SO(4)\times SO(4)$ Janus solutions to M-theory yields a two-parameter family of distinct solutions with eight Poincar\'e and, after including the eight conformal Killing spinors, the total of 16 supersymmetries. The independent parameters for this family are $0\leq a<1$ and $-(\pi/2)\leq \zeta_0\leq\pi/2$. As noted above, the third parameter in  \eqref{parameters}, which is the gauge coupling constant, $g$, merely determines the overall scale of the solution. From now on we will take $g$ as in \eqref{fixggg} and also set $\ell=1$.

\subsection{Comparison with an existing classification}

It is both instructive and surprising  to compare   our half-BPS solutions of M-theory  with the existing classification of such solutions in the literature \cite{D'Hoker:2008wc,D'Hoker:2009gg}. Indeed, the goal of this section is to show that our solutions with general $\zeta_0$ \emph{do not} fall into the classification scheme of half-BPS solutions of eleven-dimensional supergravity with $SO(2,2)\times SO(4)\times SO(4)$ global symmetry \cite{D'Hoker:2008wc}.\footnote{See \cite{Yamaguchi:2006te,Lunin:2007ab} for earlier work on half-BPS solutions of M-theory with $SO(2,2)\times SO(4)\times SO(4)$ global symmetry.} It is only for the  special values, $\zeta_0=\pm (\pi/2)$, that  our solutions have the structure predicted by the analysis of \cite{D'Hoker:2008wc} and, in fact, reproduce all solutions found in \cite{D'Hoker:2009gg}.

The metric  \eqref{ourmetr} describes  an $AdS_3\times S^3\times S^3$ fibration over a two-dimensional base space, $\Sigma$, parametrized by the coordinates $\mu$ and $\theta$ and with the metric 
\begin{equation}\label{}
ds^2_\Sigma\eql \Omega^2(d\mu^2+{2\over g^2}d\theta^2)\,.
\end{equation}
It is thus reasonable to expect that  the uplifted solutions in Section~\ref{wehavealift}  should fall within a  classification scheme of half-BPS solutions of eleven-dimensional supergravity derived in  \cite{D'Hoker:2008wc}. The  backgrounds obtained in \cite{D'Hoker:2008wc} and further studied in \cite{D'Hoker:2009gg}  have a metric of the same form as in \eqref{ourmetr}, namely, 
\begin{equation}\label{dHmet}
ds_{11}^2\eql f_1^2 ds_{AdS_3}+f_2^2 ds_{S^3_1}^2+f_3^2 ds_{S^3_2}^2+4\rho^2|dw|^2\,,
\end{equation}
where  $f_1$, $f_2$, $f_3$ and $\rho$ are functions on a Riemann surface $\Sigma$ with a complex coordinate $w$. The radii  of $AdS_3$ and  the two three spheres are normalized to one.

\def\Wsq{{4|G|^4+(G-\overline G)^2}}
For the backgrounds in \cite{D'Hoker:2008wc}, the metric functions $f_1$, $f_2$, $f_3$ and $\rho$ take a very special form in terms of a real harmonic function, $h(w,\bar w)$, and a complex function, $G(w,\bar w)$, satisfying a first order  ``master equation''
\begin{equation}\label{masteqs}
\partial_w G\eql {1\over 2}(G+\overline G\,)\,\partial_w\log h\,,
\end{equation}
and  a point-wise constraint, 
\begin{equation}\label{constG}
|G(w,\bar w)|\geq 1\,,
\end{equation}
which must hold at all points on $\Sigma$. Specifically, the metric functions are: 
\begin{align}
\label{thef1}
f_1^6 & \eql {h^2\,\big[\Wsq \big]\over 16^2(|G|^2-1)^2}\,,\\[6 pt]
\label{thef2}
f_{2 }^6 & \eql {h^2\,(|G|^2-1)\over 4\,\big[\Wsq\big]^2}\,\left[2|G|^2+ i\,(G- \overline G)\right]^3\,,\\[6 pt]
\label{thef3}
f_{3 }^6 & \eql {h^2\,(|G|^2-1)\over 4\,\big[\Wsq\big]^2}\,\left[2|G|^2- i\,(G- \overline G)\right]^3\,, 
\end{align}
and
\begin{equation}\label{therho}
\rho^6  \eql {|\partial _w h|^6\over 16^2h^4}\,(|G|^2-1)\big[\Wsq \big]\,.
\end{equation}

We will now show that a necessary condition for an arbitrary metric \eqref{dHmet} to be expressed in terms of $h$ and $G$ as in \eqref{thef1}-\eqref{therho} is that the metric functions $f_1$, $f_2$ and $f_3$ satisfy the following inequality:
\begin{equation}\label{mastineq}
{f_1^4\over f_2^2f_3^2}\geq {1\over 4}\,,
\end{equation}
where the equality is allowed only at points where $\partial_wh$ vanishes.

A direct proof is quite straightforward: After substituting  \eqref{thef1}-\eqref{thef3} in \eqref{mastineq}, the left hand side is expressed only in terms of $\Im G$ and $|G|$ such that   \eqref{mastineq} is equivalent to  a quadratic inequality for $(\Im G)^2 $ with coefficients that depend on $|G|$. The pointwise constraint \eqref{constG} guarantees then that this inequality always holds.

A more systematic way for arriving at \eqref{mastineq} is to  solve  \eqref{thef1}-\eqref{thef3} for $h$ and $G$ and then impose the condition that the resulting $\rho^2$ in \eqref{therho} is real and positive. In fact, this is how the constraint \eqref{constG} was derived in \cite{D'Hoker:2008wc} in the first place. Let us summarize the main steps:

First, we obviously have
\begin{equation}\label{htofs}
h^2\eql 16 \,f_1^2 f_2^2f_3^2\,. 
\end{equation}
Then splitting  $G$ into   real and imaginary parts, $G\eql G_r+i\,G_i$,
the ratio of \eqref{thef2} and \eqref{thef3} yields
\begin{equation}\label{rateqs}
{f_2^2\over f_3^2} \eql {G_r^2+G_i^2-G_i\over G_r^2+G_i^2+G_i}\,,
\end{equation}
which we solve for $G_r^2$. Substituting the result in \eqref{thef1} and taking the cubic root of both sides, we find that all higher powers of $G_i$ cancel and the resulting quadratic equation for $G_i$ has two solutions:
\begin{equation}\label{Gisol}
G_i^\pm\eql \mp {h^2\,(f_2^2-f_3^2)\over 16 f_2^4f_3^4\pm h^2(f_2^2+f_3^2)}\,,
\end{equation}
where we used \eqref{htofs} to eliminate $f_1$. 
The corresponding solutions of \eqref{rateqs} for the real part are:
\begin{equation}\label{Grminsq}
(G_r^{\pm})^2\eql 4f_2^2f_3^2h^2{h^2\pm 4f_2^2f_3^2(f_2^2+f_3^2)\over 
\left[16 f_2^4f_3^4\pm h^2(f_2^2+ f_3^2)\right]^2}\,.
\end{equation}
Finally, substituting the two solutions in \eqref{therho}, we get
\begin{equation}\label{rhopls}
(\rho^\pm)^2\eql -f_2^2f_3^2\,{|\partial_w h|^2\over 16 f_2^4f_3^4\pm h^2(f_2^2+f_3^2)}\,.
\end{equation}
Clearly, we can't have positive $\rho$ given by $\rho^+$ and thus  $G$ must be given by the ``$-$'' solution, with the sign of $G_r^-$ in \eqref{Grminsq} still undetermined. The freedom in choosing the sign of $G_r^-$ is then   fixed  by \eqref{masteqs}, which is sensitive to the interchange $G\leftrightarrow \overline G$.
 
Next we observe that  by being forced to chose the $G^-$ solution, we must also satisfy two inequalities. \begin{equation}\label{ineq1}
h^2\geq 4f_2^2f_3^2(f_2^2+f_3^2)\,,\qquad \text{and} \qquad h^2\geq {16 f_2^4f_3^4\over f_2^2+f_3^2}\,,
\end{equation}
that follow from the reality of $G_r^-$ in \eqref{Grminsq} and $\rho^-$ in  \eqref{rhopls}, respectively. 
 Finally, by multiplying the two inequalities sidewise and then using \eqref{htofs} to eliminate $h$, we obtain \eqref{mastineq}.
 
We will now argue using \eqref{mastineq} that, with the exception of solutions with $\zeta_0=\pm(\pi/2)$, the metrics \eqref{ourmetr} with $Y$ and $\widetilde Y$ given in  \eqref{detYs} and $|\zeta_0|<\pi/2$ {\it cannot} be recast into the form above, at least if we assume that the identification holds term by term for the parts of metrics along $AdS_3$, the two three spheres and the Riemann surface. 

Evaluating  \eqref{mastineq} for the metric \eqref{ourmetr} we get
\begin{equation}\label{}
\begin{split}
{4f_1^4\over f_2^2f_3^2} ~\equiv ~ {e^{4A}\,Y\widetilde Y\over \sin^2(2\theta)}\geq 1\,.
\end{split}
\end{equation}
Using \eqref{Asol1} and \eqref{detYs} this inequality can be rewritten as
\begin{equation}\label{}
\begin{split}
{1\over 4}\,& \left[(1-a^2){\cosh^2\mu\over\sin^2\theta}+2a\,\left(a-\kappa_\zeta\cos\zeta_0-\sqrt{1-a^2}\,\kappa\,\kappa_\zeta\sin\zeta_0\sinh\mu\right)
\right]\\
& \qquad \qquad \times \left[(1-a^2){\cosh^2\mu\over\cos^2\theta}+2a\,\left(a+\kappa_\zeta\,\cos\zeta_0+\sqrt{1-a^2}\,\kappa\,\kappa_\zeta\,\sin\zeta_0\sinh\mu\right)
\right] \geq 1\,.
\end{split}
\end{equation}

It can be shown that this inequality is obeyed for any allowed value of $(\mu,\theta)$ only for $\zeta_0=\pm (\pi/2)$. For any other value of $\zeta_0$ there is a region in $(\mu,\theta)$ space where the inequality is violated. This means that our solutions with $\zeta_0\neq\pm (\pi/2)$ \emph{cannot} be written as solutions to the BPS equations in the form studied in \cite{D'Hoker:2008wc,D'Hoker:2009gg}.

For $\zeta_0=\pm (\pi/2)$ and arbitrary $a$, one can check that the eleven-dimensional solutions  in Appendix~\ref{wehavealift} are identical to the Janus solutions found in \cite{D'Hoker:2009gg}. For those solutions, the Riemann surface is an infinite strip, $-\infty<x<\infty$, $0\leq y\leq\pi/2$, and
\begin{align}\label{hdefJ}
h(w,\bar w) & \eql {4i\over\sqrt{1+\lambda^2}}\big(\sinh(2w)-\sinh(2\bar w)\big)\,,
\\[6 pt]
\label{GdefJ}
G(w,\bar w)& \eql i\,{\cosh(w+\bar w)+\lambda\,\sinh(w-\bar w)\over \cosh(\bar w)}\,,
\end{align}
where $\lambda$ is a real parameter. The  explicit formulae for the metric functions simplify when written in terms of two real functions, $F_+(x,y) $ and $F_-(x,y) $, see formulae 
(3.16) and (3.17) of  \cite{D'Hoker:2009gg}.\footnote{There is a typo in (3.17), where $\cos(y)$ and $\sin(y)$ in $f_2$ and $f_3$, respectively,  should be interchanged.}

Suppose now that the two metrics \eqref{ourmetr} and \eqref{dHmet} are identical. Since the $AdS_3$ and the two three-sphere directions are unambigous, the product $f_1^2 f_2^2 f_3^2$ should be  equal to the corresponding product of the metric functions in \eqref{ourmetr}. Using (3.16) and (3.17) in \cite{D'Hoker:2009gg}, we then find
\begin{equation}\label{thesixes}
\begin{split}
 {4\over 1+\lambda^2}\,\cosh^2(2x)\sin^2(2y) & \eql 4\, e^{2A(\mu)}\sin^2(2\theta)\\
 & \eql (1-a^2)\,\cosh^2(\mu-\mu_0)\sin^2(2\theta)\,,
 \end{split}
\end{equation}
where in the second line we substituted the solution \eqref{Asol1} for $A(\mu)$.
From the factorized dependence  on the respective coordinates in both sides, it is clear that we must set
\begin{equation}\label{fstident}
x\eql {1\over 2}(\mu-\mu_0)\,,\qquad y\eql\theta\,,\qquad \lambda\eql  {\kappa_\lambda\, a\over \sqrt{1-a^2}}\,,\qquad \kappa_\lambda\eql \pm1\,.
\end{equation}
Next we compare the metric functions along $AdS_3$, where for  $\mu=\mu_0$ and $\theta=\pi/4$, after using \eqref{fstident}, we get that the following expression should vanish
\begin{equation}\label{cleveqs}
f_1^6-e^{6A}Y\widetilde{Y}\eql a^2(1-a^2)\cos^2\zeta_0\,.
\end{equation}
This  sets the initial angle, $\zeta_0$, in \eqref{zetasol1} to  $\zeta_0=\pm (\pi/2)$. Then by evaluating the left hand side in \eqref{cleveqs} for arbitrary $\mu$ and $\theta$ we obtain the  relation between the discrete parameters, 
\begin{equation}\label{discident}
\zeta_0\eql +{\pi\over 2}\,,\qquad \kappa_\lambda\eql \kappa_{\zeta}\kappa\qquad {\rm or}\qquad \zeta_0\eql -{\pi\over 2}\,,\qquad \kappa_\lambda\eql -\kappa_{\zeta}\kappa\,.
\end{equation}
Finally, using \eqref{fstident} and \eqref{discident}, we verify that
\begin{equation}\label{}
F_+(x,y)\eql Y(\mu,\theta)\,,\qquad F_-(x,y)\eql \widetilde{Y}(\mu,\theta)\,,
\end{equation}
and that all the metric functions agree. Since the metric \eqref{ourmetr} for $\zeta_0= +(\pi/2)$ and $\kappa$ is identical with the metric for $\zeta_0=-(\pi/2)$ and $-\kappa$, this shows that there is a one-to-one correspondence between the solutions in \cite{D'Hoker:2009gg} and the solutions in Section~\ref{wehavealift} with $\zeta_0=\kappa(\pi/2)$.

It appears that the parameter $c'$, discussed in detail in \cite{Estes:2012vm}, may offer a resolution to the puzzle that our Janus solutions with general values of $\zeta_0$ do not fall within the classification of \cite{D'Hoker:2008wc}. That parameter was fixed to a particular value, $c'=1$, in the analysis of \cite{D'Hoker:2008wc,D'Hoker:2009gg}, but as shown in \cite{Estes:2012vm} and a forthcoming paper \cite{BDEK}, there are also supergravity solutions with general values of $c'$.\footnote{We would like to thank the authors of \cite{D'Hoker:2008wc,Estes:2012vm,BDEK} for the correspondence clarifying this issue.} A preliminary analysis suggests that our solutions with generic values of $\zeta_0$ may indeed fit into that more general class of 1/2-BPS solutions with $c'\neq1$. It is also clear that a complete comparison will be quite involved and we defer it to future work. Here let us note that having such a match would be very interesting since it will imply that our Janus solutions with generic $\zeta_0$ are invariant under the superalgebra $D(2,1;c')\times D(2,1;c')$ which is \emph{not} a subalgebra of the $OSp(8|4)$ symmetry algebra of the ABJM theory.

\section{Other first order reductions}

\renewcommand{\theequation}{C.\arabic{equation}}
\renewcommand{\thetable}{C.\arabic{table}}
\setcounter{equation}{0}
\label{appendixC}

In this appendix we summarize truncations for which the Killing spinor, $\epsilon_j$,  lies in representations, $\cals R_\epsilon$, that were not considered in  Sections~\ref{sec:SU3} and \ref{sec:G2}. Those are:
\begin{itemize}
\item [(i)] $(\bfs 3,1,0) +(\overline{\bfs 3},-1,0)$ for $SU(3)\times U(1)^2$,
\item [(ii)] $\bfs 7$ for $G_2$.
\end{itemize}
Our main conclusion here is that  these representations of the Killing spinor do not allow for supersymmetric Janus-type solutions.

We start with (i) where  the spin-3/2 variations \eqref{deltagravitino} along $t$ and $x$  reduce to \eqref{gravvar}, but with $\cals W\eql \cals W_{3}$, where
\begin{equation}\label{W3pot}
\begin{split}
\cals W_3 & \eql {1\over 2\sqrt2 }\left[4 e^{-i \zeta } \sinh ^3(\alpha )+3 \cosh (\alpha )+\cosh (3 \alpha )\right]\\ & \eql \sqrt 2\,{(1+z\bar z^2)\over (1-|z|^2)^{-3/2}}\,.
\end{split}
\end{equation}
We note that $\cals W_3$ cannot be expressed in terms of a holomorphic superpotential, $\cals V$, as in \eqref{WVreln}, which appears to be a telltale of  trouble. Indeed, unlike in Section~\ref{sec:SU3}, the spin-1/2 variations split into {\it pairs} of equations for the  Killing spinors, $\epsilon^a$ and $\epsilon_a$, of opposite chirality:
\begin{equation}\label{v12su33}
\gamma^4\,\epsilon^a-{3\over g }{(\alpha'\pm{i\over 4}\sinh(2\alpha)\zeta')\over \partial_\alpha \cals W_3}\,\epsilon_a\eql 0\,,\qquad a=1,\ldots,6\,.
\end{equation}
This forces us to set 
\begin{equation}\label{}
\zeta'\eql 0\,.
\end{equation}
It is then straightforward to check that for a constant $\zeta=\zeta_0$, the consistency between the first order equations that follow from   supersymmetry variations and the equation of motion and the energy condition for the Lagrangian
\eqref{su3lag} 
yield the following equations for $A(\mu)$ and  $\alpha(\mu)$:
\begin{equation}\label{Asolutj}
{e^{-2A}\over\ell^2}\eql 2 g^2 \,\frac{\sinh ^4(2 \alpha ) \sin ^2 \zeta _0\left(\sinh (2 \alpha
   ) \cos  \zeta _0 +\cosh (2 \alpha )\right)}{\left(2 \sinh ^2(2 \alpha
   ) \cos \zeta _0 +\sinh (4 \alpha )\right)^2}\,,
\end{equation}
and
\begin{equation}\label{aleqsj}
(\alpha')^2\eql \frac{1}{2} g^2 \sinh ^2(2 \alpha ) \left(\sinh (2 \alpha ) \cos \left(\zeta
   _0\right)+\cosh (2 \alpha )\right)\,.
\end{equation}
It is clear that the latter equation does not admit solutions with a turning point, $\mu_0$, at a finite value of $\alpha(\mu_0)$ which rules out Janus-type solutions in this truncation.

The truncation in (ii) has a similar structure, where the sevenfold degenerate eigenvalue of $ A_{1}^{ij}$  in the spin-3/2 variations along $t$ and $x$ yields the superpotential
\begin{equation}\label{}
\begin{split}
\cals W_7 & \eql \sqrt{2}\,\Big[ e^{-i \zeta } \sinh ^7 \alpha  +e^{-i \zeta } \left(6+e^{4 i \zeta }\right) \sinh ^3 \alpha   \cosh
   ^4 \alpha \\ & \hskip 3 cm   +\left(6+e^{-4 i \zeta }\right) \sinh ^4 \alpha   \cosh ^3 \alpha +\cosh ^7 \alpha  \Big]\\
 & \eql {\sqrt{2}\over (1-|z|^2)^{7/2}}\left(1+z^3+6 z\bar z^2+6z^2\bar z^2+\bar z^4+z^3\bar z^4\right)\,,
\end{split}
\end{equation}
which  does not arise from any holomorphic superpotential. The spin-1/2 variations do not reduce to a simple expressions as in \eqref{v12su33}, and their consistency requires that
\begin{equation}\label{badeqs}
\begin{split}
8 \left(e^{4 i \zeta }-1\right) & \left(\sinh  \alpha  +e^{3 i \zeta } \cosh  \alpha \right) \,\alpha'
\\ &  \eql  
i \Big[2 e^{3 i \zeta } \cosh (2 \alpha ) \big[-2 e^{i \zeta } \cosh (\alpha ) \left(e^{3 i \zeta }
   \sinh (2 \alpha )+7 \cosh (2 \alpha )-3\right)\\ & \hskip 4 cm +\sinh (\alpha )-7 \sinh (3 \alpha )\big]+\cosh (3
   \alpha )-\cosh (5 \alpha )\Big]
\,\zeta'\,.
\end{split}
\end{equation}

By taking the real and imaginary parts of \eqref{badeqs}, we obtain a homogenous system of equations for $\alpha'$ and $\zeta'$, which has a non-zero solution provided
\begin{equation}\label{}
\left[1+\cosh(4\alpha)+\cos\zeta\sinh(4\alpha)\right]\,\cos\zeta\,\sin^2\zeta\eql 0\,.
\end{equation}
The first term is obviously non-zero, hence we must set
\begin{equation}\label{zetaconst}
\zeta\eql n\,{\pi\over 2}\,,\qquad n\in\ZZ\,.
\end{equation}

The resulting truncations  of the $\cN=8$ theory are  the $SO(7)^+$-invariant truncation for $n$ even and the $SO(7)^-$-invariant truncation for $n$ odd. In the former truncation, the first order system can be shown to be 
inconsistent with the equations of motion. In the latter the first order equations are consistent with the equations of motion but do not admit Janus-type solution with a finite turning point.

\end{appendices}



\end{document}